\documentclass[aps,pre,twocolumn,showpacs,superscriptaddress,groupedaddress,floatfix,10pt,
]{revtex4-1}
\pdfoutput=1
\usepackage[utf8]{inputenc}
\usepackage[english]{babel}
\usepackage{graphicx}
\usepackage{xcolor}
\colorlet{mylinkcolor}{blue!66!black!80}
\usepackage[hyperfootnotes=false,colorlinks=true,linkcolor=mylinkcolor,citecolor=mylinkcolor,filecolor=cyan,urlcolor=mylinkcolor,breaklinks=true]{hyperref}
\usepackage{amsmath,amssymb}
\usepackage{bm,bbm}
\usepackage{mathrsfs}
\usepackage{color}
\usepackage{mathtools}
\usepackage{amsfonts}
\usepackage{booktabs}
\usepackage{amsthm}
\usepackage{setspace}

\newcommand{\avg}[1]{\langle#1\rangle}
\newcommand{\bra}[1]{\langle#1|}
\newcommand{\ket}[1]{|#1\rangle}
\newcommand{\bx}{\mathbf{x}}
\newcommand{\by}{\mathbf{y}}
\newcommand{\bFF}{\mathbf{F}}
\newcommand{\bDD}{\mathbf{D}}
\newcommand{\dbW}{d\mathbf{W}_t}

\newcommand{\bsig}{\boldsymbol{\sigma}}

\newcommand{\bu}{\mathbf{u}}
\newcommand{\ii}{\mathrm{inv}}

\newcommand{\LL}{\hat{L}}
\newcommand{\LLb}{\hat{L}^{\dagger}}
\newcommand{\ee}[1]{\mathrm{e}^{#1}}
\newcommand{\tld}[1]{\tilde{#1}}
\newcommand{\ovl}[1]{\overline{#1}}
\newcommand{\lflat}{\langle -}
\newcommand{\rflat}{-\rangle}
\newcommand{\bpsi}{\boldsymbol{\psi}}

\newcommand{\fbrak}{\langle L_k|}
\newcommand{\fketk}{|R_k\rangle}
\newcommand{\bbrak}{\langle R_k|}
\newcommand{\bketk}{|L_k\rangle}
\newcommand{\OU}{\mathrm{OU}}
\newcommand{\W}{\mathrm{W}}
\newcommand{\eqqref}[1]{(\ref{#1})}
\newcommand{\llambda}{\lambda^{\dagger}}
\newcommand{\tPv}{\tilde{\mathcal{P}}^{\overline{V}}_t}
\newcommand{\tPpsi}{\tilde{\mathcal{P}}^{\psi}_t}
\newcommand{\tPvVec}{\tilde{\mathcal{P}}^{\overline{\mathbf{V}}}_t}
\newcommand{\tPpsiVec}{\tilde{\mathcal{P}}^{\boldsymbol{\psi}}_t}

\newcommand{\tPpsiVecLaplace}{\tilde{\mathcal{P}}^{\boldsymbol{\psi}}_s}

\begin{document}
\title{Spectral theory of fluctuations in time-average statistical
  mechanics of reversible and driven systems}
\author{Alessio Lapolla}
\author{David Hartich}
\author{Alja\v{z} Godec}
\email{agodec@mpibpc.mpg.de}
\affiliation{Mathematical bioPhysics group, Max Planck Institute for
  Biophysical Chemistry, G\"{o}ttingen 37077, Germany}

\begin{abstract}
We present a spectral-theoretic approach to time-average
statistical mechanics for general, non-equilibrium initial conditions.
We consider the statistics of bounded, local 
additive functionals of reversible as well as irreversible ergodic stochastic dynamics with continuous or discrete state-space.
We derive exact results for the mean, fluctuations and correlations of
time average observables from the
eigenspectrum of the underlying generator of Fokker-Planck or master equation dynamics, and discuss the results from a physical perspective. Feynman-Kac formulas are re-derived using
It\^o calculus and combined with non-Hermitian perturbation
theory. The emergence of the universal central limit law in a spectral
representation is shown explicitly on large deviation time-scales. 
For reversible dynamics with equilibrated initial conditions
we derive a general upper bound to fluctuations of occupation measures
in terms of an
integral of the return probability. Simple,
exactly solvable examples are analyzed to demonstrate how to
apply the theory. As a biophysical example we revisit the Berg-Purcell problem on the precision of concentration
measurements by a single receptor.
Our results are directly applicable
to a diverse range of phenomena underpinned by time-average
observables and additive functionals in physical, chemical,
biological, and economical systems. 
\end{abstract}
\maketitle

\section{Introduction}\label{sec0}
Many experiments on soft and biological matter, such as single-particle
tracking \cite{SPT,SPT2,SPT3,SPT4} and single-molecule spectroscopy \cite{SMS,SMS2,SMS3,SMS4,SMS5,SMS6,SMS7,SMS8,SMS9}, probe individual
trajectories. It is typically not feasible to repeat these experiments
sufficiently many times in order to allow for ensemble-averaging.
It is, however, straightforward to analyze such data by means of
time-averaging along individual realizations. However, except for
(ergodically) long observations, time-averages inferred from individual
trajectories are random with non-trivial statistics.
This naturally leads to the study of statistical properties of time-averages which formally represent functionals of stochastic processes.

The study of  functionals of stochastic processes has a long tradition
in mathematics (see
e.g. \cite{Levy,Kac,Darling,Lamperti,Feller,Bingham,Borodin,Yor}) and
finance \cite{yor_exponential_2001, geman_bessel_1993}.
In physics they were found to be relevant in the context of
diffusion-controlled chemical reactions (e.g. \cite{WFix,Szabo,Gleb}), transport
in porous media \cite{porous}, chemical inference  \cite{berg77,Weigel,Bialek_2005,endr09,mora10,lang14,mora15,bara15a,Aquino,hart16a}, astrophysical observations \cite{astro},
medical diagnostics \cite{diag}, optical imaging \cite{optical},
the study of growing surfaces \cite{growing},
blinking of colloidal quantum dots \cite{blinking,blinking2},
mesoscopic physics \cite{comtet_functionals_2005},
climate \cite{majumdar_large-deviation_2002} and computer
 science \cite{majumdar_satya_n._brownian_2005}, and most recently in
single-molecule spectroscopy \cite{SM1,SM2,Zumofen,Eli} and diffusion studies \cite{SMD}, to
name a few.

From a theoretical point of view analytical results were obtained for the occupation time statistics for discrete-
state Markov switching \cite{SM1,SM2,Zumofen,Eli}, for the local time at zero and occupation time above zero of a Brownian
particle diffusing in a simple one-dimensional potential \cite{majumdar_satya_n._brownian_2005,sabhapandit_statistical_2006,majumdar_local_2002}, the occupation time inside a spherical
domain of a Brownian particle moving in free space \cite{SMD} and for a free, uniformly biased and harmonically
bound particle undergoing subdiffusion \cite{Bel,Carmi}. Exact results were also obtained for occupation time statistics
for a general class of Markov processes \cite{dhar_residence_1999} and
a discrete stationary non-Markovian sequence \cite{Dean}. Large deviation functions for various non-linear functionals of a
class of Gaussian stationary Markov processes were studied in
\cite{majumdar_large-deviation_2002}. Numerous important results on
functionals have also been obtained in the context of persistence in
spatially extended non-equilibrium systems \cite{Bray_2013}. Exact
results were recently obtained on local times for projected
observables in stochastic many-body systems \cite{lapolla_unfolding_2018,Lapolla_2019}, which provided
insight into the emergence of memory on the level of individual non-Markovian
trajectories. Notwithstanding, a general
approach to fluctuations in time-average statistical mechanics for arbitrary
initial conditions remains elusive. 

Here, we present a spectral-theoretic approach to finite time-average
statistical mechanics of ergodic systems. In mathematical terms we
focus on the statistics of bounded, local, additive functionals of normal ergodic Markovian stochastic
processes with continuous and discrete state-spaces, incl. functionals
of their (non-Markovian) lower-dimensional projections. The paper is
organized as follows. We first provide in Sec.~\ref{Sec1} a brief introduction into
time-average statistical mechanics. In Sec.~\ref{Sec2} we re-derive the well-known
Feynman-Kac formulas for Markovian diffusion using It\^o calculus. In
Sec.~\ref{expectation derivatives} spectral theory combined with non-Hermitian perturbation
theory is applied to obtain our main result -- exact expressions for the mean, variance and
correlations of time-average observables for any non-stationary
preparation of the system, expressed explicitly in terms
of  the eigenspectrum of the underlying
generator of the dynamics, which may correspond to Fokker-Planck
diffusion or Markovian dynamics governed by a master equation.
We demonstrate explicitly the emergence of a central limit law in a
spectral representation on large deviation time-scales. In
Sec.~\ref{subsec:upper_bound} we derive our second main result -- a general upper bound on fluctuations
of occupation measures in terms of an integral of the, generally non-Markovian, return
probability that is valid for generators of overdamped dynamics
obeying detailed balance. Finally, simple analytically solvable examples are
provided in Sec.~\ref{Sec4} to demonstrate how to apply the theory. We
conclude in Sec.~\ref{Sec5}.


\section{Time-average statistical mechanics}
\label{Sec1}
\subsection{Ensemble- versus time-average observables}
Traditional (classical) ensemble statistical mechanics describes
physical observations as averages over individual realizations of the
dynamics at single (or multiple) pre-determined times. For example,
the ensemble average of an observable $V(\bx_t)$ at a time $t$ for an ergodic
stochastic process $\bx_\tau$ ($0\le\tau\le t$) starting from some non-stationary initial condition $p_0(\bx_0)$ is
defined by
\begin{equation}
\langle V (\bx_t) \rangle_{p_0}\!\equiv\!
\left\{\begin{array}{@{}ll@{}}
\displaystyle \int_\Omega \!\!d\bx\!\!\int_\Omega \!\!d\bx_0 V(\bx) P_t(\bx|\bx_0)p_0(\bx_0)&\text{$\Omega\subset\mathbb{R}^d$,}\\
\displaystyle \sum_{\bx\in\Omega} \sum_{\bx_0\in\Omega}  V(\bx) P_t(\bx|\bx_0)p_0(\bx_0)&\text{$\Omega$ discrete,}
\end{array}
\right.
  \label{ens}
\end{equation}  
where $\Omega$ is the state space of the process and $P_t(\bx|\bx_0)$ is the so-called propagator, i.e. $P_t(\bx|\bx_0)d\bx$ (upper line)
is the
probability that the process is found in $\bx\in\Omega$ within the increment $d \bx$ at time $t$
given that it started at $t=0$ at $\bx_0$.
 \begin{figure}\centering
    \includegraphics[width=0.46\textwidth]{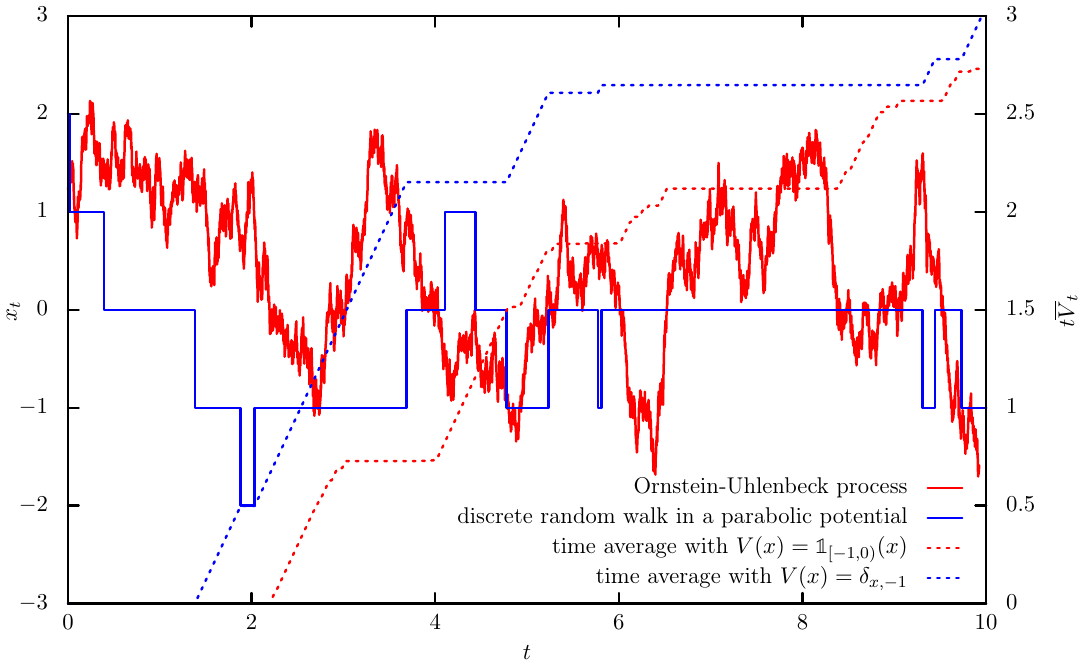}
\caption{Realization of a trajectory $\mathbf{x}_t$ (solid lines) of a continuous
Ornstein-Uhlenbeck diffusion (red) and a Markovian discrete-space continuous time random walk in a quadratic potential
(blue). The dotted lines refer to the time average, Eq.~\eqqref{TA}, with $t\overline{V}_t=\int_0^t V(x_\tau)d\tau $, where we chose $V(x)=\mathbbm{1}_{[0,1)}(x)$ for the Ornstein-Uhlenbeck and $V(x)=\delta_{-1,x}$ for the discrete random walk.}
  \label{trajs}
 \end{figure}
Note that if $\bx$ is continuously valued
($\bx\in\Omega\subset\mathbb{R}^d$; see Fig.~\ref{trajs} solid red line) then $
P_t(\bx|\bx_0)$ is a probability \emph{density}, whereas if $\bx$ is
from a discrete state space $\Omega$ (Fig.~\ref{trajs} solid blue line)
the integral in Eq.~\eqqref{ens} (upper line) becomes a sum $\int_\Omega d\bx\to\sum_{\bx\in\Omega}$
and $P_t(\bx|\bx_0)$ becomes a plain probability as shown in the lower line of Eq.~\eqqref{ens}.
If $\bx_0$ is sampled and
averaged over a stationary (invariant) measure $p(\bx_0)=P_{\ii}(\bx_0)$ or $t$
becomes sufficiently (i.e. ergodically) large $P_\infty(\bx|\bx_0)\equiv P_{\ii}(\bx)$ the ensemble average
becomes time-independent
\begin{equation}
\langle V\rangle_{\ii}\equiv \int d\bx V(\bx) P_\ii(\bx).
\label{reg}
\end{equation}

Conversely, in single-molecule dynamics, single-particle tracking and
other related experiments one probes individual realizations of
$\bx_\tau$ within the interval $0\le\tau\le t$ and instead analyzes the observation by
taking a time-average. Such time-average observables are in general
random, fluctuating quantities with non-trivial statistics. For
example, for a physical observable $V(\bx_t)$, which may correspond
to the squared displacement \cite{Gleb1,Gleb2} or local time
\cite{SMD,lapolla_unfolding_2018,Lapolla_2019,SPT_T} in single particle
tracking or the FRET efficiency \cite{SMS,SMS2,SMS3,SMS4} or distance between two
optical traps \cite{SMS5,SMS6,SMS7,SMS8,SMS9} in single-molecule
fluorescence and force spectroscopy, respectively, 
the (local) time-average is defined as
\begin{equation}
  \overline{V}_t\equiv t^{-1}\int_0^tV(\bx_\tau)d\tau
\label{TA}
\end{equation}
and depends on the entire history of $\bx_\tau$ until time $t$ (see also dotted lines in Fig.~\ref{trajs}). The
statistical evolution of $\overline{V}_t$ is therefore a
non-Markovian process characterized by the probability density that the
random observable
$\overline{V}_t$ attains, in a given realization of $\mathbf{x}_{\tau}$,
the value $\nu$
\cite{majumdar_satya_n._brownian_2005,sabhapandit_statistical_2006,majumdar_local_2002,lapolla_unfolding_2018,Lapolla_2019}
which is defined as
\begin{equation}
\mathcal{P}^{\ovl{V}}_t(\nu|\bx_0)\equiv \langle \delta(\nu-\ovl{V}_t) \rangle_{\bx_0},
\label{PA}  
\end{equation}
where $\delta(z)$ is the Dirac delta function and $\langle \cdot\rangle_{\bx_0}$ denotes the average over all
paths starting at $\bx_0$, i.e. $p_0(\bx)=\delta(\bx-\bx_0)$,  and
propagating until time $t$.
The corresponding result
  for arbitrary initial conditions $p_0(\bx_0)$, follows by
  superposition, i.e. $\mathcal{P}^{\ovl{V}}_t(\nu|p_0)\equiv\int_\Omega\mathcal{P}^{\ovl{V}}_t(\nu|\bx_0)p_0(\bx_0)d\bx_0$ (see also Sec.~\ref{subsec:backward_Feynman-Kac}).
  

The random ``empirical density'' \cite{bara15c}
$\theta_{\bx}(t)$ 
determined from a single trajectory in time-average statistical
mechanics is the so-called \emph{local time fraction}
defined as  \cite{Yor,lapolla_unfolding_2018,Lapolla_2019}
\begin{equation}
\theta_{\bx}(t)\equiv t^{-1}\int_0^td\tau\delta(\bx-\bx_{\tau}) ,
\label{ltf}  
\end{equation}
which allows to rewrite the time average \eqqref{TA} in the form
\begin{equation}
\overline{V}_t=t^{-1}\!\!\int_0^t\!d\tau\!\!\int_{\Omega}\!d\bx\delta(\bx-\bx_{\tau})V(\bx)\equiv \!\!\int_{\Omega}\!d\bx V(\bx)\theta_{\bx}(t),
\label{ltf2}  
\end{equation}
where $\delta(\bx-\bx_{\tau})$ denotes the Dirac delta function if $\bx\in \Omega$ is continuous,
whereas $\delta(\bx-\bx_{\tau})$ denotes the Kronecker delta if
$\bx\in \Omega$ is integer-valued. Note that it is often useful to generalize the
local time fraction in a point $\bx$ in Eq.~(\ref{ltf}) to the notion of
\emph{occupation time} within the
hypersurface $V(\bx)=\mathcal{V}$ defined as
\begin{equation}
\theta_{\mathcal{V}}(t)\equiv
t^{-1}\int_0^t\delta(\mathcal{V}-V(\bx_{\tau}))d\tau.
\label{localH}
\end{equation}
Accordingly, we can rewrite Eq.~(\ref{ltf2}) equivalently in terms
of $\theta_{\mathcal{V}}(t)$ as
\begin{equation}
\overline{V}_t=t^{-1}\!\!\int_0^t\!d\tau\!\!
\int\!d\mathcal{V}\delta(\mathcal{V}-V(\bx_{\tau}))\mathcal{V}\equiv
\int d\mathcal{V}\mathcal{V}\theta_{\mathcal{V}}(t).
\label{ltf3}  
\end{equation}
Because the dynamics $\bx_t$ is assumed to be ergodic we have
$\lim_{t\to\infty}\theta_{\bx}(t)=P_\ii(\bx)$ and $\lim_{t\to\infty}\theta_{\mathcal{V}}(t)=P_\ii(\mathcal{V})$
\cite{sabhapandit_statistical_2006,lapolla_unfolding_2018,Lapolla_2019} and
\begin{eqnarray}
&\lim_{t\to\infty}&\ovl{V}_t = \!\!\int \!\!d\bx V(\bx)
P_\ii(\bx)= \!\!\int\!\! d\mathcal{V} \mathcal{V}P_\ii(\mathcal{V}) \equiv  \langle V\rangle_{\ii},
\nonumber\\
&\lim_{t\to\infty}&\mathcal{P}^{\ovl{V}}_t(\nu|\bx_0)=\delta(\nu-\langle
V\rangle_{\ii}),
\label{Birkhof}
\end{eqnarray}
where we have defined the \emph{stationary (or invariant) measure of} $\ovl{V}(\bx_t)$,
i.e. $P_\ii(\mathcal{V})\equiv \int d\bx
\delta(\mathcal{V}-V(\bx_{\tau}))P_\ii(\bx)$. 

Eqs.~(\ref{Birkhof}) reflect the strong law of large numbers on
time-scales where $ \overline{V}_t$ for different values of $t$ decorrelate.  
Moreover, on the so-called large-deviation time-scale, i.e. on the
time-scale that is finite but longer that the longest relaxation time
of $\bx_t$, we find convergence in the mean, $\langle \overline{V}
_{t_{\mathrm{LD}}}\rangle=\langle V\rangle_{\ii}$, and Gaussian fluctuations around
the mean value \cite{majumdar_large-deviation_2002,LD,LD2,lapolla_unfolding_2018,Lapolla_2019}. 
For finite, and in particular sub-ergodic  (i.e. supra-large
deviation) times the statistics of $\ovl{V}_t$ is, however,
non-trivial. Below we provide intuition about the local time fraction
from a practical perspective.

\subsection{Local time fraction as a histogram inferred from a single trajectory}


\begin{figure}\centering
    \includegraphics[width=\columnwidth]{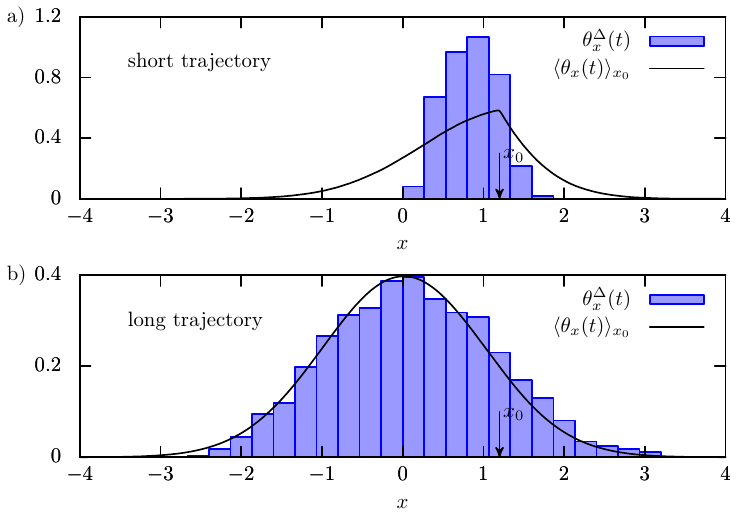}
    \caption{
     Local time fraction (or ``empirical density'')
    for a Brownian particle in a harmonic potential.
  The histogram
  $\theta_x^{\Delta}(t)$  (blue) inferred from a single trajectory starting from $x_0=1.2$
  compared the mean local time fraction
  $\langle\theta_y(t)\rangle_{x_0}=\lim_{\Delta\to 0}\langle\theta^{\Delta}_y(t)\rangle_{x_0}$.
  The histogram with bin-size $\Delta$ (here we assume $\Delta =0.3$)
  is defined by
  $\theta^{\Delta}_x(t)=\Delta^{-1}\int_{x-\Delta/2}^{x+\Delta/2}\theta_y(t)dy$,
  that is, $\theta_x^{\Delta}(t)=\overline{V}_t$ with
  $V(x)\equiv\Delta^{-1}\mathbbm{1}_{[x-\Delta/2,x+\Delta/2]}(x)$, where
  ``$\mathbbm{1}$'' denotes the indicator function being 1 if
  $x\in[x-\Delta/2, x+\Delta/2]$ and 0 otherwise. Parameters:
  $x_0=1.2$, (a) $t=1$; (b) $t=100$ (the trajectory is ten times as long as in Fig.~\ref{trajs}). 
}
  \label{histo}
\end{figure}
  To gain more intuition about the local time fraction (or ``empirical
  density'') we consider, as an example, a Brownian particle
  diffusing in a harmonic potential. 
  A single trajectory starting from $x_0=1.2$  is recorded as function
  of time (see full red line in Fig.~\ref{trajs}). We are interested
  in the distribution (i.e. a \emph{histogram}) of the particle's position $x_t$ \emph{inferred from a
    single} trajectory of length $t$ (see Fig.~\ref{histo}). Note
  that in Fig.~\ref{histo} we consider a histogram with a finite
  bin-size $\Delta$, which we denote
  explicitly as $\theta_x^{\Delta}(t)$. In this sense the local
  time fraction (\ref{ltf}) is simply a mathematical idealization
  of a histogram, i.e. $\theta_x(t)=\theta_x^{\Delta\to 0}(t)$.
  
  In case the trajectory is
  sufficiently long (i.e. $t\to\infty$) $\theta_x^{\Delta}(t)$
  converges, up to small fluctuations of order $1/\sqrt{t}$,
  to a Gaussian stationary (invariant)
  measure $P_\ii(x)\propto \ee{-x^2/2}$ (see
  Eq.~(\ref{Birkhof})). This convergence is
  depicted explicitly in Fig.~\ref{histo}b. According
  to Eq.~(\ref{Birkhof}) this result depends neither on the
  initial condition $x_0$ nor on the particular realization of the trajectory. 

We may also infer a histogram of the particle's position from a short
trajectory. The resulting histogram  $\theta_x^{\Delta}(t)$ appears
``rough" and far from Gaussian (see histogram in Fig.~\ref{histo}a). If we were to
repeat the analysis for many trajectories and infer an \emph{averaged}
histogram $\langle \theta_x^{\Delta}(t)\rangle$ we would as well find
that it deviates strongly from a Gaussian (see line in
Fig.~\ref{histo}a) with large fluctuations around the
mean, $\delta \theta_x^{\Delta}(t)\equiv |\theta_x^{\Delta}(t)-\langle
\theta_x^{\Delta}(t)\rangle|\sim\langle
\theta_x^{\Delta}(t)\rangle$ (see histogram in
Fig.~\ref{histo}a) .  Moreover, both the mean histogram $\langle
\theta_x^{\Delta}(t)\rangle$ and the fluctuations around the mean, $\delta\theta_x^{\Delta}(t)$, depend not only of $t$ but also on the
initial position $x_0$, where  we observe a persistent cusp
(Fig.~\ref{histo}a).

In the reminder of this work we will focus on the
mean values and fluctuations of entries, as well
as linear correlations between entries of such (random,
realization-dependent) histograms inferred from finite, individual
trajectories starting from general initial conditions.

%
%
  
 \subsection{Fluctuations of time averages}
In order to exploit the role of the local time fraction
$\theta_{\bx}(t)$ as a ``propagator'' in time-average statistical
mechanics (via $\ovl{V}_t=\int_{\Omega} d\bx V(\bx)\theta_{\bx}(t)$) we now relate the statistics of $\theta_{\bx}(t)$ to the
probability density of a general time-averaged
physical observable $\ovl{V}_t$ defined in Eq.~(\ref{TA}). In the
example in Fig.~\ref{histo}
$\ovl{V}_t=\theta^{\Delta}_x$ accounts for the value of the histogram
at position $x$. The characteristic function of $\ovl{V}_t$,
i.e. the Laplace transform $\tPv(u|\bx_0)\equiv \int_0^{\infty}\ee{-u\nu}\mathcal{P}^{\ovl{V}}_t(\nu|\bx_0)d\nu$ if $\nu\ge 0$, reads
\begin{equation}
\tPv(u|\bx_0)
=\langle\ee{-u\ovl{V}_t}\rangle_{\bx_0}
=\langle\ee{-u\int_\Omega V(\bx)\theta_\bx(t)d \bx}\rangle_{\bx_0},
  \label{cf}
\end{equation}
where we have used Eq.~(\ref{ltf2}) to arrive at the second equality. Eq.~\eqqref{cf} relates the statistics of the time average $\ovl{V}_t$ to the statistics of all local time fractions $\theta_\bx(t)$ ($\bx\in \Omega$).
We note that Eq.~\eqqref{cf} must be modified if $V(\bx)$ can also
become negative such that a Fourier transform is required instead, which amounts
to replacing $u\to{\rm i}\omega$ with $\omega\in\mathbb{R}$ in Eq.~\eqqref{cf}.
The probability density is obtained from Eq.~\eqqref{cf}
by Laplace inversion 
\begin{equation}
\mathcal{P}^{\ovl{V}}_t(\nu|\bx_0)=\frac{1}{2\pi
  {\rm i}}\int_{c-{\rm i}\infty}^{c+{\rm i}\infty}\ee{u\nu}\tPv(u|\bx_0)du
\label{relate}
\end{equation}
where $c\in\mathbb{R}$ lies to the right of all singularities of
$\tPv(u|\bx_0)$ and we have assumed that $\mathcal{P}^{\ovl{V}}_t(\nu|\bx_0)$ is
of exponential order for sufficiently large $\nu$ (in the
following section the conditions on $V(\bx_{\tau})$ will be made more
precise). In case the support extends to negative values of Eq.~(\ref{relate}) becomes the
inverse Fourier transform.
Here we are particularly interested in fluctuations of time averages
$ \ovl{V}_t$ and $ \ovl{W}_t$ of different physical
observables $V(\bx_\tau)$ and $W(\bx_\tau)$
which are quantified by \cite{lapolla_unfolding_2018,Lapolla_2019}
\begin{eqnarray}
\sigma^2_{\ovl{V}}(t)&\equiv& \langle \ovl{V}_t^2\rangle_{\bx_0}- \langle
\ovl{V}_t\rangle_{\bx_0}^2,\nonumber\\
C_{\ovl{V}\ovl{W}}(t)&\equiv& \langle \ovl{V}_t\ovl{W}_t\rangle_{\bx_0}-\langle \ovl{V}_t\rangle_{\bx_0}\langle\ovl{W}_t\rangle_{\bx_0},
\label{var_cov}
\end{eqnarray}
where $\sigma^2_{V}(t)=C_{\ovl{VV}}(t)$ denotes the variance of $ \ovl{V}_t$
and $C_{\ovl{V}\ovl{W}}(t)$ the covariance between $ \ovl{V}_t$ and $ \ovl{W}_t$.

In the example in Fig.~\ref{histo}, where
$\ovl{V}_t=\theta_x^{\Delta}(t)$ corresponds to the entry $x$ in the histogram, $\sigma^2_{\ovl{V}}(t)\equiv \langle
\delta \theta_x^{\Delta}(t)^2\rangle$ refers to the
variance of said entries between different realizations of the histogram (i.e. the scatter of $\theta_x^{\Delta}(t)$
around $\langle \theta_x(t)\rangle$). 
Analogously,
$C_{\ovl{V}\ovl{W}}(t)$ accounts for linear correlations
between pairs of entries at $x$ and $y$, $\theta_x^{\Delta}(t)$ and $\theta_y^{\Delta}(t)$, in a histogram inferred from a
single trajectory of length $t$.

Using Eq.~(\ref{cf}) we obtain more generally
\begin{multline}
\langle \ovl{V}_t^n\ovl{W}_t^m \rangle_{\bx_0}= \prod_{i=1}^n\int_\Omega d\bx_iV(\bx_i)\prod_{j=1}^{m}\int_\Omega
d\by_{j}W(\by_j)\\\times
\langle \theta_{\bx_1}(t)\cdots\theta_{\bx_{n}}(t)  \theta_{\by_1}(t)\cdots\theta_{\by_{m}}(t) \rangle_{\bx_0},
\label{moments}
\end{multline}
where $\bx_i,\by_j\in \Omega$, respectively. Thereby, the ensemble average 
corresponding to the last term in Eq.~\eqqref{moments} is obtained by
differentiating the characteristic function with respect to the Laplace
(or Fourier) variable.

It therefore follows, that the fluctuations
and (linear) correlations of general time-average observables are
fully specified by multi-point correlations functions of the local
time fraction. These are derived on the basis of the Feynman-Kac
formalism, which is presented below for continuous diffusion
processes. The extension to discrete state dynamics is discussed
afterwards.\vspace{0.2cm}

\section{Fluctuations of additive functionals}
\label{Sec2}

\subsection{It\^o approach to Feynman-Kac theory of additive
  functionals}
\label{Sec2F}
It seems to be customary in the physics literature to start from a
path integral approach to Feynman-Kac theory
\cite{majumdar_satya_n._brownian_2005} and in the following to derive a backward
Fokker-Planck equation for the characteristic function
\cite{sabhapandit_statistical_2006,majumdar_local_2002}. 
Here we provide a simple derivation of the ``forward'' Feynman-Kac theory based on
It\^o calculus. 

We consider a $d$-dimensional Markovian diffusion
$\bx_t\in\Omega\subset \mathbb{R}^d$ process in
the presence of a drift
$\bFF(\bx)$ driven by a $d$-dimensional
Gaussian white noise governed by the It\^o equation
\begin{equation}
  d\bx_t=\bFF(\bx_t)dt +\bsig d\mathbf{W}_t,
\label{itoe}  
\end{equation}
where $\bsig$ is a $d\times d$ noise matrix such that $\bDD=\bsig\bsig^T/2$
becomes a symmetric positive (semi)definite diffusion matrix,
$d\mathbf{W}_t$ is an increment of a $d$-dimensional Wiener process,
such that $\avg{{W}_t}=\mathbf{0}$ and
$\avg{dW_{t,i}dW_{t',j}}=\delta(t-t')\delta_{ij}dt$. We assume
throughout that
$\bFF(\bx)$ is sufficiently confining to assure that the process $\bx_t$ is
ergodic with a steady state probability density $P_{\ii}(\bx)$.
Multiplying the time average in Eq.~\eqqref{TA} by the trajectory
length $t$ we transform the time average to the additive functional
\begin{equation}
\psi_t\equiv  t\ovl{V}_t=\int_0^tV(\bx_{\tau})d\tau.
\label{funct}
\end{equation}
We now consider the joint process of $\bx_t$ and $\psi_t$. According to It\^o's lemma
any twice differentiable function $f(\bx,\psi)$ with $\bx_t$ and
$\psi_t$ defined by Eqs.~\eqqref{itoe} and \eqqref{funct}, respectively,
satisfies
\begin{multline}
  df(\bx_t,\psi_t)=[\bFF(\bx_t)\cdot\nabla_{\bx}f(\bx_t,\psi_t)+ \nabla_{\bx}\cdot \bDD\nabla_{\bx}f(\bx_t,\psi_t)]dt\\
  +\nabla_{\bx}f(\bx_t,\psi_t)\cdot\bsig\dbW+V(\bx_t)\partial_{\psi}f(\bx_t,\psi_t)d t ,
\label{ILemma} 
\end{multline}
where we inserted the diffusion matrix $\bDD=\bsig\bsig^T/2$ and for the last term we used $d\psi_t=V(\bx_t)d t$ that follows from Eq.~\eqqref{funct}.

Using It\^o's lemma \eqqref{ILemma} we derive in the following the
time evolution of the joint probability density $Q_t(\bx,\psi|\bx_0)$
to find the system in state $\bx$ and the functional $\psi_t$ in Eq.~(\ref{funct}) to
attain the 
value $\psi$ at time $t$ given that the system started from $\bx_0$. 
For convenience we first focus on positive functionals ($\psi\ge0$).
Using a test function that vanishes at the boundary $f(\bx,0)=0$
we obtain after some calculations \cite{gardiner_c.w._handbook_1985}
\begin{widetext}
\begin{eqnarray}
 \frac{d}{d
   t}\avg{f(x_t,\psi_t)}_{\bx_0}&=& \int_0^\infty d\psi\int_\Omega d \bx f(\bx,\psi)\partial_t
 Q_t(\bx,\psi|\bx_0) \nonumber\\
 &=& \int_0^\infty d\psi\int_\Omega d \bx Q_t(\bx,\psi|\bx_0)[\bFF(\bx)\cdot\nabla_{\bx}f(\bx,\psi)+ \nabla_{\bx}\cdot \bDD\nabla_{\bx}f(\bx,\psi)+V(\bx)\partial_{\psi}f(\bx,\psi)]\nonumber\\
 &=&  \int_0^\infty d\psi\int_\Omega d\bx f(\bx,\psi) [-\nabla_{\bx}\cdot\bFF(\bx)+ \nabla_{\bx}\cdot \bDD\nabla_{\bx}-V(\bx)\partial_{\psi}] Q_t(\bx,\psi|\bx_0)
 \label{almost_new}
\end{eqnarray}
\end{widetext}
which is obtained as follows. In the first line of Eq.~\eqqref{almost_new}
we differentiate both sides of the identity  $\int_0^\infty d\bx\int d\psi f(\bx,\psi)
Q_t(\bx,\psi|\bx_0)= \avg{f(\bx_t,\psi_t)}_{\bx_0}$ with respect to time
$t$. To obtain the second line we inserted
It\^o's Lemma \eqqref{ILemma} and finally performed an integration by parts.
Since Eq.~\eqqref{almost_new} holds for any function $f$ that vanishes at the boundary $\psi=0$ we obtain
\begin{equation}
\partial_tQ_t(\bx,\psi|\bx_0)=\left[\LL-V(\bx)\partial_{\psi}-V(\bx)\delta(\psi)\right]Q_t(\bx,\psi|\bx_0),
\label{FKac1}
\end{equation}
where we have defined the forward generator $\LL=\nabla_{\bx}\cdot
\bDD\nabla_{\bx}-\nabla_{\bx}\cdot\bFF(\bx)$ and further introduced a boundary term $V(\bx)\delta(\psi)Q_t(\bx,0|\bx_0)$ that vanishes for $\psi>0$ and is derived in the following two steps.
First, Eq.~\eqqref{almost_new} holds for all functions $f(\bx,\psi)$ with $f(\bx,0)=0$,
which immediately gives Eq.~\eqqref{FKac1} without the last term for
$\psi>0$ (see also \cite{gardiner_c.w._handbook_1985}). Second,  the
last term in Eq.~\eqqref{FKac1} is required to guarantee
the conservation of probability
$\int_0^\infty d\psi\int_\Omega d\bx\partial_t Q_t(\bx,\psi|\bx_0)=0$,
i.e., to correct for the fact that there is a non-zero probability
that the functional has a vanishing value $\psi=0$.
Finally, performing a Laplace transform of Eq.~\eqqref{FKac1}, 
$\tld{Q}_t(\bx,u|\bx_0)\equiv\int_0^{\infty}\ee{-u\psi}Q_t(\bx,\psi|\bx_0)d\psi=blue\avg{\delta(\bx-\bx_t)\ee{-u\psi_t}}_{\bx_0}$, 
we obtain the forward
Feynman-Kac partial differential equation for the characteristic
function of the joint density of position and $\psi$
\begin{equation}
\partial_t\tld{Q}_t(\bx,u|\bx_0)=(\LL-uV(\bx))\tld{Q}_t(\bx,u|\bx_0),
\label{FKac}
\end{equation}
where $Q_t(\bx,\psi|\bx_0)$ is the central object of
the 'forward' Feynman-Kac approach
\cite{Kac,majumdar_satya_n._brownian_2005}. 

We now relax the assumption by allowing for a \emph{negative support of}
$\psi_t$, which we denote explicitly by $\psi_t\to\Psi_t$. In this case we need not to make any additional assumptions
on $f(\bx_t,\Psi_t)$ for $\Psi_t=0$ because naturally
$\lim_{|\Psi|\to\infty}Q_t(\bx,\Psi|\bx_0)=0$. The lower boundary of integration over
$\Psi$ in Eq.~(\ref{almost_new}) is extended to
$-\infty$ and the boundary terms resulting from the partial
integration vanish as a result of the boundary conditions. The resulting Eq.~(\ref{almost_new})
implies
\begin{equation}
\partial_tQ_t(\bx,\Psi|\bx_0)=\left(\LL-V(\bx)\partial_{\Psi}\right)Q_t(\bx,\Psi|\bx_0),
\label{FKac2}
\end{equation}
which upon taking a Fourier transform
$\tld{Q}_t(\bx,\omega|\bx_0)\equiv\int_{-\infty}^{\infty}\ee{-{\rm i}\omega\Psi}Q_t(\bx,\Psi|\bx_0)d\Psi$
leads to the forward Feynman-Kac partial differential equation
\begin{equation}
\partial_t\tld{Q}_t(\bx,\omega|\bx_0)=(\LL-{\rm i}\omega V(\bx))\tld{Q}_t(\bx,\omega|\bx_0),
\label{FKacF}
\end{equation}
Note that the generalization to the case of a joint problem for multiple functionals $\psi^i_t$ with
$i=1,\ldots,p$ is straightforward. Introducing the vectorial notation
$\boldsymbol{\psi}_t\equiv\int_0^t\mathbf{V}(\bx_\tau)d\tau$ with the
corresponding Laplace images $\mathbf{u}$ the resulting equation reads
\begin{equation}
\partial_t\tld{Q}_t(\bx,\mathbf{u}|\bx_0)=(\LL-\mathbf{u}\cdot\mathbf{V}(\bx))\tld{Q}_t(\bx,\mathbf{u}|\bx_0),
\label{FKacFT}
\end{equation}
which allows for the computation of higher-order correlation functions (\ref{moments}).
What we actually seek for is the marginal probability density of
$\bpsi$ given $\bx_0$
defined by
\begin{equation}
\mathcal{P}^{\bpsi}_t(\bpsi|\bx_0)\equiv\int_{\Omega} Q_t(\bx,\bpsi|\bx_0)d\bx,
\label{marg}
\end{equation}
which is the statement of the Feynman-Kac theorem \cite{Kac}. 
Note that the corresponding characteristic function $\tPpsiVec(\mathbf{u}|\bx_0)=\int_\Omega \ee{-\mathbf{u}\cdot\bpsi}\mathcal{P}^{\bpsi}_t(\bpsi|\bx_0) d\bpsi\equiv\avg{\ee{-\mathbf{u}\cdot\bpsi_t}}_{\bx_0}$ is the solution of the 'backward'
Feynman-Kac problem \cite{sabhapandit_statistical_2006,majumdar_local_2002}. Moreover,   
the marginal probability density of $\bx$ given $\bx_0$ corresponds to the
plain propagator $P_t(\bx|\bx_0)\equiv\int
Q_t(\bx,\bpsi|\bx_0)d\bpsi$, which solves the (forward) Fokker-Planck
equation $(\partial_t-\LL)P_t(\bx|\bx_0)=0$ with initial data
$P_0(\bx|\bx_0)=\delta(\bx-\bx_0)$.

Note that the characteristic functions $\tPvVec$ and
$\tPpsiVec$ are equivalent up to a trivial rescaling of the independent variable, i.e. 
\begin{equation}
\tPvVec(\mathbf{u}|\bx_0)=\avg{\ee{-\mathbf{u}\cdot\ovl{\mathbf{V}}_t}}=\avg{\ee{-\mathbf{u}/t\cdot\bpsi_t}}=\tPpsiVec(\mathbf{u}/t|\bx_0).
\label{trivial}
\end{equation}
Therefore,
once $\tPpsiVec$ is determined according to the Feynman-Kac program a
simple change of scale of the Laplace image $\mathbf{u}\to\mathbf{u}/t$ delivers $\tPvVec$.

\subsection{From the forward to the backward Feynman-Kac equation}
\label{subsec:backward_Feynman-Kac}
For convenience we henceforth adopt the bra-ket notation,
where the ``ket'' $|h\rangle$ denotes a vector, the ``bra'' the
integral operator $\langle g|\equiv \int_{\Omega}d\bx
g^{\dagger}(\bx)$, and the scalar product is defined as $\langle
g|h\rangle\equiv \int_{\Omega}d\bx
g^{\dagger}(\bx)h(\bx)$. Introducing, moreover, the ``flat'' state
$|\rflat\equiv \int_{\Omega}d\bx|\bx\rangle$ and $\lflat|\equiv\int_{\Omega}d\bx\langle\bx|$,
Eqs.~(\ref{FKacF}) and (\ref{FKacFT})  for a general initial condition
$p_0(\bx)$, i.e. $|p_0\rangle=\int_{\Omega}d\bx_0p_0(\bx_0)|\bx_0\rangle$ and
$\langle p_0|=\int_{\Omega}d\bx_0p_0(\bx_0)\langle\bx_0|$, have the
solution \cite{lapolla_unfolding_2018}
\begin{equation}
\tPpsiVec(\mathbf{u}|p_0)\equiv \lflat|\ee{t(\LL-\mathbf{u}\cdot \mathbf{V}(\bx))}|p_0\rangle=\langle p_0|\ee{t(\LLb-\mathbf{u}\cdot \mathbf{V}(\bx))}|\rflat.
\label{braket}
\end{equation}
To  arrive at the second
equality we have used Green's identity, introduced the adjoint (or backward) Fokker-Planck
operator $\LLb=\nabla_{\bx}\cdot
\bDD\nabla_{\bx}+\bFF(\bx)\cdot\nabla_{\bx}$, and used that
the Laplace transform of a real function $f(t)$ transforms as
$\tld{f}(s^{\dagger})=\tld{f}^{\dagger}(s)$ under complex
conjugation. In the following subsection we show that for Markov jump processes \eqqref{braket} the theory can be adopted one-by-one.

\subsection{Markov-jump dynamics and additive functionals}
Markov jump processes correspond to a discrete state-space  $\Omega$
in which the system jumps with a constant rate $w_{\bx \by}$ from state $\bx\in \Omega$ to another state $\by\in\Omega$, such that the propagator $P_t(\bx|\bx_0)$ satisfies the master equation
\begin{equation}
\partial_t P_t(\bx|\bx_0)=\sum_{\by} \bra{\bx}\LL\ket{\by} P_t(\by|\bx_0),
\label{masterequation}
\end{equation}
where $ \bra{\bx}\LL\ket{\by}=w_{\by\bx}$ if $\bx\neq\by$ and $ \bra{\bx}\LL\ket{\bx}=-\sum_{\by\neq\bx}\bra{\by}\LL\ket{\bx}$ such that $-\bra{\bx}\LL\ket{\bx}>0$ is the rate of leaving state $\bx$.
According to the celebrated Gillespie algorithm
\cite{gillespie_exact_1977} a single trajectory $\bx_\tau$
($0\le\tau\le t$) consist of a sequence exponentially distributed
local waiting times $\tau_i$  in state $\bx_i\in\Omega$ followed by
an instantaneous transition to another state $\bx_{i+1}\neq \bx_i$,
with the total time being the sum of waiting times
$\sum_i\tau_i=t-\tau_R$, where $\tau_R$ is the duration of the final epoch that
contains no jump. More precisely, whenever the system is in a state $\bx_i$ at time $t_i$ the probability density to leave said state $\bx_i$ exactly at time $t_{i+1}=t_i+\tau_i$ is exponentially distributed
with a waiting time density $-\bra{\bx_i}\LL\ket{\bx_i}{\rm e}^{\bra{\bx_i}\LL\ket{\bx_i}\tau_i}$.
After the waiting time a new (accessible) state, $\bx_{i+1}$, is
randomly chosen with probability
$-\bra{\bx_{i+1}}\LL\ket{\bx_i}/\bra{\bx_i}\LL\ket{\bx_i}$. Therefore,
the joint probability density that the system, starting from  state
$\bx_i$, jumps after time $\tau_i$ and the following state is
$\bx_{i+1}$ becomes $\bra{\bx_{i+1}}\LL\ket{\bx_i}{\rm
  e}^{\bra{\bx_i}\LL\ket{\bx_i}\tau_i}$.

Denoting the number of transitions from state $\bx$
to state $\by$ until a time $t$ by $n_{\bx\by}(t)=\sum_i\delta_{\bx_i,\bx}\delta_{\bx_{i+1},\by}$ and identifying the sum of all local
waiting times in state $\bx$ up to time $t$ by $
t\theta_\bx(t)=\sum_i\delta_{\bx_i,\bx}\tau_i$
the path probability (or path weight) of $\bx_\tau$ ($0\le\tau\le t$) starting from $\bx_0$ generated by
the Markov dynamics \eqqref{masterequation} can be written as \cite{hart16a}
\begin{equation}
 P(\{\bx_\tau\}|\bx_0)=\prod_{\bx\neq\by}[\bra{\bx}\LL\ket{\by}]^{n_{\by\bx}(t)}\ee{\sum_\bx
     t\theta_\bx(t)\bra{\bx}\LL\ket{\bx}} .
 \label{pathweight}
\end{equation}

Replacing the integral in Eq.~\eqqref{ltf2} by a sum, $\overline V_t=\sum_\bx V(\bx) \theta_\bx(t)$,  allows us to identify  the characteristic function of $\psi_t=t\overline V_t$ by \cite{hart16a}
\begin{align}
  \tPpsi(u|\bx_0)&\equiv
  \avg{\ee{-ut\ovl{V}_t}}_{\bx_0}\nonumber\\
  &=
  \!\int\!d\{\bx_\tau\}P(\{\bx_\tau\}|\bx_0)\ee{-u \sum_\bx V(\bx)t\theta_\bx(t)}\nonumber\\
  &=\bra{-}\ee{\hat L(u)t}\ket{\bx_0}
  \label{generating_master}
\end{align}
where 
in the second line we inserted the path weight Eq.~\eqref{pathweight}.
While passing from the second to the third line we tilted the diagonal
of the generator in the path weight \eqref{pathweight} according to
$\bra{\bx}\hat L(u)\ket{\bx}\equiv\bra{\bx}\hat L\ket{\bx} -uV(\bx)$,
which effectively moves $\ee{-ut\ovl{V}_t}$ into the tilted path
weight. In other words, identifying $\hat L(u)$ in the second line of
Eq.~\eqref{generating_master} yields the third line. 
Note that the off-diagonal elements of the tilted generator remain unchanged, that is,
$\bra{\bx}\hat L(u)\ket{\by}\equiv\bra{\bx}\hat L\ket{\by}$ if $\bx\neq\by$.
Since all elements of $\hat L$ and $V(\bx)$ are real,
Eq.~\eqqref{braket} holds also for Markov-jump processes. 
As shown in Eq.~(\ref{trivial}) the characteristic function of
 $\overline{V}_t=\psi_t/t$ follows from a trivial
change of scale, $\tPv(u|\bx_0)=\tPpsi(u /t|\bx_0)$.
In the following we develop a spectral theory, which unifies
 diffusion processes and Markov-jump processes.

\subsection{Spectral theory of non-Hermitian generators}
We henceforth employ a spectral-theoretic approach and are thus
required to make some more specific assumptions about the underlying dynamics in order to assure
that the generator $\LL$ is diagonalizable. An excellent account of the theory for Markov-jump
dynamics governed by a discrete-state master equation can be found in
\cite{RevModPhysS}.
In the case of Fokker-Planck dynamics we
consider that $\bx_t$ is an ergodic Markovian diffusion evolving
according to Eq.~(\ref{itoe}) with the drift field $\bFF(\bx)$ not
necessarily corresponding to a potential field (which thus include
systems with a broken detailed balance) but at the same time
requiring that it is sufficiently confining, that is, it grows
sufficiently fast as $|\bx|\to\infty$ to assure that $\LL$ has a pure
point-spectrum. Moreover, we require that $\LL$ is diagonalizable and
it can be shown that any normal operator $\LL$,
satisfying $\LL\LLb=\LLb\LL$, is in fact diagonalizable 
\cite{conway_course_1985}.
A more detailed mathematical expos\'e of the
requirements for, and properties of, $\LL$ can be found in  \cite{Lapolla_2019}. In all practical examples presented
below we will in fact assume that the dynamics is
overdamped. Moreover, except for the example presented in
Sec.~\ref{uni-cyclic} where detailed balance is violated, $\LL$ will
be assumed to obey detailed balance
\cite{gardiner_c.w._handbook_1985,kampen_ng_van_stochastic_2007},
implying that it is orthogonally equivalent to a self-adjoint operator
and hence automatically diagonalizable. 

Let $-\lambda_k$ ($\mathrm{Re}(\lambda_k)\ge 0$), $\fbrak$, and
$\fketk$  denote the eigenvalue and orthonormal left and right eigenstates of $
\LL$, and $-\lambda_k^{\dagger}$, $\bbrak$ and $\bketk$ the
corresponding orthonormal eigenstates of $ \LLb$ \cite{Lapolla_2019},
i.e. $\langle
L_k|R_l\rangle=\langle R_k|L_l\rangle=\delta_{kl}$, with the
resolution of identity
$\sum_k\fketk\fbrak\equiv\sum_k\bketk\bbrak=\mathbf{1}$. Then written
in the respective eigenbases $\LL$ and $\LLb$ read
\begin{equation}
\LL=-\sum_k\lambda_k\fketk\fbrak,\quad \LLb=-\sum_k\lambda_k^{\dagger}\bketk\bbrak,
  \label{eigs}
\end{equation}
with the ground-state eigenvalue $\lambda_0=0$ and corresponding
null-space $|R_0\rangle\equiv |P_{\ii}\rangle$ and $\langle
L_0|\equiv\lflat|$. 
In the respective dual eigenbasis the propagator $P_t(\bx|\bx_0)\equiv\langle \bx|\ee{\LL t}|\bx_0\rangle=\langle \bx_0|\ee{\LLb t}|\bx\rangle$
reads
\begin{eqnarray}
P_t(\bx|\bx_0) &=& \sum_k \langle
\bx\fketk\fbrak\bx_0\rangle\ee{-\lambda_kt}\nonumber\\
&\equiv&\sum_k \langle \bx_0\bketk\bbrak\bx\rangle\ee{-\lambda^{\dagger}_kt},
  \label{propa}
\end{eqnarray}  
where $\langle \bx\fketk\equiv R_k(\bx)$ and $\fbrak\bx_0\equiv
L^{\dagger}_k(\bx_0)$, while $\langle \bx_0\bketk=L_k(\bx_0)$ and
$\bbrak\bx\rangle=R^{\dagger}_k(\bx)$. For overdamped systems with
invertible diffusion matrix $\bDD$ that obey
detailed balance, i.e. $\bDD^{-1}\bFF(\bx)=-\beta \nabla_{\bx}U(\bx)$
with inverse thermal energy $\beta=1/k_{\mathrm{B}}T$, all $\lambda_k$ are real, $|P_{\ii}\rangle=|P_{\mathrm{eq}}\rangle$ is the Boltzmann-Gibbs
equilibrium $P_{\mathrm{eq}}(\bx)=\ee{-\beta
  U(\bx)}/\int_{\Omega}\ee{-\beta U(\bx)}d\bx$, and $\bketk=\ee{\beta
  U(\bx)}\fketk$ \cite{risken_fokker-planck_1996, gardiner_c.w._handbook_1985}.

\section{Characteristic function near zero via non-Hermitian perturbation theory}
\label{Sec3}
Based on Eqs.~(\ref{cf}) and (\ref{moments}) we only require the
moment-generating function (\ref{braket}), in the limit
$|\mathbf{u}|\to 0$ to calculate moments of arbitrary order. Moreover,
recall that $\tPvVec(\mathbf{u})=\tPpsiVec(\mathbf{u}/t)$ (see Eq.~(\ref{trivial})).
To keep the treatment general we utilize the spectral
expansion of $\LL$ ($\LLb$, respectively). We employ perturbation theory
to derive the moment-generating function (\ref{braket}) in the limit
$|\mathbf{u}|\to 0$. There are (at least) two possible ways to arrive
at the result: a Dyson series approach, which is presented in
Appendix \ref{Dyson calc}, and by means of second order non-Hermitian perturbation
theory, which is detailed below. While both yield equivalent
results, the perturbation-theoretic approach is more general as it provides a
(bi)spectral expansion of the of the perturbed generator
$\LL-\mathbf{u}\cdot \mathbf{V}(\bx)$ ($\LLb-\mathbf{u}\cdot
\mathbf{V}(\bx)$, respectively) to second order in
$|\mathbf{u}|$. These perturbation-theoretic results, which in the Physics literature appear
to be new, are applicable beyond time-average statistical
mechanics in diverse problems involving perturbations of non-Hermitian
and/or non-self-adjoint eigenvalue problems.

Our aim is to
diagonalize the ``tilted'' propagator in Eq.~(\ref{braket}) in the
limit when $\mathbf{u}$ vanishes. Because $\LL$ is in general not self-adjoint we need to separately perturb left
and right eigenstates.
First we must confirm that the tilted propagator $\LL(\mathbf{u})\equiv\LL-\mathbf{u}\cdot
\mathbf{V}(\bx)$ (and $\LLb(\mathbf{u})\equiv\LLb-\mathbf{u}\cdot
\mathbf{V}(\bx)$, respectively) is actually diagonalizable in an arbitrarily small
neighborhood of $\mathbf{u}=\mathbf{0}$. We focus first on the case where
$V(\bx)\ge 0$. We Laplace transform
Eq.~(\ref{braket}) $t\to s$ yielding
\begin{equation}
\tPpsiVecLaplace(\mathbf{u}|p_0)\equiv \lflat|[s-\LL(\mathbf{u})]^{-1}|p_0\rangle=\langle p_0|[s-\LLb(\mathbf{u})]^{-1}|\rflat.
\label{LTbraket}
\end{equation}
The singularities of  Eq.~(\ref{LTbraket}) correspond to the perturbed
eigenvalue spectrum $\{-\lambda_k(\mathbf{u})\}$ of $\LL(\mathbf{u}))$
and diagonalizability is broken whenever one or more singularities are
not simple poles (see e.g. \cite{Dyson}).  
Eq.~(\ref{braket}) shows that $|\mathbf{u}|=0$ is not an
accumulation point. Moreover, $\LL$ has a pure point spectrum
therefore an arbitrarily small $|\mathbf{u}|$ cannot cause the
emergence of poles of second order in Eq.~(\ref{LTbraket}) that would
break diagonalizability, akin to the ``avoided crossing theorem''. 
Therefore, in the limit $|\mathbf{u}|\to 0$ the tilted generator
$\LL(\mathbf{u})$ is diagonalizable, $\mathbf{u}$ can be taken as real \footnote{Note
that $0$ is an eigenvalue of $\LL$ and
$\tPpsiVecLaplace(\mathbf{u}|p_0)$ is meromorphic for small
$|\mathbf{u}|$ therefore we can assume, without loss of generality,
that $\mathbf{u}$ is real.}, and the
eigenspectrum of $\LL(\mathbf{u})$ corresponds to a
regular perturbation of the original eigenvalue problem
$\LL\fketk=-\lambda_k\fketk$
($\LLb\bketk=-\lambda_k^{\dagger}\bketk$) and we seek for a
perturbative expansion of the tilted eigenspectrum, e.g. of $\LLb(\bu)$:
 \begin{eqnarray}
 - \lambda^{'\dagger'}_k(\bu)&=&-\lambda^{\dagger}_k-\sum_{i>0} \bu^i\cdot \boldsymbol{\lambda_k^{(i)}},\nonumber\\
  |L_k(\bu)\rangle&=&\bketk+\sum_{i>0} \bu^i\cdot\mathbf{|L_k^i\rangle},\nonumber\\
  \langle
  R_k(\bu)|&=&\bbrak+\sum_{i>0}\bu^i\cdot\mathbf{|R_k^i\rangle}.
  \label{ansatz}
 \end{eqnarray}
Without loss of generality we will henceforth assume that $\bu$ is real. 
Note that while the spectra of $\LL$ and $\LLb$ are complex
conjugates (see Eq.~(\ref{eigs})) the perturbation is in fact
symmetric (see Eq.~(\ref{LTbraket})). Therefore, the spectra of
$\LL(u)$ and $\LLb(u)$ are \emph{not} complex conjugates except for
the unperturbed part, which we denoted in Eq.~(\ref{ansatz}) by
quotation marks $\lambda^{'\dagger'}$.  According to Eq.~(\ref{ansatz}) multiple functionals yield additive perturbations.
It thus suffices to carry out the calculations for $\bu\to u$ and write the
corresponding general result by inspection. We are interested in up
to second order moments (\ref{moments}) and therefore need to evaluate
the perturbation up to second order in $u$:
 \begin{eqnarray}
  \!\!\!\!\!(-\LLb+uV)\sum_{n=0}^2u^{n}|L_k^n\rangle&=&\sum_{n=0}^2u^n\lambda_k^{(n)}\sum_{m=0}^2u^m|L_k^m\rangle, \\
   \!\!\!\!\! \sum_{n=0}^2 u^n\langle
  R_k^n|(-\LLb+uV)&=&\sum_{n=0}^2u^n\lambda_k^{(n)}\sum_{m=0}^2u^m\langle
  R_k^m|,
  \label{pert}
 \end{eqnarray}
where we have adopted the convention
$\lambda_k^{(0)}\equiv\llambda_k$, $\langle R_k^0|\equiv\bbrak$, and
$|L_k^0\rangle\equiv\bketk$. In Eqs.~(\ref{pert}) we only need to keep
terms up to $u^2$ and equate terms of matching order in $u$.
First we impose the \emph{preliminary} normalization $ \langle
R_k(u)|L_k\rangle=\langle R_k|L_k(u)\rangle=1$, i.e.
 \begin{eqnarray}
1&=&\langle R_k|L_k\rangle+u \langle
  R_k^1|L_k\rangle+u^2\langle R_k^2|L_k\rangle +\mathcal{O}(u^3) \nonumber\\
1&=&\langle R_k|L_k\rangle+u \langle R_k|L_k^1\rangle+u^2\langle R_k|L_k^2\rangle+\mathcal{O}(u^3), 
 \end{eqnarray}
 which implies
 \begin{equation}
  \langle R_k^n|L_k\rangle=\langle R_k|L_k^n\rangle=0, \quad \mathrm{for}\quad n>0.
  \label{eq:initialrenorm}
 \end{equation}
 The $0$-th order of the expansion gives the solution of the
 unperturbed system. For the higher orders we need to solve
 Eqs. \eqqref{pert}  matching terms of equal order. Introducing the
 \emph{coupling elements} $V_{lk}\equiv\langle R_l|V|L_k\rangle$ we
 obtain (details of the calculation are shown in Appendix A)
  \begin{eqnarray}
  \lambda_k^{(1)}&=&V_{kk},\quad \lambda_k^{(2)}=\sum_{l\neq k} \frac{V_{kl}V_{lk}}{\llambda_k-\llambda_l},\\
  |L_k^1\rangle&=&\sum_{l\neq k}\frac{V_{lk}}{\llambda_k-\llambda_l}|L_l\rangle, \quad
  \langle R_k^1|=\sum_{l\neq k} \frac{V_{kl}}{\llambda_k-\llambda_l}\langle R_l|,\nonumber\\
  |L_k^2\rangle&=&\sum_{l\neq k} \left[ \sum_{i\neq k}\frac{V_{ik}V_{li}}{(\llambda_k-\llambda_i)(\llambda_k-\llambda_l)}-\frac{V_{kk}V_{lk}}{(\llambda_k-\llambda_l)^2}\right] |L_l\rangle,\nonumber\\
  \langle R_k^2|&=&\sum_{l\neq k} \left[\sum_{i\neq k} \frac{V_{ki}V_{il}}{(\llambda_k-\llambda_i)(\llambda_k-\llambda_l)}-\frac{V_{kk}V_{kl}}{(\llambda_k-\llambda_l)^2}\right]\langle R_l|\nonumber.
\label{PPV}
  \end{eqnarray}
 However, while they are orthogonal by construction, the resulting
 perturbed eigenstates are not normalized anymore, i.e. $\langle
 R_k(u)|L_k(u)\rangle\ne 1$. Hence, we need to post-normalize them such that
 \begin{eqnarray}
\mathcal{N}_k(u)\langle R_k(u)|L_k(u)\rangle=1,
\end{eqnarray}
where from it follows that
 \begin{eqnarray}
  \mathcal{N}_k(u)&=&
  \left[(\langle R_k|+\sum_{i=1}^2u^i\langle R_k^i|)(|L_k\rangle+\sum_{i=1}^2u^i|L_k^i\rangle)\right]^{-1}\nonumber\\
  &=&\frac{1}{1+u^2\langle R_k^1|L_k^1\rangle+\mathcal{O}(u^3) }\nonumber\\
  &=&1-u^2 \sum_{i \neq
    k}\frac{V_{ki}V_{ik}}{(\llambda_k-\llambda_i)^2}+\mathcal{O}(u^3)\nonumber \\
  &\equiv&1-u^2\mathcal{M}_k+\mathcal{O}(u^3),
 \label{NN} 
 \end{eqnarray}
 where in the last line we have defined $\mathcal{M}_k$.
We now use the second-order perturbed eigenspectrum to diagonalize
the moment-generating function in Eq.~(\ref{braket})
\begin{equation}
\tPpsi(u|p_0)=\sum_k \mathcal{N}_k(u)\langle p_0|L_k(u)\rangle \langle R_k(u)|\rflat\ee{-\lambda^{'\dagger'}_k(u)t},
\label{braketD}
\end{equation}
where moreover
\begin{equation}
\ee{-\lambda^{\dagger}_k(u)t}=\ee{-\lambda_k^{\dagger}t}[1-u\lambda_k^{(1)}t+u^2(\lambda_k^{(1)2}t^2/2-\lambda_k^{(2)}t)]+\mathcal{O}(u^3)
  \label{pev}
  \end{equation}
and $\langle
R_k(u)|\rflat=\bbrak\rflat+u\langle R_k^{1}|\rflat+u^2\langle R_k^{2}|\rflat+\mathcal{O}(u^3)$
with the coefficients given by 
\begin{eqnarray}
  \langle R_k|-\rangle&=&\delta_{k0},\quad \langle
  R_k^1|-\rangle=\frac{V_{k0}}{\llambda_k}(1-\delta_{k0}),\nonumber\\
   \langle R_k^2|\rflat&=&-\frac{V_{kk}V_{k0}}{\lambda_k^{\dagger 2}}+\sum_{i\neq k}\frac{V_{ki}V_{i0}}{\llambda_k(\llambda_k-\llambda_i)}(1-\delta_{k0})
\label{expansions}  
\end{eqnarray}
and $ \langle p_0|L_k(u)\rangle=\langle p_0|L_k\rangle+u\langle
p_0|L_k^1\rangle+u^2\langle p_0|L_k^2\rangle+\mathcal{O}(u^3)$,
where $|L_k^1\rangle$ and $|L_k^2\rangle$ are given by
Eq.~(\ref{PPV}).
Using Eqs.~(\ref{PPV}),~(\ref{NN}) as well as (\ref{pev}) and
(\ref{expansions}) the tilted propagator in Eq.~(\ref{braketD}) to
second order in $u$ reads
\begin{equation}
\tPpsi(u|p_0)=1+\sum_{k>0} \left(u C^{(1)}_k+ u^2 C^{(2)}_k\right)\ee{-\lambda_k^{\dagger}t} +\mathcal{O}(u^3)
\label{expanded}
\end{equation}
where we have introduced the coefficients
\begin{eqnarray}
 C^{(1)}_k&=&\delta_{k0}\left(-V_{00}t +\langle p_0|L^{1}_0\rangle\right)+\langle p_0|L_k\rangle \langle R^{1}_k|-\rangle\nonumber\\
 C^{(2)}_k&=&\delta_{k0}\left[\langle p_0| L_0\rangle \left(\mathcal{M}_0+V_{00}^2t^2/2-\lambda_0^{(2)}t\right)+\langle p_0|L^{2}_0\rangle\right]\nonumber\\
	  &-&t\left(\langle p_0|L^{1}_0\rangle
 V_{00}\delta_{k0}-\langle p_0|L_k\rangle\langle R^{1}_k|-\rangle V_{kk}\right)\nonumber\\
	  &+&\langle p_0|L^{1}_k\rangle\langle R^{1}_k|-\rangle + \langle p_0|L_k\rangle\langle R^{2}_k|-\rangle
\end{eqnarray}
with $\mathcal{M}_0$ defined in the Eq.~\eqref{NN}.
In the special case when initial conditions $\bx_0$
are drawn from the steady state probability density, i.e. $p_0(\bx)=P_\ii(\bx)$,
these equations simplify to
 \begin{eqnarray}
 \!\!\!\!\!\! \langle P_\ii|L_k\rangle&=&\delta_{k0},\quad \langle
  P_\ii|L_k^1\rangle=\frac{V_{0k}}{\llambda_k}(1-\delta_{k0}),\nonumber\\
 \!\!\!\!\!\!  \langle
 P_\ii|L_k^2\rangle&=&-\frac{V_{kk}V_{0k}}{\lambda_k^{\dagger 2}}+\sum_{i\neq k}\frac{V_{ik}V_{0i}}{\llambda_k(\llambda_k-\llambda_i)}(1-\delta_{k0}).
\label{expansions2}  
 \end{eqnarray}
 Turning now to the case where $V(\bx)$ extends negative values we
 must simply replace $u$ by $i\omega$. In order to obtain moments up
 to second order (i.e. Eq.~(\ref{var_cov}) we are now left with
 evaluating derivatives of Eq.~(\ref{expanded}) with respect to $u$ at
 $u=0$.

 \subsection{Mean value, fluctuations and correlation functions for general initial conditions}
 \label{expectation derivatives}
We can now derive the mean, variance and covariance of time-average observables.  
We first focus on the case of a single time-average observable
$\ovl{V}_t=\psi_t/t$ (see
Eqs.~(\ref{TA}) and (\ref{funct})). According to
Eq.~(\ref{trivial}) we must make the
replacement $u\to u/t$ in Eq.~(\ref{expanded}). We will only present the result in terms of the
spectrum of the backward generator $\LLb$ since the results corresponding to
$\LL$ follow trivially from Eq.~(\ref{braket}). 
Note that $\langle p_0|\rflat =1$ for any normalized
initial condition.

To order $u^0$ Eq.~(\ref{expanded}) simply reflects the normalization
of $\mathcal{P}^{\ovl{V}}_t(\nu|p_0)$,
i.e. $\tPv(0|p_0)=\int_0^{\infty}
\mathcal{P}^{\ovl{V}}_t(\nu|p_0)d\nu=1$ (and equivalently in the case where
the support of $\ovl{V}_t$ extends to the entire real axis). 
To order $u^1$ Eq.~(\ref{expanded}) encodes the mean value
$\langle\ovl{V}_t\rangle_{p_0}$ since it follows from Eq.~(\ref{cf})
that $-\partial_u\tPv(u|p_0)|_{u=0}=\int_0^{\infty}
\nu\mathcal{P}^{\ovl{V}}_t(\nu|p_0)d\nu\equiv \langle\ovl{V}_t\rangle_{p_0}$. Thus
the mean value of a time-average observable
$\langle\ovl{V}_t\rangle_{p_0}$ evolving from an arbitrary initial
condition $p_0(\bx)$ is given by
 \begin{equation}
  \langle \ovl{V}_t\rangle_{p_0} =V_{00}+\frac{1}{t}\sum_{k> 0}\frac{V_{k0}}{\llambda_k}\langle p_0|L_k\rangle\left(1-\ee{-\llambda_kt}\right),
  \label{eq:meanfix}
 \end{equation}
with the anticipated ergodic limit $V_{00}=\langle
\ovl{V}_{t\to\infty}\rangle_{p_0}=\langle
P_\ii|V|-\rangle\equiv\int_{\Omega}V(\bx)P_\ii(\bx)d\bx$. The
result~(\ref{eq:meanfix}) is equally valid in cases where $V(\bx)$ can
become negative (as long as its is bounded). Moreover, from
Eq.~(\ref{eq:meanfix}) is easy to discern the large deviation
asymptotic
 \begin{equation}
 \lim_{t\gg1/\mathrm{Re}\llambda_1} \langle \ovl{V}_t\rangle_{p_0}\simeq V_{00}+\frac{1}{t}\sum_{k> 0}\frac{V_{k0}}{\llambda_k}\langle p_0|L_k\rangle,
  \label{eq:meanfix2}
 \end{equation}
and in the special case of steady-state initial conditions
$p_0(\bx_0)=P_\ii(\bx_0)$ Eq.~(\ref{eq:meanfix}) reduces to the
time-independent ergodic result $\langle \ovl{V}_t\rangle_{\ii}
=V_{00}$.
To order $u^2$ Eq.~(\ref{expanded}) encodes the second moment via $\partial^2_u\tPv(u|p_0)|_{u=0}=\int_0^{\infty}
\nu^2\mathcal{P}^{\ovl{V}}_t(\nu|\bx_0)d\nu\equiv
\langle\ovl{V}_t^2\rangle_{p_0}$, which  reads
 \begin{multline}
  \langle \ovl{V}_t^2\rangle_{p_0}=V_{00}^2
  +\frac{2}{t}\sum_{k>0}\frac{V_{k0}}{\llambda_k}[V_{0k}+\langle
    p_0| L_k\rangle(V_{00}+V_{kk}\ee{-\llambda_k t})]\\
+\frac{2}{t^2}\sum_{k> 0}V_{k0}\bigg\{ [\langle p_0| L_k\rangle(V_{kk}-V_{00})-V_{k0}]
\frac{(1-\mathrm{e}^{-\llambda_k t})}{\lambda_k^{\dagger 2}} 
\\+\sum_{l> 0,l\neq k} \frac{V_{lk}\langle p_0|L_l\rangle}{\llambda_k-\llambda_l} \left(\frac{1-\ee{-\llambda_lt}}{\llambda_l}-\frac{1-\ee{-\llambda_k t}}{\llambda_k}\right)\bigg\}
  \label{eq:mom2fix}
 \end{multline}
which together with Eq.~(\ref{eq:meanfix}) yields the variance
  \begin{equation}
\sigma^2_{\ovl{V},p_0}(t)\equiv  \langle \ovl{V}_t^2\rangle_{p_0} -  \langle \ovl{V}_t\rangle_{p_0}^2.
 \label{var_gen_IC}   
 \end{equation}   
We further introduce the following notational convention for localized initial conditions
$p_0(\bx)=\delta(\bx-\bx_0)$, $\sigma^2_{\ovl{V},p_0}(t)\to
\sigma^2_{\ovl{V},\bx_0}(t)$. From Eqs.~ \eqqref{eq:meanfix},
\eqqref{eq:mom2fix}, and \eqqref{var_gen_IC} follows
the anticipated ergodic result $\langle
\ovl{V}^2_{t\to\infty}\rangle_{p_0}=V_{00}^2$, proving that in the ergodic
limit $\ovl{V}_t$ becomes deterministic (i.e. the variance vanishes, $\sigma^2_{\ovl{V},p_0}(t\to\infty)=\langle
\ovl{V}^2_{t\to\infty}\rangle_{p_0}-\langle
\ovl{V}_{t\to\infty}\rangle_{p_0}^2=0$).
Conversely the large-deviation
asymptotic reads for any initial condition $p_0(\bx)$
 \begin{equation}
 \lim_{\mathrm{Re}\llambda_1t\gg 1}  \langle \ovl{V}_t^2\rangle_{p_0}=V_{00}^2
  +\frac{2}{t}\sum_{k>0}\frac{V_{k0}}{\llambda_k}\left(V_{0k}+ V_{00}\langle
    p_0| L_k\rangle\right)
  \label{eq:mom2fix2}
 \end{equation}
 yielding a large-deviation variance
 \begin{equation}
\!\lim_{\mathrm{Re}\lambda_1t\gg 1}\sigma^2_{\ovl{V}}(t)=\frac{2}{t}\sum_{k>0}\frac{V_{0k}V_{k0}}{\llambda_k}+\mathcal{O}(t^{-2}),
\label{LDvar}
 \end{equation}
 which embodies the emergence of the central limit theorem. One can
 further show that all higher cumulants decay to zero faster that
 $1/t$. Since the only distribution with a finite number of non-zero
 cumulants is the Gaussian distribution \cite{Marcinkiewicz}, the
 large deviation mean value (\ref{eq:meanfix2}) and variance (\ref{LDvar}) specify
 the entire asymptotic probability density for time-average observables along
 trajectories of length $t\gg 1/\mathrm{Re}\llambda_1$.

 In the special case of steady-state initial conditions,
$p_0(\bx)=P_\ii(\bx)$, we find the variance  satisfies (see also \cite{lapolla_unfolding_2018})
 \begin{equation}
\sigma^2_{\ovl{V},\ii}(t)=\frac{2}{t}\sum_{k>0}\frac{V_{0k}V_{k0}}{\llambda_k}\left(1-\frac{1-\ee{-\llambda_kt}}{\llambda_k
   t}\right).
\label{ss}
 \end{equation}
 Note that for overdamped systems in detailed balance we have
 $\llambda_k\in\mathbb{R}$  and $|L_k\rangle=\ee{\beta
  U(\bx)}|R_k\rangle\in\mathbb{R}$. Therefore  
 $V_{0k}V_{k0}\ge 0$,
 which implies (compare Eqs.~\eqqref{LDvar} and \eqqref{ss}) that 
the fluctuations for stationary initial conditions are bounded from
above by  $\sigma^2_{\ovl{V},\ii}(t)\le\lim_{\llambda_1t\gg 1}  \langle
    \ovl{V}_t^2\rangle_{p_0}-V_{00}^2$ irrespective of $p_0(\bx)$.
 

 We now inspect the correlation between two functionals $\ovl{V}_{1,t}$
 and $\ovl{V}_{2,t}$ (see Eq.~(\ref{var_cov})) defined as
 $C_{\ovl{V}_1\ovl{V}_2}(t)=\langle \ovl{V}_{1,t}\ovl{V}_{2,t}\rangle_{p_0}-\langle
 \ovl{V}_{1,t}\rangle_{p_0}\langle\ovl{V}_{2,t}\rangle_{p_0}$. The mean
 values were derived in Eq.~(\ref{eq:meanfix}) so we only require the mixed
 second moment $\langle \ovl{V}_{1,t}\ovl{V}_{2,t}\rangle_{p_0}$, which is
 obtained from the joint moment-generating function
 (i.e. generalization of Eq.~(\ref{braketD}) to two variables) as
 $\langle
 \ovl{V}_{1,t}\ovl{V}_{2,t}\rangle_{p_0}=\partial^2_{u_1u_2}\tPvVec(\bu|p_0)|_{\bu=\mathbf{0}}$. A
 lengthy calculation leads, upon  introducing the coupling elements
 $U^i_{kl}\equiv\langle R_k|V_i(\bx)|L_l\rangle$ and the shorthand
 notation $W_{klmn}=U^1_{kl}U^2_{mn}+U^2_{kl}U^1_{mn}$,  to the exact result
 \begin{multline}
\langle \ovl{V}_{1,t}\ovl{V}_{2,t}\rangle_{p_0}= \frac{W_{0000}}{2}\\+\frac{1}{t}\sum_{k>0}\frac{1}{\lambda_k^{\dagger}}\left( W_{k00k}+\langle p_0|L_k\rangle\left[W_{00k0}-W_{k0kk}\ee{-\llambda_kt} \right]\right)\\
  +\frac{1}{t^2}\sum_{k> 0}\bigg\{ [\langle p_0|
    L_k\rangle(W_{k0kk}-W_{00k0})-W_{k00k}]\frac{(1-\mathrm{e}^{-\llambda_k
      t})}{\lambda_k^{\dagger 2}}\\
+ \sum_{l> 0,l\neq k} \frac{W_{k0lk}\langle p_0|L_l\rangle}{\llambda_k-\llambda_l} \left(\frac{1-\ee{-\llambda_lt}}{\llambda_l}-\frac{1-\ee{-\llambda_k t}}{\llambda_k}\right)\bigg\},
   \label{eq:mom2mixfix}
 \end{multline}
and we note that $\langle
\ovl{V}_{1,t}\rangle_\ii\langle\ovl{V}_{2,t}\rangle_{\ii}=W_{0000}/2$
implying that for an ergodic system any two functionals asymptotically
decorrelate, $\lim_{t\to\infty} C_{\ovl{V}_1\ovl{V}_2}(t)=0$.
Equations (\ref{eq:meanfix}), (\ref{eq:mom2fix}), and
(\ref{eq:mom2mixfix}) expressing the mean value and
second moments (and together the variance and covariance) of the time
average of a general physical observable $V(\bx_{\tau})$ of type Eq.~(\ref{TA})
solely in terms of the eigenspectrum of the underlying generator are
the main theoretical result of this work.  

In the case of stationary initial conditions $p_0(\bx)=P_\ii(\bx)$ we
have that $\langle P_\ii|L_k\rangle=\delta_{k0}$ and as a result of
Eq.~(\ref{eq:mom2mixfix}) the covariance reduces to (note that
$W_{k00k}=W_{0kk0}$ and  $W_{0000}/2=\langle \ovl{V}_{1,t}\rangle_\ii\langle\ovl{V}_{2,t}
  \rangle_\ii$)
\begin{equation}
C_{\ovl{V}_1\ovl{V}_2}^{\ii}(t)=
  \frac{1}{t}\sum_{k>0}\frac{W_{0kk0}}{\llambda_k}\left(1-\frac{1-\ee{-\llambda_kt}}{\llambda_k
    t}\right).
    \label{eq:mom2mixeq}
\end{equation}
Finally, in the large deviation regime we recover
 \begin{equation}
   \lim_{t\gg1/\lambda_1}C_{\ovl{V}_1\ovl{V}_2}(t)=
  \frac{1}{t}\sum_{k>0}\frac{W_{0kk0}}{\llambda_k}+\mathcal{O}(t^{-2})
\label{LDcovar}
 \end{equation}
the $1/t$ scaling reflecting the emergence of the central limit
theorem. Therefore, it follows that an arbitrary set
of $m$ time-average observables $\mathbf{\ovl{V}}_t=\{\ovl{V}_{i,t}\}$ in the large deviation limit exhibits
Gaussian statistics. If we denote the vector of mean values as
$\langle\mathbf{\ovl{V}}\rangle_{\ii}$ and introduce the
symmetric covariance matrix $\mathbf{C}$ with diagonal elements
$\mathbf{C}_{ii}= t\lim_{t\gg1/\mathrm{Re}\lambda_1}\sigma^2_{\ovl{V}_i}(t)$ (see
Es.~(\ref{LDvar})) and off-diagonal elements $\mathbf{C}_{ij}=
t\lim_{t\gg1/\mathrm{Re}\lambda_1}C_{\ovl{V}_i\ovl{V}_j}(t)$ (see
Es.~(\ref{LDcovar}))
then the probability density that $\mathbf{\ovl{V}}_t$ attains a value $\boldsymbol{\nu}$
obeys
the asymptotic Gaussian limit law $ \lim_{t\gg1/\mathrm{Re}\lambda_1}\mathcal{P}^{\mathbf{\ovl{V}}}_t(\boldsymbol{\nu}|p_0)\equiv
 \mathcal{P}_t^{\mathrm{LD}}(\boldsymbol{\nu})$ where
\begin{equation}
 \mathcal{P}_t^{\mathrm{LD}}(\boldsymbol{\nu})\simeq\frac{\displaystyle{\ee{-\frac{1}{2}(\boldsymbol{\nu}-\langle\mathbf{\ovl{V}}\rangle_{\ii})^T\mathbf{C}^{-1}(\boldsymbol{\nu}-\langle\mathbf{\ovl{V}}\rangle_{\ii})t}}}{\sqrt{(2\pi)^m\mathrm{det}\mathbf{C}/t}}.
\label{Gaussian}
\end{equation}
We now introduce the rescaled variables
$\hat{\boldsymbol{\nu}}\equiv \boldsymbol{\nu}\sqrt{t}$ and scaled mean
$\boldsymbol{\mu}\equiv\langle\mathbf{\ovl{V}}_t\sqrt{t}\rangle_{\ii}$, which
upon re-normalization lead to a time-independent density. Moreover,
we define
\begin{equation}
\Xi=(\hat{\nu}_i-\mu_i)/\sqrt{\mathbf{C}_{ii}},\quad \Xi_{i|j}=(\hat{\nu}_{i}-\tilde{\mu}_{i|j})/\tilde{\sigma}_{i|j}
\label{dimls}
\end{equation}
with the shifted mean and stretched variance
\begin{eqnarray}
\tilde{\mu}_{i|j}&=&\mu_i+(\hat{\nu}_j-\mu_j)\mathbf{C}_{ij}/\mathbf{C}_{jj}\nonumber\\
\tilde{\sigma}^2_{i|j}&=&(\mathbf{C}_{ii}\mathbf{C}_{jj}-\mathbf{C}_{ij}^2)/\mathbf{C}_{jj}.
\label{streched}
\end{eqnarray}
Then the limit law (\ref{Gaussian}) implies that
the univariate large deviations $\mathcal{P}_t^{\mathrm{LD}}(\hat{\nu})$ and
conditional bivariate large deviations
$\mathcal{P}_t^{\mathrm{LD}}(\hat{\nu}_1|\hat{\nu}_2)\equiv
\mathcal{P}_t^{\mathrm{LD}}(\hat{\nu}_1,\hat{\nu}_2)/\mathcal{P}_t^{\mathrm{LD}}(\hat{\nu}_2)$
collapse, upon rescaling, onto a universal Gaussian master curve
\begin{eqnarray}
 & \sqrt{\displaystyle{\frac{\mathbf{C}_{ii}}{t^{1/2}}}}\mathcal{P}_t^{\mathrm{LD}}(\hat{\nu}_i)\to\mathcal{N}_{\Xi}(0,1),\nonumber\\
 &\displaystyle{\frac{\tilde{\sigma}_{i|j}}{\sqrt{t}}}\mathcal{P}_t^{\mathrm{LD}}(\hat{\nu}_i|\hat{\nu}_j)\to\mathcal{N}_{\Xi_{i|j}}(0,1),
\label{collapse}
\end{eqnarray}
where $\mathcal{N}_x(0,1)$ denotes the Gaussian probability density
with zero mean and unit variance. The explicit rescaling
(i.e. Eq.~(\ref{collapse}) combined with Eq.~(\ref{LDvar}) and
Eq.~(\ref{LDcovar})) leading to the collapse onto a master unit normal
density in the large deviation limit are the main practical
consequence of our large deviation result.

\subsection{Degenerate eigenspectra}
Note that if the spectrum of $\LL$ has degenerate eigenstates (such as e.g. in
single-file diffusion \cite{lapolla_unfolding_2018,Lapolla_2019}) special care is
required for initial conditions that do not correspond to the steady
state, i.e. $p_0(\bx)\ne P_\ii(\bx)$, as a result of the singularities the degeneracy
causes in Eqs. (\ref{eq:mom2fix}) and
(\ref{eq:mom2mixfix}). As it is customary in regular perturbation
theory (see e.g. \cite{Sakurai}) one must first post-diagonalize all
the respective degenerate sub-spaces prior to using Eqs. (\ref{eq:mom2fix}) and
(\ref{eq:mom2mixfix}). Once this has been taken care of (using any of the
many possible methods \cite{Klein}) and the degenerate
eigenstates are replaced by their appropriate linear combinations, Eqs. (\ref{eq:mom2fix}-\ref{eq:mom2mixfix}) can be used as they stand.  

\subsection{A general upper bound for occupation measures for overdamped reversible dynamics}
\label{subsec:upper_bound}
When $\LL$ corresponds to reversible overdamped dynamics
(i.e. $\bDD^{-1}\bFF(\bx)=-\beta \nabla_{\bx}U(\bx)$ is a gradient field), or to a reversible Markov jump
process (i.e. the transition matrix elements in Eq.~\eqqref{masterequation} satisfy the symmetry $\bra{\by}\LL\ket{\bx}/\bra{\bx}\LL\ket{\by}=\ee{\beta U(\bx)-\beta U(\by)}$)
the large deviation
asymptotic (\ref{LDvar}) provides an upper bound for fluctuations of
the \emph{occupation time fraction} in any subdomain $\mathcal{V}\subseteq\Omega$, $V(\bx)=\mathcal{V}$,  for
any duration of the trajectory (which naturally includes the local time
fraction when $V(\bx)=\bx$).

Let us define the projection operator
$\hat{\Gamma}_\bx(\mathcal{V};V)\equiv\int_{\Omega}d\bx \delta(\mathcal{V}-V(\bx))$
\cite{Lapolla_2019}, which projects the full dynamics $\bx\subset\mathbb{R}^d$ onto the
hypersurface compatible with a given value of the observable
$V(\bx)=\mathcal{V}$. Then
the (generally non-Markovian) joint
probability density that the observable $V(\bx)$ starts from $\mathcal{V}$ and returns to
the initial value $\mathcal{V}$ at time $t$ in
an ensemble of trajectories $\bx_t$ starting from the equilibrium
probability density $P_\ii(\bx_0)=P_{\mathrm{eq}}(\bx_0)$ is defined as
\cite{Lapolla_2019}
\begin{equation}
G_t^{\mathrm{eq}}(\mathcal{V},\mathcal{V})\equiv \hat{\Gamma}_\bx(\mathcal{V};V)\hat{\Gamma}_{\bx_0}(\mathcal{V};V)P_t(\bx|\bx_0)P_{\mathrm{eq}}(\bx_0).
\label{NMP}  
\end{equation}
We now recall the definition of  the \emph{occupation time fraction} of $\bx_{\tau}$ within the
hypersurface $V(\bx)=\mathcal{V}$ in Eq.~(\ref{ltf3}). 
Then Eq.~(\ref{LDvar}) and the spectral decomposition of
$P_t(\bx|\bx_0)$ in Eq.~(\ref{propa}) imply the general upper bound
on $\theta_{\mathcal{V}}(t)$
\begin{equation}
  t\sigma_{\theta_{\mathcal{V}},\ii}^2(t)\le 2\int_0^{\infty}[G_t^{\mathrm{eq}}(\mathcal{V},\mathcal{V})-G_\infty^{\mathrm{eq}}(\mathcal{V},\mathcal{V})]dt,
\label{bound}  
\end{equation}
where equality holds in the limit $t\to\infty$. Note that in the
special case when 
$V(\bx)=\bx$, Eq.~(\ref{bound}) bounds the local time fraction defined
in Eq.~(\ref{ltf}).

To prove the bound ~(\ref{bound}) let us express Eq.~(\ref{NMP}) using the spectral expansion
of $\LLb$ (or equivalently $\LL$). Since we are considering systems in
detailed balance the eigenspectrum is real. Introducing the elements
$V_{kl}(\mathcal{V})\equiv\langle
R_l|\delta(\mathcal{V}-V(\bx))|L_k\rangle$ the spectral representation
of Eq.~(\ref{NMP}) reads (see also \cite{Lapolla_2019})
\begin{equation}
G_t^{\mathrm{eq}}(\mathcal{V},\mathcal{V})=\sum_{k}V_{0k}(\mathcal{V})V_{k0}(\mathcal{V})\ee{-\lambda_kt}
\label{NMPs}  
\end{equation}
such that $\lim_{t\to\infty}G_t^{\mathrm{eq}}\equiv
G_\infty^{\mathrm{eq}}(\mathcal{V},\mathcal{V})=V_{00}(\mathcal{V})^2$. Therefore
\begin{equation}
  \int_0^{\infty}[G_t^{\mathrm{eq}}(\mathcal{V},\mathcal{V})-V_{00}(\mathcal{V})^2]dt=\sum_{k>0}V_{0k}(\mathcal{V})V_{k0}(\mathcal{V})/\lambda_k.
  \label{Qproof}
\end{equation}
Multiplying Eq.~(\ref{Qproof}) by 2 and dividing by $t$ we obtain
Eq.~(\ref{LDvar}) for the case when
$\ovl{V}_t=\theta_{\mathcal{V}}(t)$ defined in Eq.~(\ref{ltf3}) which in
turn proves asymptotic equality as $t\to\infty$. Because
for systems obeying detailed we further have $|L_k\rangle=\ee{\beta
  U(\bx)}|R_k\rangle$ each coefficient is positive,
$V_{0k}(\mathcal{V})V_{k0}(\mathcal{V})\ge 0$, because it corresponds to $\ee{-\beta
  U(\bx)}>0$ multiplied by the square of a real number. Together with
Eq.~(\ref{ss}) this proves that the inequality holds for any $t$ and
completes the proof of the existence and tightness of the bound (\ref{bound}).

Eq.~(\ref{bound}) enables to obtain an upper bound on fluctuations of
$\theta_{\mathcal{V}}(t)$ -- the (generally non-Markovian) \emph{occupation measure}
that the full dynamics $\bx_{\tau}$ along \emph{a single trajectory}
is found within the hypersurface
$V(\bx)=\mathcal{V}$  --  from the integral over the return probability
(\ref{NMP}). It thereby also bounds the fluctuations of random time-average ``empirical
densities'', that is, local time fractions (see
Eq.(\ref{ltf})), by means of the
corresponding deterministic (ensemble) joint return probability
density (\ref{NMP}). Eq.~(\ref{bound}) is the main practical result of this work. 
Interestingly, a similar bound involving the integral of the return
probability has been found in Ref.~\cite{hart19a} in the study of
large deviation asymptotics of the first passage times.

\subsection{Physical interpretation of the results}
We now provide some intuition about the developed theory. As time
evolves the value of $\ovl{V}_t$ for an ergodic process eventually becomes
only weakly correlated.
The statistics of $\ovl{V}_t$ passes first through the large deviation
regime (\ref{Gaussian}), where the central limit theorem kicks in with
Gaussian statistics, and finally ends up in Khinchin's law of large
numbers where it becomes deterministic and equal to $\langle
\ovl{V}_t\rangle_\ii$ \cite{Khinchin}. For simplicity we start in the
large deviation regime (\ref{LDvar}).  

By using spectral theory we map fluctuations of $\ovl{V}_t$ onto
the eigenmodes of $\LL$ (and/or $\LLb$ respectively), with
the ``similarity'' to a given eigenmode reflected by the overlaps
$V_{0k},V_{k0}$. Since on these time-scales all memory of the
initial condition is lost, which is equivalent to imposing stationary initial conditions, only overlaps from and to the ground state are
relevant. 
Moreover, due to the orthogonality of eigenmodes these
projections are statistically independent. Each eigenmode has a finite lifetime or
correlation time $1/\lambda_k$. Therefore, in a time $t\gg \lambda_k^{-1}$
any $k$-th projection acts as shot-noise and 
there will be $t\lambda_k$ independent realizations of such a
projection reducing the (co)variance by a factor $1/t\lambda_k$ (see
Eqs.~(\ref{LDvar}) and (\ref{LDcovar})). In the limit $t\to\infty$ the
Gaussian converges to a Dirac delta, i.e. $\lim_{t\to\infty}\mathcal{P}^{\mathrm{LD}}_t(\boldsymbol{\nu})=\delta(\boldsymbol{\nu}-\langle\mathbf{\ovl{V}}\rangle_{\rm
inv})$. 

At shorter times non-trivial corrections  to these large
deviation results arise due to strong correlations between the values of $\ovl{V}_t$ at different
times $t$. As a result of these correlations the 'completely decorrelated' large deviation
results in Eqs.~(\ref{LDvar}) and (\ref{LDcovar})) becomes reduced by
a term that seems to reflect the 'effective probability of mode $k$ to persist
until $t$',
$t^{-1}\int_0^t\ee{-\lambda_k\tau}d\tau=(1-\ee{-\lambda_kt})/\lambda_kt$
(see Eqs.~(\ref{ss}) and (\ref{eq:mom2mixeq}) as well as
Eqs.~(\ref{dys0}) and ~(\ref{dys})). In the case of general
initial conditions $p_0(\bx)$ additional terms arise (see
Eqs.~(\ref{eq:mom2fix2}) and (\ref{eq:mom2mixfix})) that reflect the
memory of the initial condition. These terms, however, are difficult
to interpret beyond the point that they reflect projections that couple
different excited eigenstates and thus describe fluctuation-modes that are
more complicated than simple excursions starting and ending in the
steady state.


\section{Applications of the theory}
\label{Sec4}
We now apply the theory to a collection of simple illustrative examples. Due to the
fundamental role played by the local time fraction $\theta_{\bx}(t)$
and because it determines the dynamics of other time-average observables 
(see Eq.~(\ref{ltf2})) we focus on
$\theta_{\bx}(t)$ alone. The coupling elements are therefore simply
given by $
V_{lk}=\int_{\Omega}d\by R_l(\by)\delta(\bx-\by)L_k(\by)\equiv R_l(\bx)L_k(\bx)$. We first present explicit results for
local times for
continuous space-time Markovian diffusion processes and an
irreversible (i.e. driven) three-state uni-cyclic network.
Next, we apply
the theory to a simple two-state Markov model of the celebrated Berg-Purcell
problem \cite{berg77,Weigel,Bialek_2005,Aquino}, i.e.
the physical limit to the precision of receptor-mediated measurement of the concentration of ligand
molecules.

As minimal, exactly solvable models of
continuous-space Markovian diffusion we consider a Wiener process confined to
a unit interval with reflective boundaries and the Ornstein-Uhlenbeck
process. To demonstrate the theory for Markov-jump dynamics we
consider a random walk in a finite harmonic potential
and a simple 3-state uni-cyclic network.

\subsection{Local time fraction of the Wiener process in the unit interval}
\begin{figure}
   \includegraphics[width=0.49\textwidth]{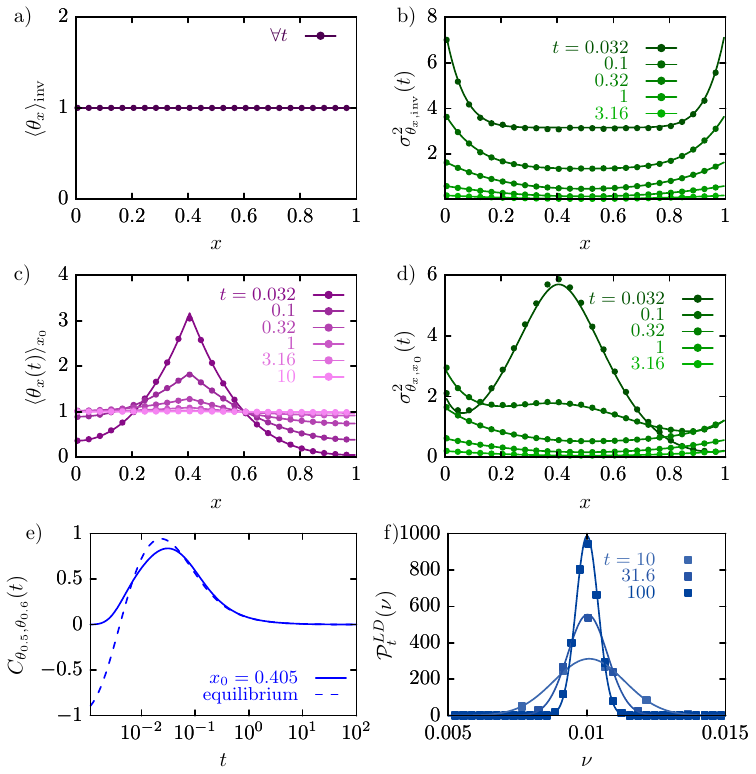}
  \caption{Statistics of the fraction of local-time $\theta_x(t)$ as a
    function of $x$ at different times $t$ for
    equilibrium initial conditions, $p_0(x_0)=P_\ii(x_0)$, (a and b) and localized initial
    condition at $x_0=0.405$ (c and d);
    a) $\langle \theta_x(t)\rangle_\ii$ is constant for all times; b)
    $\sigma_{\theta_x,\ii}^2(t)$ as a function of $x$ for equilibrium
    initial conditions; c) $\langle \theta_x(t)\rangle_{x_0}$ and d)
    $\sigma_{\theta_x}^2(t)$; e) covariance starting from equilibrated
    initial conditions, $C_{\theta_x\theta_y}^{\ii}(t)$, (dashed
    line) and localized initial conditions, $C_{\theta_x\theta_y}(t)$ (solid line). f) Probability density of occupation time fraction
$\theta_{\mathcal{V}}(t)=t^{-1}\int_0^t\mathbbm{1}_{[0.45,0.55]}(x_\tau)d\tau$
    (see also Eq.~(\ref{localH})) on
large deviation time-scales
(symbols) and corresponding theoretical result (\ref{Gaussian})
(lines). Symbols were obtained from Brownian dynamics simulations of $10^{5}$
    trajectories simulated with a time-step $dt=10^{-4}$.}
 \label{MeanVarianceFlat}
\end{figure}

The propagator of the Wiener process confined to a unit
interval (i.e. $L=1$) is the solution of 
\begin{equation}
(\partial_t-\partial^2_x)P^{\W}_t(x|x_0)=0,\,\, \partial_xP^{\W}_t|_{x=0}=\partial_xP^{\W}_t|_{x=1}=0,
 \label{Wiener} 
\end{equation}
with initial condition $P^{\W}_0(x|x_0)=\delta(x-x_0)$.
The eigenvalues of $\partial^2_x$ in a unit interval are given by
$\lambda^\W_k=k^2\pi^2$ (time is expressed in units of
$\tau=L^2/D$) and the eigenvectors read \cite{risken_fokker-planck_1996,
  gardiner_c.w._handbook_1985}, 
\begin{equation}
L^\W_k(x)=R^\W_k(x)=\delta_{k0}+(1-\delta_{k0})\sqrt{2}\cos(k\pi x),
\label{Wexpansion}
\end{equation}
since $\partial^2_x$ is self-adjoint. The
mean local time-fraction, $\langle \theta_x(t)\rangle$, the variance
$\sigma_\theta^2(t)$ and covariance $C_{\theta_1\theta_2}(t)$ for the
confined Wiener process are
shown in Fig.~\ref{MeanVarianceFlat}. In the case of equilibrium
initial conditions $\langle \theta_x(t)\rangle_\ii$ is constant and equal
to $P_\ii(x)$, and the fluctuations of $\theta_x(t)$ are largest at
the boundaries as a result of repeated collisions with the walls. 
Notably, starting from localized conditions $\langle
\theta_x(t)\rangle_{x_0}$ as a function of $x$, in contrast to the ensemble
propagator $P_t^\W(x|x_0)$, displays a persistent cusp
located at the initial condition $x_0$ (see
Fig.~\ref{MeanVarianceFlat} c). The fluctuations of $\theta_x(t)$ are larger near the
initial condition and at the boundaries. Note that the fluctuations
are always larger for equilibrium initial conditions (compare
Fig.~\ref{MeanVarianceFlat}b and  Fig.~\ref{MeanVarianceFlat}d).

\subsection{Local time fraction of the Ornstein-Uhlenbeck process}
\begin{figure}
   \includegraphics[width=0.49\textwidth]{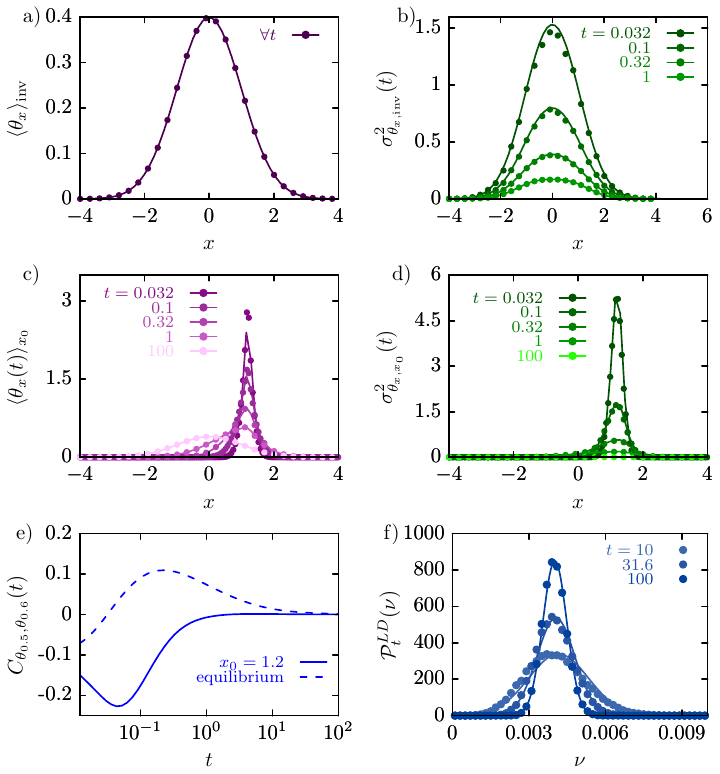}
   \caption{Statistics of the fraction of local-time $\theta_x(t)$ as
     a function of $x$ for
     the Ornstein Uhlenbeck process (symbols) and corresponding
     lattice random walk (\ref{markovHarmonic}) with $10^3$ states in the interval
    $x\in [-5,5]$ as a
    function of $x$ at different times $t$ for
    equilibrium initial conditions, i.e. $p_0(x_0)= P_{\ii}(x_0)$ (a
    and b), and initial conditions localized at $x_0=1.2$
    i.e. $p_0(x_0)=\delta(x_0-1.2)$ (c and d).
    a) $\langle \theta_x(t)\rangle_{\ii}=P_{\ii}(x)$ is a time-independent Gaussian; b)
    $\sigma_{\theta_x,\ii}^2(t)$ at various times as a function of
    $x$;
    c) $\langle \theta_x(t)\rangle_{x_0}$ and d)
    $\sigma_{\theta_x,x_0}^2(t)$ 
     at various times as a function of
    $x$; e) $C_{\theta_{0.5}\theta_{0.6}}(t)$
    for equilibrium (dashed line) and localized (full lines) initial
    conditions; f) occupation time fraction
    $\theta_{\mathcal{V}}(t)=t^{-1}\int_0^t\mathbbm{1}_{[0,0.01]}[x(\tau)]d\tau$
    (see also Eq.~(\ref{localH})) with symbols derived from simulations and
    the solid line representing the theoretical result (\ref{Gaussian}) for 
   the lattice random walk (\ref{markovHarmonic}).
    To obtain each simulation point we generated $10^{5}$ Brownian
    dynamics trajectories using $D=\gamma=1$ with a time-step $dt=10^{-4}$.}
  \label{MeanVarianceHarmonic}
 \end{figure}

Trajectories of the one-dimensional Ornstein-Uhlenbeck process are solutions of the
It\^o equation
\begin{equation}
d x_t=-\gamma x_t+\sqrt{2D}dW_t
  \label{ItoOU}
\end{equation}
and on the level or probability density correspond to the
Fokker-Planck equation
$(\partial_t-D[\partial^2_x+\gamma\partial_xx])P^{\OU}_t(x|x_0)=0$
with initial condition $P^{\OU}_0(x|x_0)=\delta(x-x_0)$ and natural
boundary conditions $\lim_{|x|\to\infty}P^{\OU}_t(x|x_0)=0$.
To connect continuous processes to discrete ones we translate the
Fokker-Planck equation of the Ornstein-Uhlenbeck process to a random
walk on a lattice with spacing $\Delta x$ and the
harmonic potential $\gamma x^2$ entering transition rates according to \cite{holu19}
\begin{eqnarray}
 \bra{x+\Delta x}\LL\ket{x}&=&\frac{D}{\Delta x^2}\ee{\frac14\gamma
   [x^2-(x+\Delta x)^2]},\nonumber\\
 \bra{x}\LL\ket{x+\Delta x}&=&\frac{D}{\Delta x^2}\ee{\frac14\gamma [(x+\Delta x)^2-x^2]}
 \label{markovHarmonic},
\end{eqnarray}
in a confined domain $\Omega_{\rm conf}=\{-l,-l+\Delta x,\ldots,l-\Delta x,l\}\subset\Omega$.
The matrix $\LL$ is tri-diagonal and satisfies $\sum_y \bra{y}\LL\ket{x}=0$ for all $\bx\in \Omega$ and $x\in\mathbb{Z}\Delta x$. We diagonalized $\LL$
numerically using the library from Ref.~\cite{eigenweb}. The mean, variance and correlation function
for the continuous-space Ornstein-Uhlenbeck process (\ref{ItoOU}) obtained
from Brownian dynamics simulations are depicted in
Fig. \ref{MeanVarianceHarmonic}  (symbols) and are in excellent
agreement with the spectral-theoretic results for the corresponding lattice
random walk approximation (\ref{markovHarmonic}) (lines). In Fig.~\ref{MeanVarianceHarmonic}f we also investigate the
full probability density function of the 
fraction of occupation time in the interval $x\in [0,0.01]$,
i.e. $\theta_{\mathcal{V}}(t)=t^{-1}\int_0^t\mathbbm{1}_{[0,0.01]}[x(\tau)]d\tau$
(see Eq.~(\ref{localH}), on large
deviation time scales and compare it to the theoretical Gaussian
prediction Eq.~(\ref{LDvar}). 
Note that while the eigenspectrum of the generator of continuous
Ornstein-Uhlenbeck dynamics is unbounded, implying that the
spectral-theoretic result would require the summation of a large
number of terms, the summation in the lattice approximation is limited
by the number of lattice points. Therefore, except for very short
times, where the lattice approximation naturally breaks down,
this example demonstrates that our formalism applies equally well for Markov jump processes and diffusion dynamics.
Note that the results in Fig.~\ref{MeanVarianceHarmonic}c,d
for times
$t=1$ and $t=100$ correspond to the ``short'' and ``long'' trajectory in Fig.~\ref{histo}, respectively.

\subsection{Local time fraction in a driven uni-cyclic network}
\label{uni-cyclic}
\begin{figure}
  \includegraphics[width=0.48\textwidth]{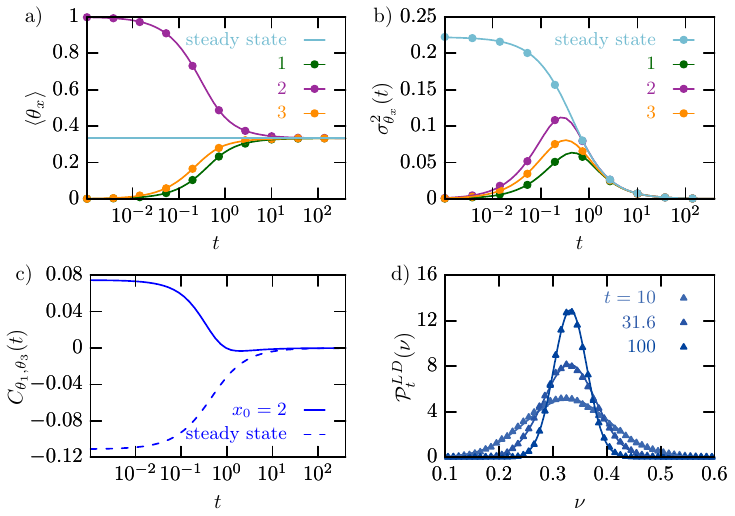}
  \caption{a) Mean local time fraction $\langle \theta_x(t)\rangle$ in
    states $x=1,2,3$ respectively, for a
    driven uni-cyclic network starting from steady-state initial
    conditions ($p_0(x_0)=P_\ii(x_0)$, light blue line; result
    identical for any state) and an initial condition localized at
    $x_0=2$  (lines are theory and points simulations); b)
    variance of the local time fraction $\sigma^2_{\theta,\ii}(t)$ starting from steady-state initial
    conditions (light blue line; result identical for any state) and $\sigma^2_{\theta,2}(t)$
    in states $x=1,2,3$ starting from $x_0=2$; c) covariance of local time
    fraction between states $x=1$ and $x=2$, $C_{\theta_1\theta_2}(t)$, as a function of time for
    stationary (dashed line) and localized (full line) initial
    conditions; d) probability density function of the local time
    fraction in state $x=1$, $\ovl{V}_t\equiv\theta_{1}(t)$, on large deviation time-scales alongside
    theoretical prediction of Eq.~(\ref{Gaussian}).}
 \label{MeanVarianceMM}
\end{figure}

Let us in the following address a simple 3-state model with broken
detailed balance to also address driven systems. The model corresponds
to a simple cycle with states $1,2$ and $3$, where all rates in a given direction are equal but
each of them has the same forward/backward asymmetry. The model may represent,
for example, a molecular motor such as the F1-ATPase driven by ATP
hydrolysis \cite{toya11}. The corresponding transition matrix
of the model reads
 \begin{equation}
 \LL=\left(\begin{array}{ccc}
                -3 & 1  & 2\\
		2  & -3 & 1\\
		1  &   2  & -3
              \end{array}\right),
 \label{markov3}
 \end{equation}
and has eigenvalues $\lambda_0=0$, $\lambda_{1,2}=-9/2\pm
{\rm i}\sqrt{3}/2$ and eigenvectors $|R_0\rangle=3^{-1}(1,1,1)^T$ and
\begin{equation}
|R_{1,2}\rangle=\frac{1}{3}\left(\frac{-3\sqrt{3}\pm {\rm
    i}}{\sqrt{3}\mp 5{\rm i}},\frac{2\sqrt{3}\pm 2{\rm i}}{\sqrt{3}\mp 5{\rm i}},1\right)^T.
\end{equation}
As a result of broken detailed balance the eigenspectrum is
complex. In Fig. \ref{MeanVarianceMM} we analyze the mean (panel a),
fluctuations (panel b) and correlation
function (panel c) of the local time fraction $\theta_x(t)$ in the various
states for non-equilibrium steady-state initial conditions (light blue
lines) and conditions initially localized in state $x_0=2$, i.e. $|p_0\rangle=(0,1,0)^T$. The theoretical results (lines) show an excellent
agreement with simulations (symbols) carried out using the Gillespie algorithm
\cite{gillespie_exact_1977}. We also confirm the Gaussian statistics of
the local time fraction $\theta_x(t)$ from Eq.~(\ref{Gaussian}) in Fig. \ref{MeanVarianceMM}d. 

\subsubsection{Generic behavior of local time fraction in ergodic systems}
\begin{figure}
    \includegraphics[width=0.48\textwidth]{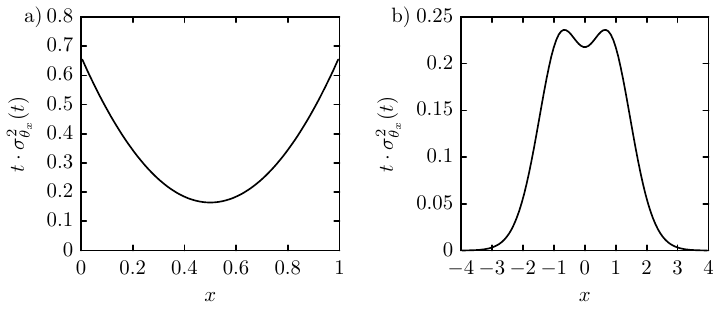}
  \caption{Long time behavior  (i.e. $t\gg \lambda_1^{-1}$) of the
    scaled variance of the local-time, $t\sigma^2_{\theta_x}(t)$, that
    is time-independent; 
    (a) results for the Wiener process and (b) the Ornstein-Uhlenbeck processes.}
  \label{longtimevariance}
 \end{figure}

Note that an exhaustive study of the statistics of the local-time
fraction is beyond the scope of this work. Nevertheless, we here discuss
some general features of $\theta_x(t)$. The manner in which $\langle
\theta_x(t)\rangle$, starting from some non-equilibrium initial
condition, approaches the ergodic invariant measure $P_\ii(x)$ can
be highly non-trivial and even non-monotonic (see e.g.
Figs.~\ref{longtimevariance}a). Even when the ergodic limit is reached,
where the variance ceases to depend on time, i.e. $t\sigma_{\theta_x}^2(t)\ne f(t)$, the
fluctuations display a non-trivial behavior (see
Fig.~\ref{longtimevariance}). For example, in the case of the Wiener process fluctuations are
enhanced close to the boundaries, while for the Ornstein-Uhlenbeck
process they become depressed near the minimum. Both results may be
interpreted in terms of random 'oscillations' around a typical
position and confined by a boundary that amplifies fluctuations.

\begin{figure}
   \includegraphics[width=0.48\textwidth]{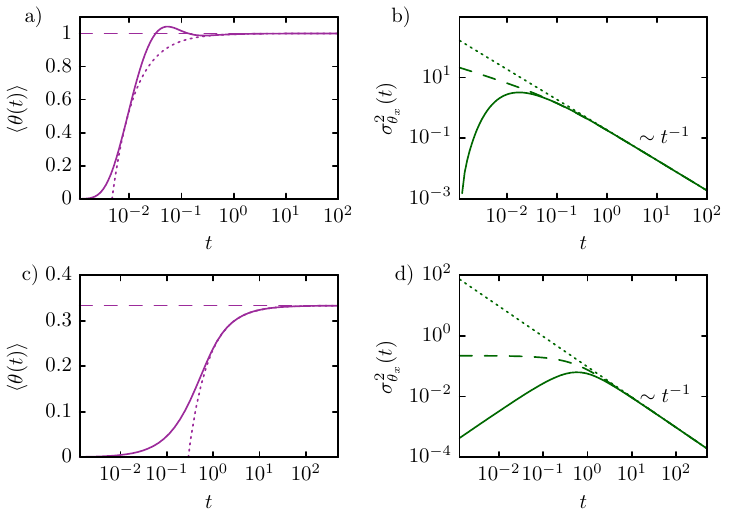}
   \caption{(a) Mean local-time and (b) variance at $x=0.6$ for a
     Wiener process starting from equilibrium (dashed line) and from a
     localized initial condition $p_0(x)=\delta(x-0.405)$ (solid
     line). (a) Mean local-time $\langle \theta_x(t)\rangle$ and (b)
     variance $\sigma^2_{\theta_x}(t)$ at
     $x=0.6$ for the driven 3-state cycle Eq.~(\ref{markov3}) starting
     from stationary (dashed line) and from a localized initial condition $p_0(x)=\delta(x-1)$
     (solid line); The dotted lines correspond to the large
     deviation limit (\ref{LDvar}) and depict the validity and
     long-time saturation of the upper bound Eq.~(\ref{bound}).}
  \label{EqvsFix}
 \end{figure}

 Moreover, the time-dependence of $\langle \theta_x(t)\rangle$ for
 non-stationary initial conditions is often non-monotonic or has a
 non-monotonic derivative (see
 Fig.~\ref{EqvsFix}a and c). A comparison between $\langle
 \theta_x(t)\rangle$  starting from stationary (dashed lines) and
 localized (full lines) initial
 conditions illustrates the two coexisting decorrelation mechanisms of
 $\theta_x(t)$ at different times, one corresponding to self-averaging
 and emergence of the central limit theorem (compare dashed and dotted
 lines), the other additionally reflecting the loss of
 memory of the initial condition (full lines). Stationary initial conditions always give rise to larger
fluctuations than non-stationary initial conditions (compare dashed
and full lines in Fig.~\ref{EqvsFix}b and d), and
in the particular case of equilibrium initial conditions for systems
obeying detailed balance,
$\sigma^2_{\theta}(t)$ is a monotonically decaying function of time
$t$ (see Eq.~(\ref{ss}) and Fig.~\ref{EqvsFix}b and d) with an upper bound given by the large
deviation asymptotic (see Eq.~(\ref{bound}) with a $\propto 1/t$
scaling dictated by the central limit theorem (dotted lines in Fig.~\ref{EqvsFix}b and d).

The covariance of the local time fraction between a pair of points
$x_1$ and $x_2$ in the continuous setting (Figs.~\ref{MeanVarianceFlat}e and \ref{MeanVarianceHarmonic}e) or $x=1$ and $x=2$ in the
discrete setting (Fig.~\ref{MeanVarianceMM}c), $C_{\theta_1\theta_2}(t)$, displays a similarly non-trivial and non-monotonic dependence on time and initial
conditions $p_0(x_0)$ as shown in Figs.~\ref{MeanVarianceFlat}e,
\ref{MeanVarianceHarmonic}e, and \ref{MeanVarianceMM}c. The striking dependence on the tagged points reflects a directional persistence of
individual trajectories in-between said points and can therefore be
used as a robust indicator of directional persistence and thus
'temporally correlated exploration' on the level of
a single trajectory. 

\subsubsection{Universal asymptotic Gaussian limit law for time-average physical observables}
\begin{figure}
   \includegraphics[width=0.45\textwidth]{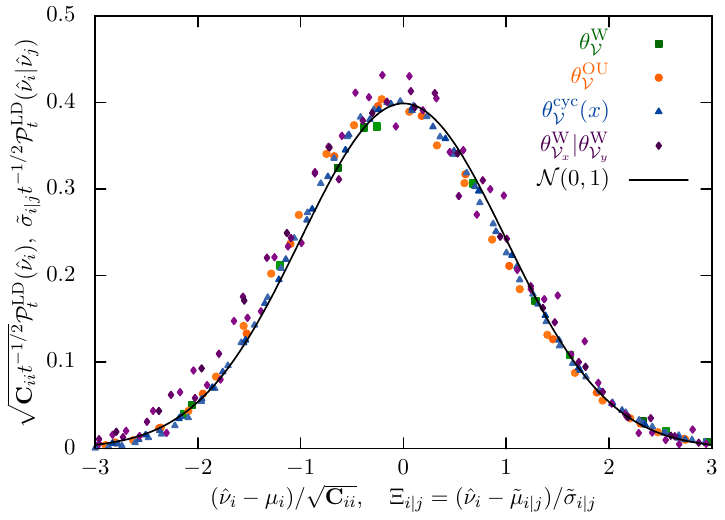}
   \caption{Collapse of probability density functions of all studied
     models at long time onto a unit normal probability density. The
     symbols correspond to rescaled probability density of occupation
     time fraction $\theta_{\mathcal{V}}(t)^{\mathrm{W,OU,cyc}}$ (see
     also Eq.~(\ref{localH})) of the
     Wiener process \eqqref{Wiener}, the Ornstein-Uhlenbeck process \eqqref{ItoOU}, the driven
     3-state cycle \eqqref{markov3}, respectively, and the conditional probability
     density function of occupation time fraction
     $\theta_{\mathcal{V}_x}(t)^{\mathrm{W}}|\theta_{\mathcal{V}_y}(t)^{\mathrm{W}}$ of
     the Wiener process for $\mathcal{V}_x=x\in[0.1,0.4]$ and $\mathcal{V}_y=y\in[0.6,0.9]$. The lines is a zero
     mean unit
   normal probability density function $\mathcal{N}(0,1)$.}
  \label{collapsed}
 \end{figure}

Finally, we comment on the universal asymptotic Gaussian limit law
Eq.~\eqqref{Gaussian} for Markovian as well as non-Markovian time-average physical observables
of type \eqqref{TA} of ergodic stochastic dynamics of the form
given in Eqs.~(\ref{itoe}) and \eqqref{masterequation}. Namely, using
the asymptotic results \eqqref{eq:meanfix2}, \eqqref{LDvar}, and
\eqqref{LDcovar} in the
large-deviation probability density function \eqqref{Gaussian}, and
rescaling to the centered and time-independent variables $\Xi$ and
$\Xi_{i|j}$ defined in Eqs.~\eqqref{dimls} and \eqqref{streched}, we
can rescale the
probability density of any time-average physical observable
$\mathcal{P}_t^{\mathrm{LD}}(\hat{\nu})$, and the conditional probability density of a time-average physical observable
given another time-average physical observable
$\mathcal{P}_t^{\mathrm{LD}}(\hat{\nu}_i|\hat{\nu}_j)$, to collapse at long
times onto a unit normal probability density \eqqref{collapse}. For
the three models studied here, Figs.~\ref{MeanVarianceFlat}f,
\ref{MeanVarianceHarmonic}f, and \ref{MeanVarianceMM}d, and
additionally for the
conditional probability density function of occupation time fraction in $x\in[0.1,0.4]$ given
occupation time fraction in $y\in[0.6,0.9]$ for the Wiener
process, we demonstrate this collapse explicitly in Fig.~\ref{collapsed}.

\subsection{Precision limit of concentration measurement by a single receptor}
Let us now investigate the physical limit to the precision of
concentrations measurements by means of the
simplest two-state Markov jump process with states $\Omega=\{0,1\}$
\cite{berg77}. The receptor can either be occupied by a ligand ($x=1$)
 or be empty ($x=0$). Let the background ligand concentration be $c$ and
 assume that the ligand binds with a rate $kc$ and unbinds
 with rate $k$, ignoring for simplicity any spatial variations of
 concentration. The generator and its eigenvectors are given by
 \begin{equation}
  \LL=\begin{pmatrix}
   -kc&k\\
   kc&-k
  \end{pmatrix},
  \ket{R_0}=
  \frac{1}{1+c}\begin{pmatrix}
   1\\c
  \end{pmatrix},\ket{R_1}=
  \frac{1}{1+c}
  \begin{pmatrix}
   1\\-1
  \end{pmatrix}
  \label{receptor1}
 \end{equation}
with $\lambda_1=-k(1+c)$ being the only non-zero eigenvalue. The left
eigenvectors corresponding to Eq.~\eqqref{receptor1} are
$\bra{L_0}=\bra{-}=(1,1)$ and $\bra{L_1}=(c,-1)$. Moreover, since the
entire state space has only two states we have
$\theta_0(t)=1-\theta_1(t)$. Assuming that the system was initially in equilibrium
$\ket{p_0}=\ket{R_0}$ this implies that the mean values of the
respective local time fractions are given by
$\avg{\theta_1(t)}=(1+c)^{-1}c$ and $\avg{\theta_0(t)}=(1+c)^{-1}$.

If the receptor estimates the concentration $c$ by reading out and
averaging the fraction of time the ligand is bound, $\theta_1$, over an interval of duration
$t$, the precision of the estimate is bounded from above by the
variance of the local time fraction given by
Eq.~(\ref{ss}) and reads explicitly
\begin{equation}
\sigma_{\theta_1}^2(t)=\frac{2c}{(1+c)^3kt}\bigg[1-\frac{1-\ee{-k(1+c)t}}{k(1+c)t}\bigg].
\label{prec_exact}
\end{equation}
Typically one assumes that the measurement $t$ is longer 
than any correlation time \cite{berg77,Bialek_2005,GodecMetz,GodecMetz2}, which in the
present setting implies $t\gg 1/[k(1+c)]$, i.e. much longer than the
correlation time of two-state Markov switching noise, $\tau_c\equiv\lambda_1^{-1}=1/(k+kc)$
\cite{Bialek_2005}. In this regime the
averaging-noise corresponds to shot-noise such that the variance decreases with the number of statistically
independent receptor measurements $\#_t$
\cite{berg77,Bialek_2005,GodecMetz,GodecMetz2}, where $\#_t\sim
t/\tau_c$ is the number of statistically independent realizations of the two-state
process. Therefore $\sigma_{\theta_1}^2(t)\propto 1/\#_t=\tau_c/t$,
according to the central limit theorem.

Based on the bound derived in Eq.~(\ref{bound}) the shot-noise limit is in fact an
upper bound to fluctuations of receptor occupancy at any duration of
measurement, and saturates only in the limit $t\gg
\tau_c$. Namely, a direct application of the bound
\eqqref{bound} indeed yields, using $G^{\rm
  eq}_t(1,1)=P_t(1|1)P_\infty(1|1)$, 
\begin{eqnarray}
 B&\equiv&2\int_0^\infty [G^{\rm eq}_t(1,1)-G^{\rm
     eq}_\infty(1,1)]dt\nonumber\\
 &=&
 2\int_0^\infty \frac{c}{(1+c)^2}\ee{-k(1+c)t}d t =\frac{2c}{(1+c)^3k},
 \label{receptor2}
\end{eqnarray}
implying, according to Eq.~\eqqref{bound} 
\begin{equation}
\sigma_{\theta_1}^2(t)\le
\frac{B}{t}=\frac{2c}{(1+c)^2}\frac{1}{\#_t}\equiv \lim_{t\gg \tau_c}\sigma_{\theta_1}^2(t).
\label{shot}
\end{equation}
Therefore, for short, and particularly finite measurements the shot-noise limit of fluctuations for long receptor read-out
\cite{berg77,Bialek_2005,GodecMetz,GodecMetz2} gives only an upper bound
to the uncertainty of the estimate, whereas the inequality becomes
sharp at long times.

Fundamental bounds on the precision of inferring $c$ from
$\theta_1(t)$ can be found in \cite{berg77}.
Using the entire time trace of the receptor occupancy $x_\tau$ ($0\le \tau\le t$) instead
of the occupation time $\theta_1(t)$ alone, and employing a maximum likelihood estimate of the concentration $c$, the error of the resulting estimate (i.e., its variance) is found to be reduced
further by a factor of $1/2$ \cite{endr09,mora10}. A
detailed discussion of the precision of inferring kinetic parameters
by means of non-local functionals can be found in \cite{hart16a}.

\section{Concluding perspective}
\label{Sec5}
 We developed a general spectral-theoretic approach to time-average
 statistical mechanics, i.e. to the statistics of bounded, local 
additive functionals of (normal)
 ergodic stochastic
processes with continuous and discrete state-spaces. In particular, we have shown how to obtain exactly the
mean, variance and correlations of time-average observables from the
eigenspectrum of the underlying forward or backward
generator. We re-derived the famous Feynman-Kac formulas using
It\^o calculus and included a brief derivation for
Markov-jump processes. We 
combined Feynman-Kac formulas with non-Hermitian perturbation
theory to derive an exact spectral representation of the results. We
demonstrated explicitly, and quantitatively, the emergence of the universal
central limit law in a spectral representation on large deviation time-scales. For
the special case of equilibrated initial conditions and dynamics
obeying detailed balance
we derived a general upper bound on
fluctuations of occupation measures inferred from individual trajectories.
We discussed our theoretical results from a physical
perspective and provided simple but instructive practical examples
to demonstrate how the theory is to be applied. Our work is applicable
to continuous as well as discrete state-space processes, reversible as
well as irreversible, encompassing a wide and diverse range of phenomena involving time-average
observables and additive functionals in physical, chemical, and biological
systems as well as financial mathematics and econophysics.

  \begin{acknowledgments}
The financial support from the German Research Foundation (DFG) through the Emmy Noether Program GO 2762/1-1 to AG is gratefully acknowledged.
  \end{acknowledgments}


 \appendix
 \section{The perturbative calculation \label{pert calc}}
We carry out all calculations with the spectrum of the backward
generator $\LLb$. Equivalent results can be derived using the
forward generator instead. We carry out perturbative calculations (\ref{pert}) up to second
order to derive the results in Eq.~(\ref{PPV}).
 
 \subsection*{Terms of $1^{st}$ order in $u$}
 Starting with a perturbation of the backward 'kets' and collecting
 terms of first order in Eq.~(\ref{pert})  we find
  \begin{equation}
   -\LLb|L_k^1\rangle+V|L_k\rangle=\lambda_k^{(1)} |L_k\rangle+\llambda_k|L_k^1\rangle
\label{SE1}
  \end{equation}
and multiply Eq.~(\ref{SE1}) by $\langle R_l|$ from the left to obtain
  \begin{equation}
  - \langle R_l|\LLb|L_k^1\rangle+\langle R_l|V|L_k\rangle=\lambda_k^{(1)} \delta_{kl}+\llambda_k\langle R_l|L_k^1\rangle.
  \end{equation}
  Therefore, if $k=l$ we find
  \begin{equation}
    \lambda_k^{(1)}=\langle R_k|V|L_k\rangle\equiv V_{kk}
   \label{correction} 
  \end{equation}
 while for $k\neq l$ we obtain
 \begin{equation}
  \langle R_l|L_k^1\rangle=\frac{V_{lk}}{\llambda_k-\llambda_l}
 \end{equation}
 and therefore 
 \begin{equation}
  |L_k^1\rangle=\sum_{l\neq k}\frac{V_{lk}}{\llambda_k-\llambda_l}|L_l\rangle.
 \end{equation}

 We now turn  to the perturbation of $\LLb$ acting on the bra's from
 the right and multiply the resulting first order equation by
 $|L_l\rangle$ from the right to obtain
 \begin{equation}
  -\langle R_k^1|\LLb|L_l\rangle+\langle R_k|V|L_l\rangle=\llambda_k\langle R_k^1|L_l\rangle+\lambda_k^{(1)}\delta_{kl}.
 \end{equation}
 For $k=l$ we obtain the eigenvalue-corrections (\ref{correction})
 while for $k\neq l$ we have 
 \begin{equation}
  \langle R_k^1|=\sum_{l\neq k} \frac{V_{kl}}{\llambda_k-\llambda_l}\langle R_l|
 \end{equation}

 \subsection*{Terms of $2^{nd}$ order in $u$}

 Collecting in Eq.~(\ref{pert}) corrections of second order to the kets we find, upon
 multiplying by $\langle R_l|$ from the left, 
 \begin{eqnarray}
  && -\langle R_l|\LLb|L_k^2\rangle+\langle R_l|V|L_k^1\rangle\nonumber\\
  && =\lambda_k^{(2)}\delta_{lk}+\lambda_k^{(1)}\langle R_l|L_k^1\rangle+\llambda_k\langle R_l|L_k^2\rangle,
 \end{eqnarray}
 yielding, for $k=l$
 \begin{equation}
  \lambda_k^{(2)}=\langle R_k^0|\hat{V}|L_k^1\rangle,
 \end{equation}
because $\langle R_k|L_k^1\rangle=0$ due to the Eq.~(\ref{eq:initialrenorm}) and thus
 \begin{equation}
  \lambda_k^{(2)}=\sum_{l\neq k}
  \frac{V_{kl}V_{lk}}{\llambda_k-\llambda_l}\langle R_l|V|L_k\rangle.
  \label{EV2}
 \end{equation}
 Conversely, if $k\neq l$ we obtain the second order correction
 \begin{equation}
 |L_k^2\rangle=\sum_{l\neq k} \left[ \sum_{i\neq k}\frac{V_{ik}V_{li}}{(\llambda_i-\llambda_k)(\llambda_l-\llambda_k)}-\frac{V_{kk}V_{lk}}{(\llambda_l-\llambda_k)^2}\right] |L_l\rangle.
 \end{equation}
Collecting in Eq.~(\ref{pert}) corrections of second order to the bra's we find, upon
 multiplying from the right by $|L_l\rangle$
 \begin{eqnarray}
 &-&\langle R_k^2|\LLb|L_l\rangle+\langle
   R_k^1|V|L_l\rangle\nonumber\\
 &=&  \lambda_k^{(2)}\delta_{kl}+\lambda_k^{(1)}\langle R_k^1|L_l\rangle+\llambda_k\langle R_k^2|L_l\rangle.
 \end{eqnarray}
When $k=l$ we obtain Eq.~(\ref{EV2}) while in the case when $k\neq l$
we find the second correction to the bra
 \begin{equation}
  \langle R_k^2|=\sum_{l\neq k} \left[\sum_{i\neq k} \frac{V_{ki}V_{il}}{(\llambda_i-\llambda_k)(\llambda_l-\llambda_k)}-\frac{V_{kk}V_{kl}}{(\llambda_l-\llambda_k)^2}\right]\langle R_l|,
 \end{equation}
which completes the derivation of Eq.~(\ref{PPV}).

\section{Derivation via the Dyson series \label{Dyson calc}}

 In a previous publication \cite{lapolla_unfolding_2018} we showed how
 to derive equations for the moments of $\psi_t$ for stationary
 initial conditions, $p_0(\bx)=P_\ii(\bx)$, (Eq.~\eqqref{ss} and \eqqref{eq:mom2mixeq}) using a Dyson series
 approach. Here we sketch how to obtain the moments of $\psi_t$ for
 generic initial conditions $p_0(\bx)$. In contrast to
 Ref. \cite{lapolla_unfolding_2018}  we here use the forward-approach
 and expand the tilted propagator up to the second order in $u$
 \begin{multline}
\langle -|\ee{(\LL-uV)t}|p_0\rangle= 1- u\langle-|\int_0^t dt' \ee{\LL (t-t')}V\ee{\LL t'}|p_0\rangle\\
+  u^2\langle-|\int_0^t dt'\int_0^{t'} dt''
  \ee{\LL (t-t')}V \ee{\LL (t'-t'')}V\ee{\LL t''}|p_0\rangle\\+ \mathcal{O}(u^3),
  \label{eq:dysonseries}
 \end{multline}
 assuming $t>t'>t''>0$. Here
 we confirm that the Dyson series gives results identical to the perturbation-calculation.

 \subsection*{Mean, fluctuations and correlations}
 We now derive $\langle \ovl{V}_t\rangle$,
 $\sigma^2_{\ovl{V}}(t)$ and $C_{\ovl{V}_1\ovl{V}_2}(t)$
 presented in Eqs.~(\ref{eq:meanfix})-(\ref{eq:mom2mixeq}) using the
 Dyson series. Note that this calculation does not diagonalize the
 tilted generator $\LLb-uV(\bx)$ ($\LLb-uV(\bx)$, respectively).
 Starting form Eq. \eqqref{eq:dysonseries} we can carry out all
 integrations analytically  for arbitrary initial conditions $p_0(\bx)$.
  To first order in $u$ we obtain 
  \begin{eqnarray}
 && t^{-1}\langle-|\int_0^t d t'\ee{\LL (t-t')}V\ee{\LL t'}|p_0\rangle \nonumber\\&=&
  t^{-1}\int_0^t d t' \sum_k \sum_l \langle -|R_{l}\rangle \langle L_l|V|R_k\rangle \langle L_k|p_0\rangle
  \ee{-\lambda_l(t-t') -\lambda_k t'}\nonumber\\&=&
 t^{-1} \int_0^t dt' \sum_k \langle -|V|R_k\rangle \langle L_k|p_0\rangle \ee{-\lambda_k t'}\nonumber\\&=&
  V_{00} +\frac{1}{t}\sum_k \frac{V_{k0}}{\lambda_k}\langle L_k|p_0\rangle (1-\ee{-\lambda_k t}),
  \end{eqnarray}
where for $p_0(\bx)=P_\ii(\bx)$ we have $\langle
L_k|p_0\rangle=\delta_{k0}$ leading to $\langle \ovl{V}\rangle_\ii=V_{00}=P_\ii(\bx)$.

To second order in $u$ we find for an arbitrary $p_0(\bx)$
\begin{widetext}
  \begin{eqnarray}
&&t^{-2} \langle-|\int_0^t d t' \int_0^{t'} d t'' \ee{\LL(t-t')}V\ee{\LL(t'-t'')}V\ee{\LL
   t''}|p_0\rangle\nonumber\\
&=&t^{-2} \int_0^t d t' \int_0^{t'} d t''  \langle-| \sum_m |R_m\rangle\langle L_m| \ee{-\lambda_m(t-t')}V
    \sum_k |R_k\rangle\langle L_k| \ee{-\lambda_k(t'-t'')}V\sum_l |R_i\rangle\langle L_l| \ee{-\lambda_l t''} |p_0\rangle\nonumber\\
&=&t^{-2}\sum_k \sum_l V_{0k}V_{kl} \langle L_l|p_0\rangle \int_0^t d
    t' \int_0^{t'} d t'' \ee{-\lambda_k(t'-t'')-\lambda_l t''};
  \label{dys0}  
  \end{eqnarray}
  \end{widetext}
  when $k=l=0$ only $V_{00}^2/2$ survives, while for $k\neq 0$ and $l=0$ we find
  \begin{equation}
  \frac{1}{t}\sum_{k>0}
  \frac{V_{k0}V_{0k}}{\lambda_k}\left(1-\frac{1-\ee{-\lambda_kt}}{t\lambda_k}\right).
  \label{dys}
  \end{equation}
 Conversely,  when $k=0$ and $l\neq 0$ we end up with
  \begin{equation}
 \frac{1}{t} \sum_{l>0} \frac{V_{00}V_{l0}}{\lambda_l}\langle L_l|p_0\rangle \left(1-\frac{1-\ee{-\lambda_lt}}{t\lambda_l}\right),
  \end{equation}
  while for $k=l$ and $k\neq0$ and $l\neq0$ we have
  \begin{equation}
  \frac{1}{t}\sum_{k>0} \frac{V_{k0}V_{kk}}{\lambda_k}\langle L_k|p_0\rangle\left(\ee{-\lambda_kt}-\frac{1-\ee{-\lambda_kt}}{t\lambda_k}\right).
  \end{equation}
 Finally, when $k\neq l\neq0$ we obtain 
  \begin{equation}
  \frac{1}{t^2}\sum_{k>0}\sum_{l>0,l\neq k} \frac{V_{k0} V_{lk}}{\lambda_k-\lambda_l} \langle L_l|p_0\rangle\left(
  \frac{1-\mathrm{e}^{-\lambda_l
      t}}{\lambda_l}
-\frac{1-\mathrm{e}^{-\lambda_k t}}{\lambda_k}
  \right)
  \end{equation}
  The sum of these terms yields the result sought for. The
  corresponding result for stationary initial conditions, $p_0(\bx)=P_\ii(\bx)$, is obtained using $\langle
  L_l|P_\ii\rangle=\delta_{l0}$, which leads to Eq.~(\ref{ss}) and (\ref{eq:mom2mixeq}).
  When  considering correlations we make the replacement $uV\to
  u_1V_1+u_2V_2$ and replace $\partial^2_{u_1}\to\partial^2_{u_1u_2}$ to compute the
  covariance. The formulas above thereby generalize in a  straightforward manner. 

 \bibliography{local_time.bib}

\begin{thebibliography}{84}%
\makeatletter
\providecommand \@ifxundefined [1]{%
 \@ifx{#1\undefined}
}%
\providecommand \@ifnum [1]{%
 \ifnum #1\expandafter \@firstoftwo
 \else \expandafter \@secondoftwo
 \fi
}%
\providecommand \@ifx [1]{%
 \ifx #1\expandafter \@firstoftwo
 \else \expandafter \@secondoftwo
 \fi
}%
\providecommand \natexlab [1]{#1}%
\providecommand \enquote  [1]{``#1''}%
\providecommand \bibnamefont  [1]{#1}%
\providecommand \bibfnamefont [1]{#1}%
\providecommand \citenamefont [1]{#1}%
\providecommand \href@noop [0]{\@secondoftwo}%
\providecommand \href [0]{\begingroup \@sanitize@url \@href}%
\providecommand \@href[1]{\@@startlink{#1}\@@href}%
\providecommand \@@href[1]{\endgroup#1\@@endlink}%
\providecommand \@sanitize@url [0]{\catcode `\\12\catcode `\$12\catcode
  `\&12\catcode `\#12\catcode `\^12\catcode `\_12\catcode `\%12\relax}%
\providecommand \@@startlink[1]{}%
\providecommand \@@endlink[0]{}%
\providecommand \url  [0]{\begingroup\@sanitize@url \@url }%
\providecommand \@url [1]{\endgroup\@href {#1}{\urlprefix }}%
\providecommand \urlprefix  [0]{URL }%
\providecommand \Eprint [0]{\href }%
\providecommand \doibase [0]{http://dx.doi.org/}%
\providecommand \selectlanguage [0]{\@gobble}%
\providecommand \bibinfo  [0]{\@secondoftwo}%
\providecommand \bibfield  [0]{\@secondoftwo}%
\providecommand \translation [1]{[#1]}%
\providecommand \BibitemOpen [0]{}%
\providecommand \bibitemStop [0]{}%
\providecommand \bibitemNoStop [0]{.\EOS\space}%
\providecommand \EOS [0]{\spacefactor3000\relax}%
\providecommand \BibitemShut  [1]{\csname bibitem#1\endcsname}%
\let\auto@bib@innerbib\@empty
\bibitem [{\citenamefont {Saxton}(2008)}]{SPT}%
  \BibitemOpen
  \bibfield  {author} {\bibinfo {author} {\bibfnamefont {M.~J.}\ \bibnamefont
  {Saxton}},\ }\href {\doibase 10.1038/nmeth0808-671} {\bibfield  {journal}
  {\bibinfo  {journal} {Nat. Methods}\ }\textbf {\bibinfo {volume} {5}},\
  \bibinfo {pages} {671} (\bibinfo {year} {2008})}\BibitemShut {NoStop}%
\bibitem [{\citenamefont {Metzler}\ \emph {et~al.}(2014)\citenamefont
  {Metzler}, \citenamefont {Jeon}, \citenamefont {Cherstvy},\ and\
  \citenamefont {Barkai}}]{SPT2}%
  \BibitemOpen
  \bibfield  {author} {\bibinfo {author} {\bibfnamefont {R.}~\bibnamefont
  {Metzler}}, \bibinfo {author} {\bibfnamefont {J.-H.}\ \bibnamefont {Jeon}},
  \bibinfo {author} {\bibfnamefont {A.~G.}\ \bibnamefont {Cherstvy}}, \ and\
  \bibinfo {author} {\bibfnamefont {E.}~\bibnamefont {Barkai}},\ }\href
  {\doibase 10.1039/c4cp03465a} {\bibfield  {journal} {\bibinfo  {journal}
  {Phys. Chem. Chem. Phys.}\ }\textbf {\bibinfo {volume} {16}},\ \bibinfo
  {pages} {24128} (\bibinfo {year} {2014})}\BibitemShut {NoStop}%
\bibitem [{\citenamefont {Ernst}\ \emph {et~al.}(2014)\citenamefont {Ernst},
  \citenamefont {Köhler},\ and\ \citenamefont {Weiss}}]{SPT3}%
  \BibitemOpen
  \bibfield  {author} {\bibinfo {author} {\bibfnamefont {D.}~\bibnamefont
  {Ernst}}, \bibinfo {author} {\bibfnamefont {J.}~\bibnamefont {Köhler}}, \
  and\ \bibinfo {author} {\bibfnamefont {M.}~\bibnamefont {Weiss}},\ }\href
  {\doibase 10.1039/c4cp00292j} {\bibfield  {journal} {\bibinfo  {journal}
  {Phys. Chem. Chem. Phys.}\ }\textbf {\bibinfo {volume} {16}},\ \bibinfo
  {pages} {7686} (\bibinfo {year} {2014})}\BibitemShut {NoStop}%
\bibitem [{\citenamefont {Shen}\ \emph {et~al.}(2017)\citenamefont {Shen},
  \citenamefont {Tauzin}, \citenamefont {Baiyasi}, \citenamefont {Wang},
  \citenamefont {Moringo}, \citenamefont {Shuang},\ and\ \citenamefont
  {Landes}}]{SPT4}%
  \BibitemOpen
  \bibfield  {author} {\bibinfo {author} {\bibfnamefont {H.}~\bibnamefont
  {Shen}}, \bibinfo {author} {\bibfnamefont {L.~J.}\ \bibnamefont {Tauzin}},
  \bibinfo {author} {\bibfnamefont {R.}~\bibnamefont {Baiyasi}}, \bibinfo
  {author} {\bibfnamefont {W.}~\bibnamefont {Wang}}, \bibinfo {author}
  {\bibfnamefont {N.}~\bibnamefont {Moringo}}, \bibinfo {author} {\bibfnamefont
  {B.}~\bibnamefont {Shuang}}, \ and\ \bibinfo {author} {\bibfnamefont {C.~F.}\
  \bibnamefont {Landes}},\ }\href {\doibase 10.1021/acs.chemrev.6b00815}
  {\bibfield  {journal} {\bibinfo  {journal} {Chem. Rev.}\ }\textbf {\bibinfo
  {volume} {117}},\ \bibinfo {pages} {7331} (\bibinfo {year}
  {2017})}\BibitemShut {NoStop}%
\bibitem [{\citenamefont {Hughes}\ and\ \citenamefont {Dougan}(2016)}]{SMS}%
  \BibitemOpen
  \bibfield  {author} {\bibinfo {author} {\bibfnamefont {M.~L.}\ \bibnamefont
  {Hughes}}\ and\ \bibinfo {author} {\bibfnamefont {L.}~\bibnamefont
  {Dougan}},\ }\href {\doibase 10.1088/0034-4885/79/7/076601} {\bibfield
  {journal} {\bibinfo  {journal} {Rep. Prog. Phys.}\ }\textbf {\bibinfo
  {volume} {79}},\ \bibinfo {pages} {076601} (\bibinfo {year}
  {2016})}\BibitemShut {NoStop}%
\bibitem [{\citenamefont {Xie}(1996)}]{SMS2}%
  \BibitemOpen
  \bibfield  {author} {\bibinfo {author} {\bibfnamefont {X.~S.}\ \bibnamefont
  {Xie}},\ }\href {\doibase 10.1021/ar950246m} {\bibfield  {journal} {\bibinfo
  {journal} {Acc. Chem. Res.}\ }\textbf {\bibinfo {volume} {29}},\ \bibinfo
  {pages} {598} (\bibinfo {year} {1996})}\BibitemShut {NoStop}%
\bibitem [{\citenamefont {Ambrose}\ \emph {et~al.}(1999)\citenamefont
  {Ambrose}, \citenamefont {Goodwin}, \citenamefont {Jett}, \citenamefont
  {Van~Orden}, \citenamefont {Werner},\ and\ \citenamefont {Keller}}]{SMS3}%
  \BibitemOpen
  \bibfield  {author} {\bibinfo {author} {\bibfnamefont {W.~P.}\ \bibnamefont
  {Ambrose}}, \bibinfo {author} {\bibfnamefont {P.~M.}\ \bibnamefont
  {Goodwin}}, \bibinfo {author} {\bibfnamefont {J.~H.}\ \bibnamefont {Jett}},
  \bibinfo {author} {\bibfnamefont {A.}~\bibnamefont {Van~Orden}}, \bibinfo
  {author} {\bibfnamefont {J.~H.}\ \bibnamefont {Werner}}, \ and\ \bibinfo
  {author} {\bibfnamefont {R.~A.}\ \bibnamefont {Keller}},\ }\href {\doibase
  10.1021/cr980132z} {\bibfield  {journal} {\bibinfo  {journal} {Chem. Rev.}\
  }\textbf {\bibinfo {volume} {99}},\ \bibinfo {pages} {2929} (\bibinfo {year}
  {1999})}\BibitemShut {NoStop}%
\bibitem [{\citenamefont {Plakhotnik}\ \emph {et~al.}(1997)\citenamefont
  {Plakhotnik}, \citenamefont {Donley},\ and\ \citenamefont {Wild}}]{SMS4}%
  \BibitemOpen
  \bibfield  {author} {\bibinfo {author} {\bibfnamefont {T.}~\bibnamefont
  {Plakhotnik}}, \bibinfo {author} {\bibfnamefont {E.~A.}\ \bibnamefont
  {Donley}}, \ and\ \bibinfo {author} {\bibfnamefont {U.~P.}\ \bibnamefont
  {Wild}},\ }\href {\doibase 10.1146/annurev.physchem.48.1.181} {\bibfield
  {journal} {\bibinfo  {journal} {Annu. Rev. Phys. Chem.}\ }\textbf {\bibinfo
  {volume} {48}},\ \bibinfo {pages} {181} (\bibinfo {year} {1997})}\BibitemShut
  {NoStop}%
\bibitem [{\citenamefont {Neuman}\ and\ \citenamefont {Nagy}(2008)}]{SMS5}%
  \BibitemOpen
  \bibfield  {author} {\bibinfo {author} {\bibfnamefont {K.~C.}\ \bibnamefont
  {Neuman}}\ and\ \bibinfo {author} {\bibfnamefont {A.}~\bibnamefont {Nagy}},\
  }\href {\doibase 10.1038/nmeth.1218} {\bibfield  {journal} {\bibinfo
  {journal} {Nat. Methods}\ }\textbf {\bibinfo {volume} {5}},\ \bibinfo {pages}
  {491} (\bibinfo {year} {2008})}\BibitemShut {NoStop}%
\bibitem [{\citenamefont {Rief}\ and\ \citenamefont
  {Grubm\"uller}(2002)}]{SMS6}%
  \BibitemOpen
  \bibfield  {author} {\bibinfo {author} {\bibfnamefont {M.}~\bibnamefont
  {Rief}}\ and\ \bibinfo {author} {\bibfnamefont {H.}~\bibnamefont
  {Grubm\"uller}},\ }\href {\doibase
  10.1002/1439-7641(20020315)3:3<255::AID-CPHC255>3.0.CO;2-M} {\bibfield
  {journal} {\bibinfo  {journal} {ChemPhysChem}\ }\textbf {\bibinfo {volume}
  {3}},\ \bibinfo {pages} {255} (\bibinfo {year} {2002})}\BibitemShut {NoStop}%
\bibitem [{\citenamefont {Woodside}\ and\ \citenamefont {Block}(2014)}]{SMS7}%
  \BibitemOpen
  \bibfield  {author} {\bibinfo {author} {\bibfnamefont {M.~T.}\ \bibnamefont
  {Woodside}}\ and\ \bibinfo {author} {\bibfnamefont {S.~M.}\ \bibnamefont
  {Block}},\ }\href {\doibase 10.1146/annurev-biophys-051013-022754} {\bibfield
   {journal} {\bibinfo  {journal} {Annu. Rev. Biophys.}\ }\textbf {\bibinfo
  {volume} {43}},\ \bibinfo {pages} {19} (\bibinfo {year} {2014})}\BibitemShut
  {NoStop}%
\bibitem [{\citenamefont {Ritort}(2006)}]{SMS8}%
  \BibitemOpen
  \bibfield  {author} {\bibinfo {author} {\bibfnamefont {F.}~\bibnamefont
  {Ritort}},\ }\href {\doibase 10.1088/0953-8984/18/32/r01} {\bibfield
  {journal} {\bibinfo  {journal} {J. Phys.: Condens. Matter}\ }\textbf
  {\bibinfo {volume} {18}},\ \bibinfo {pages} {R531} (\bibinfo {year}
  {2006})}\BibitemShut {NoStop}%
\bibitem [{\citenamefont {Camunas-Soler}\ \emph {et~al.}(2016)\citenamefont
  {Camunas-Soler}, \citenamefont {Ribezzi-Crivellari},\ and\ \citenamefont
  {Ritort}}]{SMS9}%
  \BibitemOpen
  \bibfield  {author} {\bibinfo {author} {\bibfnamefont {J.}~\bibnamefont
  {Camunas-Soler}}, \bibinfo {author} {\bibfnamefont {M.}~\bibnamefont
  {Ribezzi-Crivellari}}, \ and\ \bibinfo {author} {\bibfnamefont
  {F.}~\bibnamefont {Ritort}},\ }\href {\doibase
  10.1146/annurev-biophys-062215-011158} {\bibfield  {journal} {\bibinfo
  {journal} {Annu. Rev. Biophys.}\ }\textbf {\bibinfo {volume} {45}},\ \bibinfo
  {pages} {65} (\bibinfo {year} {2016})}\BibitemShut {NoStop}%
\bibitem [{\citenamefont {L\'evy}(1940)}]{Levy}%
  \BibitemOpen
  \bibfield  {author} {\bibinfo {author} {\bibfnamefont {P.}~\bibnamefont
  {L\'evy}},\ }\href {http://www.numdam.org/item/CM_1940__7__283_0} {\bibfield
  {journal} {\bibinfo  {journal} {Compositio Mathematica}\ }\textbf {\bibinfo
  {volume} {7}},\ \bibinfo {pages} {283} (\bibinfo {year} {1940})}\BibitemShut
  {NoStop}%
\bibitem [{\citenamefont {Kac}(1949)}]{Kac}%
  \BibitemOpen
  \bibfield  {author} {\bibinfo {author} {\bibfnamefont {M.}~\bibnamefont
  {Kac}},\ }\href {\doibase 10.1090/s0002-9947-1949-0027960-x} {\bibfield
  {journal} {\bibinfo  {journal} {Trans. Amer. Math. Soc.}\ }\textbf {\bibinfo
  {volume} {65}},\ \bibinfo {pages} {1} (\bibinfo {year} {1949})}\BibitemShut
  {NoStop}%
\bibitem [{\citenamefont {Darling}\ and\ \citenamefont {Kac}(1957)}]{Darling}%
  \BibitemOpen
  \bibfield  {author} {\bibinfo {author} {\bibfnamefont {D.~A.}\ \bibnamefont
  {Darling}}\ and\ \bibinfo {author} {\bibfnamefont {M.}~\bibnamefont {Kac}},\
  }\href {\doibase 10.1090/s0002-9947-1957-0084222-7} {\bibfield  {journal}
  {\bibinfo  {journal} {Trans. Am. Math. Soc.}\ }\textbf {\bibinfo {volume}
  {84}},\ \bibinfo {pages} {444} (\bibinfo {year} {1957})}\BibitemShut
  {NoStop}%
\bibitem [{\citenamefont {Lamperti}(1958)}]{Lamperti}%
  \BibitemOpen
  \bibfield  {author} {\bibinfo {author} {\bibfnamefont {J.}~\bibnamefont
  {Lamperti}},\ }\href {\doibase 10.1090/s0002-9947-1958-0094863-x} {\bibfield
  {journal} {\bibinfo  {journal} {Trans. Amer. Math. Soc.}\ }\textbf {\bibinfo
  {volume} {88}},\ \bibinfo {pages} {380} (\bibinfo {year} {1958})}\BibitemShut
  {NoStop}%
\bibitem [{\citenamefont {Feller}(1949)}]{Feller}%
  \BibitemOpen
  \bibfield  {author} {\bibinfo {author} {\bibfnamefont {W.}~\bibnamefont
  {Feller}},\ }\href {\doibase 10.1090/s0002-9947-1949-0032114-7} {\bibfield
  {journal} {\bibinfo  {journal} {Trans. Am. Math. Soc.}\ }\textbf {\bibinfo
  {volume} {67}},\ \bibinfo {pages} {98} (\bibinfo {year} {1949})}\BibitemShut
  {NoStop}%
\bibitem [{\citenamefont {Bingham}(1975)}]{Bingham}%
  \BibitemOpen
  \bibfield  {author} {\bibinfo {author} {\bibfnamefont {N.~H.}\ \bibnamefont
  {Bingham}},\ }\href {\doibase 10.2307/1426397} {\bibfield  {journal}
  {\bibinfo  {journal} {Adv. Appl. Probab.}\ }\textbf {\bibinfo {volume} {7}},\
  \bibinfo {pages} {705} (\bibinfo {year} {1975})}\BibitemShut {NoStop}%
\bibitem [{\citenamefont {Borodin}(1989)}]{Borodin}%
  \BibitemOpen
  \bibfield  {author} {\bibinfo {author} {\bibfnamefont {A.~N.}\ \bibnamefont
  {Borodin}},\ }\href {\doibase 10.1070/rm1989v044n02abeh002050} {\bibfield
  {journal} {\bibinfo  {journal} {Russ. Math. Surv.}\ }\textbf {\bibinfo
  {volume} {44}},\ \bibinfo {pages} {1} (\bibinfo {year} {1989})}\BibitemShut
  {NoStop}%
\bibitem [{\citenamefont {Yen}\ and\ \citenamefont {Yor}(2013)}]{Yor}%
  \BibitemOpen
  \bibfield  {author} {\bibinfo {author} {\bibfnamefont {J.-Y.}\ \bibnamefont
  {Yen}}\ and\ \bibinfo {author} {\bibfnamefont {M.}~\bibnamefont {Yor}},\
  }\href {\doibase 10.1007/978-3-319-01270-4} {\bibfield  {journal} {\bibinfo
  {journal} {Lect. Notes Math.}\ } (\bibinfo {year} {2013}),\
  10.1007/978-3-319-01270-4}\BibitemShut {NoStop}%
\bibitem [{\citenamefont {Yor}(2001)}]{yor_exponential_2001}%
  \BibitemOpen
  \bibfield  {author} {\bibinfo {author} {\bibfnamefont {M.}~\bibnamefont
  {Yor}},\ }\href {\doibase 10.1007/978-3-642-56634-9} {{\selectlanguage
  {english}\emph {\bibinfo {title} {Exponential {Functionals} of {Brownian}
  {Motion} and {Related} {Processes}}}}},\ Springer {Finance} {Lecture}
  {Notes}\ (\bibinfo  {publisher} {Springer-Verlag},\ \bibinfo {address}
  {Berlin Heidelberg},\ \bibinfo {year} {2001})\BibitemShut {NoStop}%
\bibitem [{\citenamefont {Geman}\ and\ \citenamefont
  {Yor}(1993)}]{geman_bessel_1993}%
  \BibitemOpen
  \bibfield  {author} {\bibinfo {author} {\bibfnamefont {H.}~\bibnamefont
  {Geman}}\ and\ \bibinfo {author} {\bibfnamefont {M.}~\bibnamefont {Yor}},\
  }\href {\doibase 10.1111/j.1467-9965.1993.tb00092.x} {\bibfield  {journal}
  {\bibinfo  {journal} {Mathematical Finance}\ }\textbf {\bibinfo {volume}
  {3}},\ \bibinfo {pages} {349} (\bibinfo {year} {1993})}\BibitemShut {NoStop}%
\bibitem [{\citenamefont {Wilemski}\ and\ \citenamefont {Fixman}(1973)}]{WFix}%
  \BibitemOpen
  \bibfield  {author} {\bibinfo {author} {\bibfnamefont {G.}~\bibnamefont
  {Wilemski}}\ and\ \bibinfo {author} {\bibfnamefont {M.}~\bibnamefont
  {Fixman}},\ }\href {\doibase 10.1063/1.1679757} {\bibfield  {journal}
  {\bibinfo  {journal} {J. Chem. Phys.}\ }\textbf {\bibinfo {volume} {58}},\
  \bibinfo {pages} {4009} (\bibinfo {year} {1973})}\BibitemShut {NoStop}%
\bibitem [{\citenamefont {Szabo}(1989)}]{Szabo}%
  \BibitemOpen
  \bibfield  {author} {\bibinfo {author} {\bibfnamefont {A.}~\bibnamefont
  {Szabo}},\ }\href {\doibase 10.1021/j100356a011} {\bibfield  {journal}
  {\bibinfo  {journal} {J. Phys. Chem.}\ }\textbf {\bibinfo {volume} {93}},\
  \bibinfo {pages} {6929} (\bibinfo {year} {1989})}\BibitemShut {NoStop}%
\bibitem [{\citenamefont {Bénichou}\ \emph {et~al.}(2005)\citenamefont
  {Bénichou}, \citenamefont {Coppey}, \citenamefont {Moreau},\ and\
  \citenamefont {Oshanin}}]{Gleb}%
  \BibitemOpen
  \bibfield  {author} {\bibinfo {author} {\bibfnamefont {O.}~\bibnamefont
  {Bénichou}}, \bibinfo {author} {\bibfnamefont {M.}~\bibnamefont {Coppey}},
  \bibinfo {author} {\bibfnamefont {M.}~\bibnamefont {Moreau}}, \ and\ \bibinfo
  {author} {\bibfnamefont {G.}~\bibnamefont {Oshanin}},\ }\href {\doibase
  10.1063/1.2109967} {\bibfield  {journal} {\bibinfo  {journal} {J. Chem.
  Phys.}\ }\textbf {\bibinfo {volume} {123}},\ \bibinfo {pages} {194506}
  (\bibinfo {year} {2005})}\BibitemShut {NoStop}%
\bibitem [{\citenamefont {Grebenkov}(2007)}]{porous}%
  \BibitemOpen
  \bibfield  {author} {\bibinfo {author} {\bibfnamefont {D.~S.}\ \bibnamefont
  {Grebenkov}},\ }\href {\doibase 10.1103/PhysRevE.76.041139} {\bibfield
  {journal} {\bibinfo  {journal} {Phys. Rev. E}\ }\textbf {\bibinfo {volume}
  {76}},\ \bibinfo {pages} {041139} (\bibinfo {year} {2007})}\BibitemShut
  {NoStop}%
\bibitem [{\citenamefont {Berg}\ and\ \citenamefont {Purcell}(1977)}]{berg77}%
  \BibitemOpen
  \bibfield  {author} {\bibinfo {author} {\bibfnamefont {H.~C.}\ \bibnamefont
  {Berg}}\ and\ \bibinfo {author} {\bibfnamefont {E.~M.}\ \bibnamefont
  {Purcell}},\ }\href {\doibase 10.1016/S0006-3495(77)85544-6} {\bibfield
  {journal} {\bibinfo  {journal} {Biophys. J.}\ }\textbf {\bibinfo {volume}
  {20}},\ \bibinfo {pages} {193} (\bibinfo {year} {1977})}\BibitemShut
  {NoStop}%
\bibitem [{\citenamefont {Wiegel}(1983)}]{Weigel}%
  \BibitemOpen
  \bibfield  {author} {\bibinfo {author} {\bibfnamefont {F.}~\bibnamefont
  {Wiegel}},\ }\href {\doibase https://doi.org/10.1016/0370-1573(83)90078-9}
  {\bibfield  {journal} {\bibinfo  {journal} {Phys. Rep.}\ }\textbf {\bibinfo
  {volume} {95}},\ \bibinfo {pages} {283} (\bibinfo {year} {1983})}\BibitemShut
  {NoStop}%
\bibitem [{\citenamefont {Bialek}\ and\ \citenamefont
  {Setayeshgar}(2005)}]{Bialek_2005}%
  \BibitemOpen
  \bibfield  {author} {\bibinfo {author} {\bibfnamefont {W.}~\bibnamefont
  {Bialek}}\ and\ \bibinfo {author} {\bibfnamefont {S.}~\bibnamefont
  {Setayeshgar}},\ }\href {\doibase 10.1073/pnas.0504321102} {\bibfield
  {journal} {\bibinfo  {journal} {Proc. Natl. Acad. Sci. USA}\ }\textbf
  {\bibinfo {volume} {102}},\ \bibinfo {pages} {10040} (\bibinfo {year}
  {2005})}\BibitemShut {NoStop}%
\bibitem [{\citenamefont {Endres}\ and\ \citenamefont
  {Wingreen}(2009)}]{endr09}%
  \BibitemOpen
  \bibfield  {author} {\bibinfo {author} {\bibfnamefont {R.~G.}\ \bibnamefont
  {Endres}}\ and\ \bibinfo {author} {\bibfnamefont {N.~S.}\ \bibnamefont
  {Wingreen}},\ }\href {\doibase 10.1103/PhysRevLett.103.158101} {\bibfield
  {journal} {\bibinfo  {journal} {Phys. Rev. Lett.}\ }\textbf {\bibinfo
  {volume} {103}},\ \bibinfo {pages} {158101} (\bibinfo {year}
  {2009})}\BibitemShut {NoStop}%
\bibitem [{\citenamefont {Mora}\ and\ \citenamefont {Wingreen}(2010)}]{mora10}%
  \BibitemOpen
  \bibfield  {author} {\bibinfo {author} {\bibfnamefont {T.}~\bibnamefont
  {Mora}}\ and\ \bibinfo {author} {\bibfnamefont {N.~S.}\ \bibnamefont
  {Wingreen}},\ }\href {\doibase 10.1103/PhysRevLett.104.248101} {\bibfield
  {journal} {\bibinfo  {journal} {Phys. Rev. Lett.}\ }\textbf {\bibinfo
  {volume} {104}},\ \bibinfo {pages} {248101} (\bibinfo {year}
  {2010})}\BibitemShut {NoStop}%
\bibitem [{\citenamefont {Lang}\ \emph {et~al.}(2014)\citenamefont {Lang},
  \citenamefont {Fisher}, \citenamefont {Mora},\ and\ \citenamefont
  {Mehta}}]{lang14}%
  \BibitemOpen
  \bibfield  {author} {\bibinfo {author} {\bibfnamefont {A.~H.}\ \bibnamefont
  {Lang}}, \bibinfo {author} {\bibfnamefont {C.~K.}\ \bibnamefont {Fisher}},
  \bibinfo {author} {\bibfnamefont {T.}~\bibnamefont {Mora}}, \ and\ \bibinfo
  {author} {\bibfnamefont {P.}~\bibnamefont {Mehta}},\ }\href {\doibase
  10.1103/PhysRevLett.113.148103} {\bibfield  {journal} {\bibinfo  {journal}
  {Phys. Rev. Lett.}\ }\textbf {\bibinfo {volume} {113}},\ \bibinfo {pages}
  {148103} (\bibinfo {year} {2014})}\BibitemShut {NoStop}%
\bibitem [{\citenamefont {Mora}(2015)}]{mora15}%
  \BibitemOpen
  \bibfield  {author} {\bibinfo {author} {\bibfnamefont {T.}~\bibnamefont
  {Mora}},\ }\href {\doibase 10.1103/PhysRevLett.115.038102} {\bibfield
  {journal} {\bibinfo  {journal} {Phys. Rev. Lett.}\ }\textbf {\bibinfo
  {volume} {115}},\ \bibinfo {pages} {038102} (\bibinfo {year}
  {2015})}\BibitemShut {NoStop}%
\bibitem [{\citenamefont {Barato}\ and\ \citenamefont
  {Seifert}(2015)}]{bara15a}%
  \BibitemOpen
  \bibfield  {author} {\bibinfo {author} {\bibfnamefont {A.~C.}\ \bibnamefont
  {Barato}}\ and\ \bibinfo {author} {\bibfnamefont {U.}~\bibnamefont
  {Seifert}},\ }\href {\doibase 10.1103/PhysRevE.92.032127} {\bibfield
  {journal} {\bibinfo  {journal} {Phys. Rev. E}\ }\textbf {\bibinfo {volume}
  {92}},\ \bibinfo {pages} {032127} (\bibinfo {year} {2015})}\BibitemShut
  {NoStop}%
\bibitem [{\citenamefont {Aquino}\ \emph {et~al.}(2015)\citenamefont {Aquino},
  \citenamefont {Wingreen},\ and\ \citenamefont {Endres}}]{Aquino}%
  \BibitemOpen
  \bibfield  {author} {\bibinfo {author} {\bibfnamefont {G.}~\bibnamefont
  {Aquino}}, \bibinfo {author} {\bibfnamefont {N.~S.}\ \bibnamefont
  {Wingreen}}, \ and\ \bibinfo {author} {\bibfnamefont {R.~G.}\ \bibnamefont
  {Endres}},\ }\href {\doibase 10.1007/s10955-015-1412-9} {\bibfield  {journal}
  {\bibinfo  {journal} {J. Stat. Phys.}\ }\textbf {\bibinfo {volume} {162}},\
  \bibinfo {pages} {1353} (\bibinfo {year} {2015})}\BibitemShut {NoStop}%
\bibitem [{\citenamefont {Hartich}\ and\ \citenamefont
  {Seifert}(2016)}]{hart16a}%
  \BibitemOpen
  \bibfield  {author} {\bibinfo {author} {\bibfnamefont {D.}~\bibnamefont
  {Hartich}}\ and\ \bibinfo {author} {\bibfnamefont {U.}~\bibnamefont
  {Seifert}},\ }\href {\doibase 10.1103/PhysRevE.94.042416} {\bibfield
  {journal} {\bibinfo  {journal} {Phys. Rev. E}\ }\textbf {\bibinfo {volume}
  {94}},\ \bibinfo {pages} {042416} (\bibinfo {year} {2016})}\BibitemShut
  {NoStop}%
\bibitem [{\citenamefont {Ferraro}\ and\ \citenamefont
  {Zaninetti}(2004)}]{astro}%
  \BibitemOpen
  \bibfield  {author} {\bibinfo {author} {\bibfnamefont {M.}~\bibnamefont
  {Ferraro}}\ and\ \bibinfo {author} {\bibfnamefont {L.}~\bibnamefont
  {Zaninetti}},\ }\href {\doibase https://doi.org/10.1016/j.physa.2004.01.062}
  {\bibfield  {journal} {\bibinfo  {journal} {Physica A}\ }\textbf {\bibinfo
  {volume} {338}},\ \bibinfo {pages} {307} (\bibinfo {year}
  {2004})}\BibitemShut {NoStop}%
\bibitem [{\citenamefont {Gandjbakhche}\ and\ \citenamefont
  {Weiss}(2000)}]{diag}%
  \BibitemOpen
  \bibfield  {author} {\bibinfo {author} {\bibfnamefont {A.~H.}\ \bibnamefont
  {Gandjbakhche}}\ and\ \bibinfo {author} {\bibfnamefont {G.~H.}\ \bibnamefont
  {Weiss}},\ }\href {\doibase 10.1103/PhysRevE.61.6958} {\bibfield  {journal}
  {\bibinfo  {journal} {Phys. Rev. E}\ }\textbf {\bibinfo {volume} {61}},\
  \bibinfo {pages} {6958} (\bibinfo {year} {2000})}\BibitemShut {NoStop}%
\bibitem [{\citenamefont {Weiss}\ and\ \citenamefont
  {Calabrese}(1996)}]{optical}%
  \BibitemOpen
  \bibfield  {author} {\bibinfo {author} {\bibfnamefont {G.~H.}\ \bibnamefont
  {Weiss}}\ and\ \bibinfo {author} {\bibfnamefont {P.~P.}\ \bibnamefont
  {Calabrese}},\ }\href {\doibase
  https://doi.org/10.1016/S0378-4371(96)00362-7} {\bibfield  {journal}
  {\bibinfo  {journal} {Physica A}\ }\textbf {\bibinfo {volume} {234}},\
  \bibinfo {pages} {443} (\bibinfo {year} {1996})}\BibitemShut {NoStop}%
\bibitem [{\citenamefont {Toroczkai}\ \emph {et~al.}(1999)\citenamefont
  {Toroczkai}, \citenamefont {Newman},\ and\ \citenamefont
  {Das~Sarma}}]{growing}%
  \BibitemOpen
  \bibfield  {author} {\bibinfo {author} {\bibfnamefont {Z.}~\bibnamefont
  {Toroczkai}}, \bibinfo {author} {\bibfnamefont {T.~J.}\ \bibnamefont
  {Newman}}, \ and\ \bibinfo {author} {\bibfnamefont {S.}~\bibnamefont
  {Das~Sarma}},\ }\href {\doibase 10.1103/PhysRevE.60.R1115} {\bibfield
  {journal} {\bibinfo  {journal} {Phys. Rev. E}\ }\textbf {\bibinfo {volume}
  {60}},\ \bibinfo {pages} {R1115} (\bibinfo {year} {1999})}\BibitemShut
  {NoStop}%
\bibitem [{\citenamefont {Brokmann}\ \emph {et~al.}(2003)\citenamefont
  {Brokmann}, \citenamefont {Hermier}, \citenamefont {Messin}, \citenamefont
  {Desbiolles}, \citenamefont {Bouchaud},\ and\ \citenamefont
  {Dahan}}]{blinking}%
  \BibitemOpen
  \bibfield  {author} {\bibinfo {author} {\bibfnamefont {X.}~\bibnamefont
  {Brokmann}}, \bibinfo {author} {\bibfnamefont {J.-P.}\ \bibnamefont
  {Hermier}}, \bibinfo {author} {\bibfnamefont {G.}~\bibnamefont {Messin}},
  \bibinfo {author} {\bibfnamefont {P.}~\bibnamefont {Desbiolles}}, \bibinfo
  {author} {\bibfnamefont {J.-P.}\ \bibnamefont {Bouchaud}}, \ and\ \bibinfo
  {author} {\bibfnamefont {M.}~\bibnamefont {Dahan}},\ }\href {\doibase
  10.1103/PhysRevLett.90.120601} {\bibfield  {journal} {\bibinfo  {journal}
  {Phys. Rev. Lett.}\ }\textbf {\bibinfo {volume} {90}},\ \bibinfo {pages}
  {120601} (\bibinfo {year} {2003})}\BibitemShut {NoStop}%
\bibitem [{\citenamefont {Stefani}\ \emph {et~al.}(2009)\citenamefont
  {Stefani}, \citenamefont {Hoogenboom},\ and\ \citenamefont
  {Barkai}}]{blinking2}%
  \BibitemOpen
  \bibfield  {author} {\bibinfo {author} {\bibfnamefont {F.~D.}\ \bibnamefont
  {Stefani}}, \bibinfo {author} {\bibfnamefont {J.~P.}\ \bibnamefont
  {Hoogenboom}}, \ and\ \bibinfo {author} {\bibfnamefont {E.}~\bibnamefont
  {Barkai}},\ }\href {\doibase 10.1063/1.3086100} {\bibfield  {journal}
  {\bibinfo  {journal} {Phys. Today}\ }\textbf {\bibinfo {volume} {62}},\
  \bibinfo {pages} {34} (\bibinfo {year} {2009})}\BibitemShut {NoStop}%
\bibitem [{\citenamefont {Comtet}\ \emph {et~al.}(2005)\citenamefont {Comtet},
  \citenamefont {Desbois},\ and\ \citenamefont
  {Texier}}]{comtet_functionals_2005}%
  \BibitemOpen
  \bibfield  {author} {\bibinfo {author} {\bibfnamefont {A.}~\bibnamefont
  {Comtet}}, \bibinfo {author} {\bibfnamefont {J.}~\bibnamefont {Desbois}}, \
  and\ \bibinfo {author} {\bibfnamefont {C.}~\bibnamefont {Texier}},\ }\href
  {\doibase 10.1088/0305-4470/38/37/R01} {\bibfield  {journal} {\bibinfo
  {journal} {J. Phys. A: Math. Gen.}\ }\textbf {\bibinfo {volume} {38}},\
  \bibinfo {pages} {R341} (\bibinfo {year} {2005})}\BibitemShut {NoStop}%
\bibitem [{\citenamefont {Majumdar}\ and\ \citenamefont
  {Bray}(2002)}]{majumdar_large-deviation_2002}%
  \BibitemOpen
  \bibfield  {author} {\bibinfo {author} {\bibfnamefont {S.~N.}\ \bibnamefont
  {Majumdar}}\ and\ \bibinfo {author} {\bibfnamefont {A.~J.}\ \bibnamefont
  {Bray}},\ }\href {\doibase 10.1103/PhysRevE.65.051112} {\bibfield  {journal}
  {\bibinfo  {journal} {Phys. Rev. E}\ }\textbf {\bibinfo {volume} {65}},\
  \bibinfo {pages} {051112} (\bibinfo {year} {2002})}\BibitemShut {NoStop}%
\bibitem [{\citenamefont {{Majumdar, Satya
  N.}}(2005)}]{majumdar_satya_n._brownian_2005}%
  \BibitemOpen
  \bibfield  {author} {\bibinfo {author} {\bibnamefont {{Majumdar, Satya
  N.}}},\ }\href@noop {} {\bibfield  {journal} {\bibinfo  {journal} {Curr.
  Sci.}\ }\textbf {\bibinfo {volume} {89}},\ \bibinfo {pages} {2079} (\bibinfo
  {year} {2005})}\BibitemShut {NoStop}%
\bibitem [{\citenamefont {Wennmalm}\ \emph {et~al.}(1997)\citenamefont
  {Wennmalm}, \citenamefont {Edman},\ and\ \citenamefont {Rigler}}]{SM1}%
  \BibitemOpen
  \bibfield  {author} {\bibinfo {author} {\bibfnamefont {S.}~\bibnamefont
  {Wennmalm}}, \bibinfo {author} {\bibfnamefont {L.}~\bibnamefont {Edman}}, \
  and\ \bibinfo {author} {\bibfnamefont {R.}~\bibnamefont {Rigler}},\ }\href
  {\doibase 10.1073/pnas.94.20.10641} {\bibfield  {journal} {\bibinfo
  {journal} {Proc. Natl. Acad. Sci. USA}\ }\textbf {\bibinfo {volume} {94}},\
  \bibinfo {pages} {10641} (\bibinfo {year} {1997})}\BibitemShut {NoStop}%
\bibitem [{\citenamefont {Gopich}\ and\ \citenamefont {Szabo}(2012)}]{SM2}%
  \BibitemOpen
  \bibfield  {author} {\bibinfo {author} {\bibfnamefont {I.~V.}\ \bibnamefont
  {Gopich}}\ and\ \bibinfo {author} {\bibfnamefont {A.}~\bibnamefont {Szabo}},\
  }\href {\doibase 10.1073/pnas.1205120109} {\bibfield  {journal} {\bibinfo
  {journal} {Proc. Natl. Acad. Sci. USA}\ }\textbf {\bibinfo {volume} {109}},\
  \bibinfo {pages} {7747} (\bibinfo {year} {2012})}\BibitemShut {NoStop}%
\bibitem [{\citenamefont {Fleury}\ \emph {et~al.}(2000)\citenamefont {Fleury},
  \citenamefont {Segura}, \citenamefont {Zumofen}, \citenamefont {Hecht},\ and\
  \citenamefont {Wild}}]{Zumofen}%
  \BibitemOpen
  \bibfield  {author} {\bibinfo {author} {\bibfnamefont {L.}~\bibnamefont
  {Fleury}}, \bibinfo {author} {\bibfnamefont {J.-M.}\ \bibnamefont {Segura}},
  \bibinfo {author} {\bibfnamefont {G.}~\bibnamefont {Zumofen}}, \bibinfo
  {author} {\bibfnamefont {B.}~\bibnamefont {Hecht}}, \ and\ \bibinfo {author}
  {\bibfnamefont {U.~P.}\ \bibnamefont {Wild}},\ }\href {\doibase
  10.1103/PhysRevLett.84.1148} {\bibfield  {journal} {\bibinfo  {journal}
  {Phys. Rev. Lett.}\ }\textbf {\bibinfo {volume} {84}},\ \bibinfo {pages}
  {1148} (\bibinfo {year} {2000})}\BibitemShut {NoStop}%
\bibitem [{\citenamefont {Barkai}\ \emph {et~al.}(2004)\citenamefont {Barkai},
  \citenamefont {Jung},\ and\ \citenamefont {Silbey}}]{Eli}%
  \BibitemOpen
  \bibfield  {author} {\bibinfo {author} {\bibfnamefont {E.}~\bibnamefont
  {Barkai}}, \bibinfo {author} {\bibfnamefont {Y.}~\bibnamefont {Jung}}, \ and\
  \bibinfo {author} {\bibfnamefont {R.}~\bibnamefont {Silbey}},\ }\href
  {\doibase 10.1146/annurev.physchem.55.111803.143246} {\bibfield  {journal}
  {\bibinfo  {journal} {Annu. Rev. Phys. Chem.}\ }\textbf {\bibinfo {volume}
  {55}},\ \bibinfo {pages} {457} (\bibinfo {year} {2004})}\BibitemShut
  {NoStop}%
\bibitem [{\citenamefont {Agmon}(2010)}]{SMD}%
  \BibitemOpen
  \bibfield  {author} {\bibinfo {author} {\bibfnamefont {N.}~\bibnamefont
  {Agmon}},\ }\href {\doibase https://doi.org/10.1016/j.cplett.2010.08.019}
  {\bibfield  {journal} {\bibinfo  {journal} {Chem. Phys. Lett.}\ }\textbf
  {\bibinfo {volume} {497}},\ \bibinfo {pages} {184} (\bibinfo {year}
  {2010})}\BibitemShut {NoStop}%
\bibitem [{\citenamefont {Sabhapandit}\ \emph {et~al.}(2006)\citenamefont
  {Sabhapandit}, \citenamefont {Majumdar},\ and\ \citenamefont
  {Comtet}}]{sabhapandit_statistical_2006}%
  \BibitemOpen
  \bibfield  {author} {\bibinfo {author} {\bibfnamefont {S.}~\bibnamefont
  {Sabhapandit}}, \bibinfo {author} {\bibfnamefont {S.~N.}\ \bibnamefont
  {Majumdar}}, \ and\ \bibinfo {author} {\bibfnamefont {A.}~\bibnamefont
  {Comtet}},\ }\href {\doibase 10.1103/PhysRevE.73.051102} {\bibfield
  {journal} {\bibinfo  {journal} {Phys. Rev. E}\ }\textbf {\bibinfo {volume}
  {73}},\ \bibinfo {pages} {051102} (\bibinfo {year} {2006})}\BibitemShut
  {NoStop}%
\bibitem [{\citenamefont {Majumdar}\ and\ \citenamefont
  {Comtet}(2002)}]{majumdar_local_2002}%
  \BibitemOpen
  \bibfield  {author} {\bibinfo {author} {\bibfnamefont {S.~N.}\ \bibnamefont
  {Majumdar}}\ and\ \bibinfo {author} {\bibfnamefont {A.}~\bibnamefont
  {Comtet}},\ }\href {\doibase 10.1103/PhysRevLett.89.060601} {\bibfield
  {journal} {\bibinfo  {journal} {Phys. Rev. Lett.}\ }\textbf {\bibinfo
  {volume} {89}},\ \bibinfo {pages} {060601} (\bibinfo {year}
  {2002})}\BibitemShut {NoStop}%
\bibitem [{\citenamefont {Bel}\ and\ \citenamefont {Barkai}(2005)}]{Bel}%
  \BibitemOpen
  \bibfield  {author} {\bibinfo {author} {\bibfnamefont {G.}~\bibnamefont
  {Bel}}\ and\ \bibinfo {author} {\bibfnamefont {E.}~\bibnamefont {Barkai}},\
  }\href {\doibase 10.1103/PhysRevLett.94.240602} {\bibfield  {journal}
  {\bibinfo  {journal} {Phys. Rev. Lett.}\ }\textbf {\bibinfo {volume} {94}},\
  \bibinfo {pages} {240602} (\bibinfo {year} {2005})}\BibitemShut {NoStop}%
\bibitem [{\citenamefont {Carmi}\ and\ \citenamefont {Barkai}(2011)}]{Carmi}%
  \BibitemOpen
  \bibfield  {author} {\bibinfo {author} {\bibfnamefont {S.}~\bibnamefont
  {Carmi}}\ and\ \bibinfo {author} {\bibfnamefont {E.}~\bibnamefont {Barkai}},\
  }\href {\doibase 10.1103/PhysRevE.84.061104} {\bibfield  {journal} {\bibinfo
  {journal} {Phys. Rev. E}\ }\textbf {\bibinfo {volume} {84}},\ \bibinfo
  {pages} {061104} (\bibinfo {year} {2011})}\BibitemShut {NoStop}%
\bibitem [{\citenamefont {Dhar}\ and\ \citenamefont
  {Majumdar}(1999)}]{dhar_residence_1999}%
  \BibitemOpen
  \bibfield  {author} {\bibinfo {author} {\bibfnamefont {A.}~\bibnamefont
  {Dhar}}\ and\ \bibinfo {author} {\bibfnamefont {S.~N.}\ \bibnamefont
  {Majumdar}},\ }\href {\doibase 10.1103/PhysRevE.59.6413} {\bibfield
  {journal} {\bibinfo  {journal} {Phys. Rev. E}\ }\textbf {\bibinfo {volume}
  {59}},\ \bibinfo {pages} {6413} (\bibinfo {year} {1999})}\BibitemShut
  {NoStop}%
\bibitem [{\citenamefont {Majumdar}\ and\ \citenamefont {Dean}(2002)}]{Dean}%
  \BibitemOpen
  \bibfield  {author} {\bibinfo {author} {\bibfnamefont {S.~N.}\ \bibnamefont
  {Majumdar}}\ and\ \bibinfo {author} {\bibfnamefont {D.~S.}\ \bibnamefont
  {Dean}},\ }\href {\doibase 10.1103/PhysRevE.66.041102} {\bibfield  {journal}
  {\bibinfo  {journal} {Phys. Rev. E}\ }\textbf {\bibinfo {volume} {66}},\
  \bibinfo {pages} {041102} (\bibinfo {year} {2002})}\BibitemShut {NoStop}%
\bibitem [{\citenamefont {Bray}\ \emph {et~al.}(2013)\citenamefont {Bray},
  \citenamefont {Majumdar},\ and\ \citenamefont {Schehr}}]{Bray_2013}%
  \BibitemOpen
  \bibfield  {author} {\bibinfo {author} {\bibfnamefont {A.~J.}\ \bibnamefont
  {Bray}}, \bibinfo {author} {\bibfnamefont {S.~N.}\ \bibnamefont {Majumdar}},
  \ and\ \bibinfo {author} {\bibfnamefont {G.}~\bibnamefont {Schehr}},\ }\href
  {\doibase 10.1080/00018732.2013.803819} {\bibfield  {journal} {\bibinfo
  {journal} {Adv. Phys.}\ }\textbf {\bibinfo {volume} {62}},\ \bibinfo {pages}
  {225} (\bibinfo {year} {2013})}\BibitemShut {NoStop}%
\bibitem [{\citenamefont {Lapolla}\ and\ \citenamefont
  {Godec}(2018)}]{lapolla_unfolding_2018}%
  \BibitemOpen
  \bibfield  {author} {\bibinfo {author} {\bibfnamefont {A.}~\bibnamefont
  {Lapolla}}\ and\ \bibinfo {author} {\bibfnamefont {A.}~\bibnamefont
  {Godec}},\ }\href {\doibase 10.1088/1367-2630/aaea1b} {\bibfield  {journal}
  {\bibinfo  {journal} {New J. Phys.}\ }\textbf {\bibinfo {volume} {20}},\
  \bibinfo {pages} {113021} (\bibinfo {year} {2018})}\BibitemShut {NoStop}%
\bibitem [{\citenamefont {Lapolla}\ and\ \citenamefont
  {Godec}(2019)}]{Lapolla_2019}%
  \BibitemOpen
  \bibfield  {author} {\bibinfo {author} {\bibfnamefont {A.}~\bibnamefont
  {Lapolla}}\ and\ \bibinfo {author} {\bibfnamefont {A.}~\bibnamefont
  {Godec}},\ }\href {\doibase 10.3389/fphy.2019.00182} {\bibfield  {journal}
  {\bibinfo  {journal} {Front. Phys.}\ }\textbf {\bibinfo {volume} {7}},\
  \bibinfo {pages} {182} (\bibinfo {year} {2019})}\BibitemShut {NoStop}%
\bibitem [{\citenamefont {Boyer}\ \emph
  {et~al.}(2012{\natexlab{a}})\citenamefont {Boyer}, \citenamefont {Dean},
  \citenamefont {Mej\'{\i}a-Monasterio},\ and\ \citenamefont
  {Oshanin}}]{Gleb1}%
  \BibitemOpen
  \bibfield  {author} {\bibinfo {author} {\bibfnamefont {D.}~\bibnamefont
  {Boyer}}, \bibinfo {author} {\bibfnamefont {D.~S.}\ \bibnamefont {Dean}},
  \bibinfo {author} {\bibfnamefont {C.}~\bibnamefont {Mej\'{\i}a-Monasterio}},
  \ and\ \bibinfo {author} {\bibfnamefont {G.}~\bibnamefont {Oshanin}},\ }\href
  {\doibase 10.1103/PhysRevE.86.060101} {\bibfield  {journal} {\bibinfo
  {journal} {Phys. Rev. E}\ }\textbf {\bibinfo {volume} {86}},\ \bibinfo
  {pages} {060101} (\bibinfo {year} {2012}{\natexlab{a}})}\BibitemShut
  {NoStop}%
\bibitem [{\citenamefont {Boyer}\ \emph
  {et~al.}(2012{\natexlab{b}})\citenamefont {Boyer}, \citenamefont {Dean},
  \citenamefont {Mej\'{\i}a-Monasterio},\ and\ \citenamefont
  {Oshanin}}]{Gleb2}%
  \BibitemOpen
  \bibfield  {author} {\bibinfo {author} {\bibfnamefont {D.}~\bibnamefont
  {Boyer}}, \bibinfo {author} {\bibfnamefont {D.~S.}\ \bibnamefont {Dean}},
  \bibinfo {author} {\bibfnamefont {C.}~\bibnamefont {Mej\'{\i}a-Monasterio}},
  \ and\ \bibinfo {author} {\bibfnamefont {G.}~\bibnamefont {Oshanin}},\ }\href
  {\doibase 10.1103/PhysRevE.85.031136} {\bibfield  {journal} {\bibinfo
  {journal} {Phys. Rev. E}\ }\textbf {\bibinfo {volume} {85}},\ \bibinfo
  {pages} {031136} (\bibinfo {year} {2012}{\natexlab{b}})}\BibitemShut
  {NoStop}%
\bibitem [{\citenamefont {Jørgensen}\ \emph {et~al.}(2005)\citenamefont
  {Jørgensen}, \citenamefont {Mann}, \citenamefont {Ott}, \citenamefont
  {Pécseli},\ and\ \citenamefont {Trulsen}}]{SPT_T}%
  \BibitemOpen
  \bibfield  {author} {\bibinfo {author} {\bibfnamefont {J.~B.}\ \bibnamefont
  {Jørgensen}}, \bibinfo {author} {\bibfnamefont {J.}~\bibnamefont {Mann}},
  \bibinfo {author} {\bibfnamefont {S.}~\bibnamefont {Ott}}, \bibinfo {author}
  {\bibfnamefont {H.~L.}\ \bibnamefont {Pécseli}}, \ and\ \bibinfo {author}
  {\bibfnamefont {J.}~\bibnamefont {Trulsen}},\ }\href {\doibase
  10.1063/1.1863259} {\bibfield  {journal} {\bibinfo  {journal} {Phys. Fluids}\
  }\textbf {\bibinfo {volume} {17}},\ \bibinfo {pages} {035111} (\bibinfo
  {year} {2005})}\BibitemShut {NoStop}%
\bibitem [{\citenamefont {Barato}\ and\ \citenamefont
  {Chetrite}(2015)}]{bara15c}%
  \BibitemOpen
  \bibfield  {author} {\bibinfo {author} {\bibfnamefont {A.}~\bibnamefont
  {Barato}}\ and\ \bibinfo {author} {\bibfnamefont {R.}~\bibnamefont
  {Chetrite}},\ }\href {\doibase 10.1007/s10955-015-1283-0} {\bibfield
  {journal} {\bibinfo  {journal} {J. Stat. Phys.}\ }\textbf {\bibinfo {volume}
  {160}},\ \bibinfo {pages} {1154} (\bibinfo {year} {2015})}\BibitemShut
  {NoStop}%
\bibitem [{\citenamefont {Touchette}(2009)}]{LD}%
  \BibitemOpen
  \bibfield  {author} {\bibinfo {author} {\bibfnamefont {H.}~\bibnamefont
  {Touchette}},\ }\href {\doibase
  https://doi.org/10.1016/j.physrep.2009.05.002} {\bibfield  {journal}
  {\bibinfo  {journal} {Phys. Rep.}\ }\textbf {\bibinfo {volume} {478}},\
  \bibinfo {pages} {1} (\bibinfo {year} {2009})}\BibitemShut {NoStop}%
\bibitem [{\citenamefont {Touchette}(2018)}]{LD2}%
  \BibitemOpen
  \bibfield  {author} {\bibinfo {author} {\bibfnamefont {H.}~\bibnamefont
  {Touchette}},\ }\href {\doibase https://doi.org/10.1016/j.physa.2017.10.046}
  {\bibfield  {journal} {\bibinfo  {journal} {Physica A}\ }\textbf {\bibinfo
  {volume} {504}},\ \bibinfo {pages} {5} (\bibinfo {year} {2018})},\ \bibinfo
  {note} {lecture Notes of the 14th International Summer School on Fundamental
  Problems in Statistical Physics}\BibitemShut {NoStop}%
\bibitem [{\citenamefont {Gardiner}(2004)}]{gardiner_c.w._handbook_1985}%
  \BibitemOpen
  \bibfield  {author} {\bibinfo {author} {\bibfnamefont {C.~W.}\ \bibnamefont
  {Gardiner}},\ }\href {\doibase 10.1007/978-3-662-05389-8} {\emph {\bibinfo
  {title} {Handbook of Stochastic Methods}}},\ \bibinfo {edition} {3rd}\ ed.\
  (\bibinfo  {publisher} {Springer},\ \bibinfo {address} {Berlin},\ \bibinfo
  {year} {2004})\BibitemShut {NoStop}%
\bibitem [{\citenamefont {Gillespie}(1977)}]{gillespie_exact_1977}%
  \BibitemOpen
  \bibfield  {author} {\bibinfo {author} {\bibfnamefont {D.~T.}\ \bibnamefont
  {Gillespie}},\ }\href {\doibase 10.1021/j100540a008} {\bibfield  {journal}
  {\bibinfo  {journal} {J. Phys. Chem.}\ }\textbf {\bibinfo {volume} {81}},\
  \bibinfo {pages} {2340} (\bibinfo {year} {1977})}\BibitemShut {NoStop}%
\bibitem [{\citenamefont {Schnakenberg}(1976)}]{RevModPhysS}%
  \BibitemOpen
  \bibfield  {author} {\bibinfo {author} {\bibfnamefont {J.}~\bibnamefont
  {Schnakenberg}},\ }\href {\doibase 10.1103/RevModPhys.48.571} {\bibfield
  {journal} {\bibinfo  {journal} {Rev. Mod. Phys.}\ }\textbf {\bibinfo {volume}
  {48}},\ \bibinfo {pages} {571} (\bibinfo {year} {1976})}\BibitemShut
  {NoStop}%
\bibitem [{\citenamefont {Conway}(1985)}]{conway_course_1985}%
  \BibitemOpen
  \bibfield  {author} {\bibinfo {author} {\bibfnamefont {J.~B.}\ \bibnamefont
  {Conway}},\ }\href {\doibase 10.1007/978-1-4757-3828-5} {{\selectlanguage
  {english}\emph {\bibinfo {title} {A Course in Functional Analysis}}}},\
  Graduate Texts in Mathematics\ (\bibinfo  {publisher} {Springer-Verlag},\
  \bibinfo {address} {New York},\ \bibinfo {year} {1985})\BibitemShut {NoStop}%
\bibitem [{\citenamefont {van Kampen}(2007)}]{kampen_ng_van_stochastic_2007}%
  \BibitemOpen
  \bibfield  {author} {\bibinfo {author} {\bibfnamefont {N.~G.}\ \bibnamefont
  {van Kampen}},\ }\href {\doibase 10.1016/B978-044452965-7/50006-4} {\emph
  {\bibinfo {title} {Stochastic Processes in Physics and Chemistry}}},\
  \bibinfo {edition} {3rd}\ ed.,\ North-Holland Personal Library\ (\bibinfo
  {publisher} {Elsevier},\ \bibinfo {address} {Amsterdam},\ \bibinfo {year}
  {2007})\BibitemShut {NoStop}%
\bibitem [{\citenamefont {Risken}(1996)}]{risken_fokker-planck_1996}%
  \BibitemOpen
  \bibfield  {author} {\bibinfo {author} {\bibfnamefont {H.}~\bibnamefont
  {Risken}},\ }\href {\doibase 10.1007/978-3-642-61544-3} {{\selectlanguage
  {english}\emph {\bibinfo {title} {The {Fokker}-{Planck} {Equation}: {Methods}
  of {Solution} and {Applications}}}}},\ \bibinfo {edition} {2nd}\ ed.,\
  Springer {Series} in {Synergetics}\ (\bibinfo  {publisher}
  {Springer-Verlag},\ \bibinfo {address} {Berlin Heidelberg},\ \bibinfo {year}
  {1996})\BibitemShut {NoStop}%
\bibitem [{\citenamefont {Stoller}\ \emph {et~al.}(1991)\citenamefont
  {Stoller}, \citenamefont {Happer},\ and\ \citenamefont {Dyson}}]{Dyson}%
  \BibitemOpen
  \bibfield  {author} {\bibinfo {author} {\bibfnamefont {S.~D.}\ \bibnamefont
  {Stoller}}, \bibinfo {author} {\bibfnamefont {W.}~\bibnamefont {Happer}}, \
  and\ \bibinfo {author} {\bibfnamefont {F.~J.}\ \bibnamefont {Dyson}},\ }\href
  {\doibase 10.1103/PhysRevA.44.7459} {\bibfield  {journal} {\bibinfo
  {journal} {Phys. Rev. A}\ }\textbf {\bibinfo {volume} {44}},\ \bibinfo
  {pages} {7459} (\bibinfo {year} {1991})}\BibitemShut {NoStop}%
\bibitem [{Note1()}]{Note1}%
  \BibitemOpen
  \bibinfo {note} {Note that $0$ is an eigenvalue of $\protect \hat {L}$ and
  $\protect \tilde {\protect \mathcal {P}}^{\protect \bm {\psi }}_s(\protect
  \mathbf {u}|p_0)$ is meromorphic for small $|\protect \mathbf {u}|$ therefore
  we can assume, without loss of generality, that $\protect \mathbf {u}$ is
  real.}\BibitemShut {Stop}%
\bibitem [{\citenamefont {Marcinkiewicz}(1939)}]{Marcinkiewicz}%
  \BibitemOpen
  \bibfield  {author} {\bibinfo {author} {\bibfnamefont {J.}~\bibnamefont
  {Marcinkiewicz}},\ }\href {http://eudml.org/doc/168829} {\bibfield  {journal}
  {\bibinfo  {journal} {Math. Z.}\ }\textbf {\bibinfo {volume} {44}},\ \bibinfo
  {pages} {612} (\bibinfo {year} {1939})}\BibitemShut {NoStop}%
\bibitem [{\citenamefont {Sakurai}\ and\ \citenamefont
  {Napolitano}(2017)}]{Sakurai}%
  \BibitemOpen
  \bibfield  {author} {\bibinfo {author} {\bibfnamefont {J.~J.}\ \bibnamefont
  {Sakurai}}\ and\ \bibinfo {author} {\bibfnamefont {J.}~\bibnamefont
  {Napolitano}},\ }\href {\doibase 10.1017/9781108499996} {\emph {\bibinfo
  {title} {Modern Quantum Mechanics}}}\ (\bibinfo  {publisher} {Cambridge
  University Press},\ \bibinfo {year} {2017})\BibitemShut {NoStop}%
\bibitem [{\citenamefont {Klein}(1974)}]{Klein}%
  \BibitemOpen
  \bibfield  {author} {\bibinfo {author} {\bibfnamefont {D.~J.}\ \bibnamefont
  {Klein}},\ }\href {\doibase 10.1063/1.1682018} {\bibfield  {journal}
  {\bibinfo  {journal} {J. Chem. Phys.}\ }\textbf {\bibinfo {volume} {61}},\
  \bibinfo {pages} {786} (\bibinfo {year} {1974})}\BibitemShut {NoStop}%
\bibitem [{\citenamefont {Hartich}\ and\ \citenamefont
  {Godec}(2019)}]{hart19a}%
  \BibitemOpen
  \bibfield  {author} {\bibinfo {author} {\bibfnamefont {D.}~\bibnamefont
  {Hartich}}\ and\ \bibinfo {author} {\bibfnamefont {A.}~\bibnamefont
  {Godec}},\ }\href {\doibase 10.1088/1751-8121/ab1eca} {\bibfield  {journal}
  {\bibinfo  {journal} {J. Phys. A: Math. Theor.}\ }\textbf {\bibinfo {volume}
  {52}},\ \bibinfo {pages} {244001} (\bibinfo {year} {2019})}\BibitemShut
  {NoStop}%
\bibitem [{\citenamefont {Khinchin}(1929)}]{Khinchin}%
  \BibitemOpen
  \bibfield  {author} {\bibinfo {author} {\bibfnamefont {A.~I.}\ \bibnamefont
  {Khinchin}},\ }\href {https://ci.nii.ac.jp/naid/10017470734/en/} {\bibfield
  {journal} {\bibinfo  {journal} {Comptes rendus de l'Academie des Sciences}\
  }\textbf {\bibinfo {volume} {189}},\ \bibinfo {pages} {477} (\bibinfo {year}
  {1929})}\BibitemShut {NoStop}%
\bibitem [{\citenamefont {Holubec}\ \emph {et~al.}(2019)\citenamefont
  {Holubec}, \citenamefont {Kroy},\ and\ \citenamefont {Steffenoni}}]{holu19}%
  \BibitemOpen
  \bibfield  {author} {\bibinfo {author} {\bibfnamefont {V.}~\bibnamefont
  {Holubec}}, \bibinfo {author} {\bibfnamefont {K.}~\bibnamefont {Kroy}}, \
  and\ \bibinfo {author} {\bibfnamefont {S.}~\bibnamefont {Steffenoni}},\
  }\href {\doibase 10.1103/PhysRevE.99.032117} {\bibfield  {journal} {\bibinfo
  {journal} {Phys. Rev. E}\ }\textbf {\bibinfo {volume} {99}},\ \bibinfo
  {pages} {032117} (\bibinfo {year} {2019})}\BibitemShut {NoStop}%
\bibitem [{\citenamefont {Guennebaud}\ \emph {et~al.}(2010)\citenamefont
  {Guennebaud}, \citenamefont {Jacob} \emph {et~al.}}]{eigenweb}%
  \BibitemOpen
  \bibfield  {author} {\bibinfo {author} {\bibfnamefont {G.}~\bibnamefont
  {Guennebaud}}, \bibinfo {author} {\bibfnamefont {B.}~\bibnamefont {Jacob}},
  \emph {et~al.},\ }\href@noop {} {\enquote {\bibinfo {title} {Eigen v3},}\
  }\bibinfo {howpublished} {http://eigen.tuxfamily.org} (\bibinfo {year}
  {2010})\BibitemShut {NoStop}%
\bibitem [{\citenamefont {Toyabe}\ \emph {et~al.}(2011)\citenamefont {Toyabe},
  \citenamefont {Watanabe-Nakayama}, \citenamefont {Okamoto}, \citenamefont
  {Kudo},\ and\ \citenamefont {Muneyuki}}]{toya11}%
  \BibitemOpen
  \bibfield  {author} {\bibinfo {author} {\bibfnamefont {S.}~\bibnamefont
  {Toyabe}}, \bibinfo {author} {\bibfnamefont {T.}~\bibnamefont
  {Watanabe-Nakayama}}, \bibinfo {author} {\bibfnamefont {T.}~\bibnamefont
  {Okamoto}}, \bibinfo {author} {\bibfnamefont {S.}~\bibnamefont {Kudo}}, \
  and\ \bibinfo {author} {\bibfnamefont {E.}~\bibnamefont {Muneyuki}},\ }\href
  {\doibase 10.1073/pnas.1106787108} {\bibfield  {journal} {\bibinfo  {journal}
  {Proc. Natl. Acad. Sci. USA}\ }\textbf {\bibinfo {volume} {108}},\ \bibinfo
  {pages} {17951} (\bibinfo {year} {2011})}\BibitemShut {NoStop}%
\bibitem [{\citenamefont {Godec}\ and\ \citenamefont
  {Metzler}(2015)}]{GodecMetz}%
  \BibitemOpen
  \bibfield  {author} {\bibinfo {author} {\bibfnamefont {A.}~\bibnamefont
  {Godec}}\ and\ \bibinfo {author} {\bibfnamefont {R.}~\bibnamefont
  {Metzler}},\ }\href {\doibase 10.1103/PhysRevE.92.010701} {\bibfield
  {journal} {\bibinfo  {journal} {Phys. Rev. E}\ }\textbf {\bibinfo {volume}
  {92}},\ \bibinfo {pages} {010701} (\bibinfo {year} {2015})}\BibitemShut
  {NoStop}%
\bibitem [{\citenamefont {Godec}\ and\ \citenamefont
  {Metzler}(2016)}]{GodecMetz2}%
  \BibitemOpen
  \bibfield  {author} {\bibinfo {author} {\bibfnamefont {A.}~\bibnamefont
  {Godec}}\ and\ \bibinfo {author} {\bibfnamefont {R.}~\bibnamefont
  {Metzler}},\ }\href {\doibase 10.1088/1751-8113/49/36/364001} {\bibfield
  {journal} {\bibinfo  {journal} {J. Phys. A: Math. Theor.}\ }\textbf {\bibinfo
  {volume} {49}},\ \bibinfo {pages} {364001} (\bibinfo {year}
  {2016})}\BibitemShut {NoStop}%
\end{thebibliography}%

\end{document}